\documentclass[preprint,12pt,onecolumn]{emulateapj}

\usepackage{amsmath}
\usepackage{ulem}
\usepackage[usenames]{color}
\usepackage{multirow}
\usepackage{bm}
\usepackage{hyperref}
\shorttitle{Prelude to a double degenerate merger}
\shortauthors{Dan, Rosswog, Guillochon and Ramirez-Ruiz}

\begin{document}
\title{Prelude to a double degenerate merger: the onset of mass transfer and
  its impact on gravitational waves and surface detonations}
\author{Marius Dan\altaffilmark{1}, Stephan Rosswog\altaffilmark{1}, James
  Guillochon\altaffilmark{2}, Enrico Ramirez-Ruiz\altaffilmark{2}} 
\altaffiltext{1}{School of Engineering and Science, Jacobs University
Bremen, Campus Ring 1, 28759 Bremen, Germany; m.dan@jacobs-university.de, rosswog@jacobs-university.de}
\altaffiltext{2}{TASC, Department of Astronomy and Astrophysics, University
of California, Santa Cruz, CA 95064; jfg@ucolick.org, enrico@ucolick.org}
\newcommand{\Nifs}{\ensuremath{^{56}\mathrm{Ni}}}
\def\paren#1{\left( #1 \right)}
\def\Mesz{M\'esz\'aros~}
\def\Pacz{Paczy\'nski~}
\def\Kluz{Klu\'zniak~}
\def\p{$e^\pm \;$}
\def\msun{$M_{\odot}$}
\def\Msun{$M_{\odot}$ }
\def\be{\begin{equation}}
\def\ee{\end{equation}}
\def\ben{\begin{enumerate}}
\def\een{\end{enumerate}}
\def\bea{\begin{eqnarray}}
\def\eea{\end{eqnarray}}
\def\gcc{g cm$^{-3}$ }
\def\edo{\end{document}}
\def\red{\textcolor{red}}
\begin{abstract}
We present the results of a systematic numerical study of the onset of mass
transfer in double degenerate binary systems and its impact on the subsequent
evolution. All investigated systems belong to the regime of direct impact, unstable
mass transfer. In all of the investigated cases, even those considered
unstable by conventional stability analysis, we find a long-lived mass
transfer phase continuing for as many as several dozen orbital periods. This
settles a recent debate sparked by a discrepancy between earlier SPH
calculations that showed disruptions after a few orbital periods and newer
grid-based studies in which mass transfer continued for tens of orbits.  
The number of orbits a binary survives sensitively depends on the exact
initial conditions. We find that the approximate initial conditions that have
been used in most previous SPH calculations have a serious impact on all
stages of the evolution from the onset of mass transfer up to the final
structure of the remnant. We compare ``approximate'' initial conditions where
spherical stars are placed at an initial separation obtained from an estimate
of the Roche lobe size with ``accurate'' initial conditions that were
constructed by carefully driving the binary system to equilibrium by a
relaxation scheme. 
Simulations that use the approximate initial conditions underestimate the
initial separation when mass transfer sets in, which yields a binary that only
survives for only few orbits and thus a rapidly fading gravitational wave
signal. Conversely, the accurate initial conditions produce a binary system in
which the mass transfer phase is extended by almost two orders of magnitude in
time, resulting in a gravitational wave signal with amplitude and frequency
that remain essentially constant up until merger. The mass transfer also shows
a unique oscillatory signal that is most pronounced for binary systems that
are marginally unstable.
As we show that these binaries can survive at small separation for hundreds of
orbital periods, their associated gravitational wave 
signal should be included when calculating the gravitational wave foreground
(although expected to below LISA's sensitivity at  these high frequencies).
We also show that the inclusion of the entropy increase associated with
shock-heating of the accreted material reduces the number of orbits a binary
survives given the same initial conditions, although the effect is not as
pronounced when using the appropriate initial conditions. The use of accurate
initial conditions and a correct treatment of shock heating allows for a reliable
 time evolution of the temperature, density, and
angular momentum, which are important when considering thermonuclear events
that may occur during the mass transfer phase and/or after merger. 
Our treatment allows us to accurately identify when surface detonations may
occur in the lead-up to the merger, as well as the properties of final merger
products. 
\end{abstract}
\section{Introduction}
White dwarfs (WDs) are the end point of the evolution of the vast majority of the stars in the universe.
Our Galaxy harbours of order $10^{10}$ WDs \citep{napiwotzki09} and about 
$2.5 \times 10^8$ close binary WDs \citep{nelemans01a}, often referred to as double degenerates (DD). 
For about half of them a Hubble time is long enough so that gravitational wave emission can drive them 
to a phase of mass transfer. Such interacting DDs are suspected to produce observable objects 
such as sdb stars \citep{webbink84,iben86,saio00,han03}, R Corona Borealis stars \citep{webbink84,clayton07},
AM CVn stars \citep{paczynski67,nelemans01b,solheim10} and, maybe most spectacularly, type Ia supernovae
\citep{iben84,webbink84}.
While for many years there was a strong inclination towards the single degenerate scenario, the recent 
discovery of ``super-Chandrasekhar'' events \citep{howell06,hicken07,yuan07,silverman11} has re-strengthened 
the interest in DD mergers as possible type Ia progenitors.
The lead-up to a merger includes a phase of mass transfer that is crucial for
understanding the evolution and fates of double WD systems. This phase will
affect the orbital evolution, the angular momentum and thermodynamical state
of the final merger product, the associated gravitational wave signal, the
environment into which any potential ejecta produced by an thermonuclear
explosion will expand into, and whether or not a particular binary will merge
at all. The numerical study of the onset of this mass transfer phase is the
main topic of this investigation. 

The merger process of two WDs has been modeled by a number of groups 
\citep{benz90b,rasio95,segretain97,guerrero04,yoon07a,pakmor10}, all of which
use the smoothed particle hydrodynamics (SPH) method. In all of these
simulations 
the mass transfer phase lasted for very few orbital time scales. Recently, grid-based 
simulations \citep{motl02,dsouza06,motl07} were performed in which the authors carefully
tried to reduce the angular momentum non-conservation due to advection errors 
that often plague grid-based codes \citep{new97}. The latter results differed
from the previous SPH calculations in that they showed long-lived mass transfer for
many orbital periods and this has sparked some discussion about the stability of mass 
transfer in double degenerate systems during these late stages. Based on our experience 
in modeling the dynamics of neutron star black hole systems \citep{rosswog04b,rosswog05a}, 
we had suspected that double WD binaries may be very sensitive to the exact initial 
conditions. As we will demonstrate below, the initial conditions used have a
decisive impact and are one of the major reasons for the disagreement between
various mass transfer calculations. 

This is the companion paper of \cite{guillochon10}, in which we focused on the
fate of accreted helium in unstable, direct impact systems. The work showed
that, contrary to prior belief, the mass transfer phase between two WDs may be
eventful, and can be accompanied by Kelvin-Helmholtz-triggered thermonuclear
surface explosions. These explosions may lead to the ignition of the accreting
CO-WD in a double detonation scenario, which could be the long sought-after
mechanism to initiate a type Ia supernova from WD binary merger. For
the described work we had chosen a hybrid approach by combining results using
the SPH method for 
the orbital evolution and the adaptive mesh refinement code FLASH
\citep{fryxell00} to focus on the evolution 
of the low-density helium material. We consider this a ``best-of-both-worlds'' approach, because it combines
the major strengths of both Lagrangian and Eulerian methods. In this paper we present the details of the mass 
transfer/orbital evolution calculations that have been obtained with SPH, as
well as a detailed comparison with the previously published FLASH
calculations. 

The main focus of this paper is the onset of the mass transfer and its impact on the orbital evolution; the
structure of the resulting remnants will be discussed elsewhere. In Section \ref{sec:MT_stability} we briefly 
summarize conditions for the stability of mass transfer, in Section \ref{sec:numerics} we focus on the 
numerical aspects of the mass transfer modeling. We describe in detail how we construct our initial 
conditions and demonstrate their impact on the outcome of essentially every aspect of double degenerate 
merger simulations. We also briefly summarize the essential features of the SPH code that we are using. 
In a set of test calculations we explore to which extent the numerical resolution, the form of artificial 
viscosity, and the exact numerical initial conditions affect the results. In
Section \ref{sec:results} we highlight results from two systems representative
of a set of binary evolution calculations. In Sections \ref{sec:gwave} and
\ref{sec:surfdet} we discuss the implications of the extended mass transfer
phase on the gravitational wave signal and the triggering of surface
detonations prior to merger, respectively. We conclude the paper in Section
\ref{sec:summary} with a summary. 

\section{Stability of mass transfer}
\label{sec:MT_stability}
In what follows we briefly collect the standard arguments on the stability of mass transfer \citep[see e.g.][]{frank02}.
Once stars are close enough that mass transfer via Roche lobe overflow can set in, the further evolution
of the binary system depends on how both the orbit/Roche lobe size and the star react to the re-arrangement 
of mass and angular momentum. Due to their particular mass radius relation, WDs react to mass loss 
by expansion which tends to enhance the mass transfer rate. If mass is transferred to the heavier star, the
orbit has to widen to ensure the conservation of the centre of mass and, therefore, mass transfer is tendentially 
quenched. The interplay of both effects determines the detailed orbital evolution.
The total angular momentum of a point mass binary is given by
\be
J= M_1 M_2 \left(\frac{G a}{M_{\rm tot}}\right)^{1/2},
\ee
where the $M_i$ are the masses, $G$ is the gravitational constant, $M_{\rm tot}$ is the total mass and $a$ the 
separation of the binary system. In the remainder of this work we will always denote the donor by label 2, the 
accretor by 1. On logarithmic differentiation one finds 
\be
\frac{\dot{a}}{2 a}= \frac{\dot{J}}{J} - \frac{\dot{M_2}}{M_2} \left( 1 - q \right),\label{eq:a_dot}
\ee
where $q= M_2/M_1$. As expected, angular momentum loss tends ($\dot{J}<0$) to shrink the orbit, while mass 
transfer ($\dot{M}_2<0$) tends to increase the orbital separation. The change of the donor Roche lobe size 
can be related to the change in orbital separation by logarithmically differentiating a simple Roche lobe 
size estimate \citep{paczynski71}
\be
R_{\rm L,2}= 0.462 \; a \; \left( \frac{q}{1+q}\right)^{1/3} ,
\ee
which yields
\be
\frac{\dot{R}_{\rm L,2}}{R_{\rm L,2}}= \frac{\dot{a}}{a} + \frac{1}{3} \frac{\dot{M}_2}{M_2}.
\ee
Using this to eliminate $\dot{a}$ from Equation (\ref{eq:a_dot}) then yields
\be
\frac{\dot{R}_{\rm L,2}}{R_{\rm L,2}}= \frac{2 \dot{J}}{J} + \frac{2 \dot{M}_2}{M_2} \left(q-\frac{5}{6}\right),
\ee
and therefore for a mass ratio $q>5/6$ a non-zero mass loss will trigger an expansion of the donor 
WD and at the same time shrink the Roche lobe size $R_{\rm L,2}$. This will be further accelerated by 
any additional angular momentum loss, so with the used assumptions the mass transfer is clearly
unstable in such cases.
\begin{figure}[!h]
\centerline{
\includegraphics[height=4.2in]{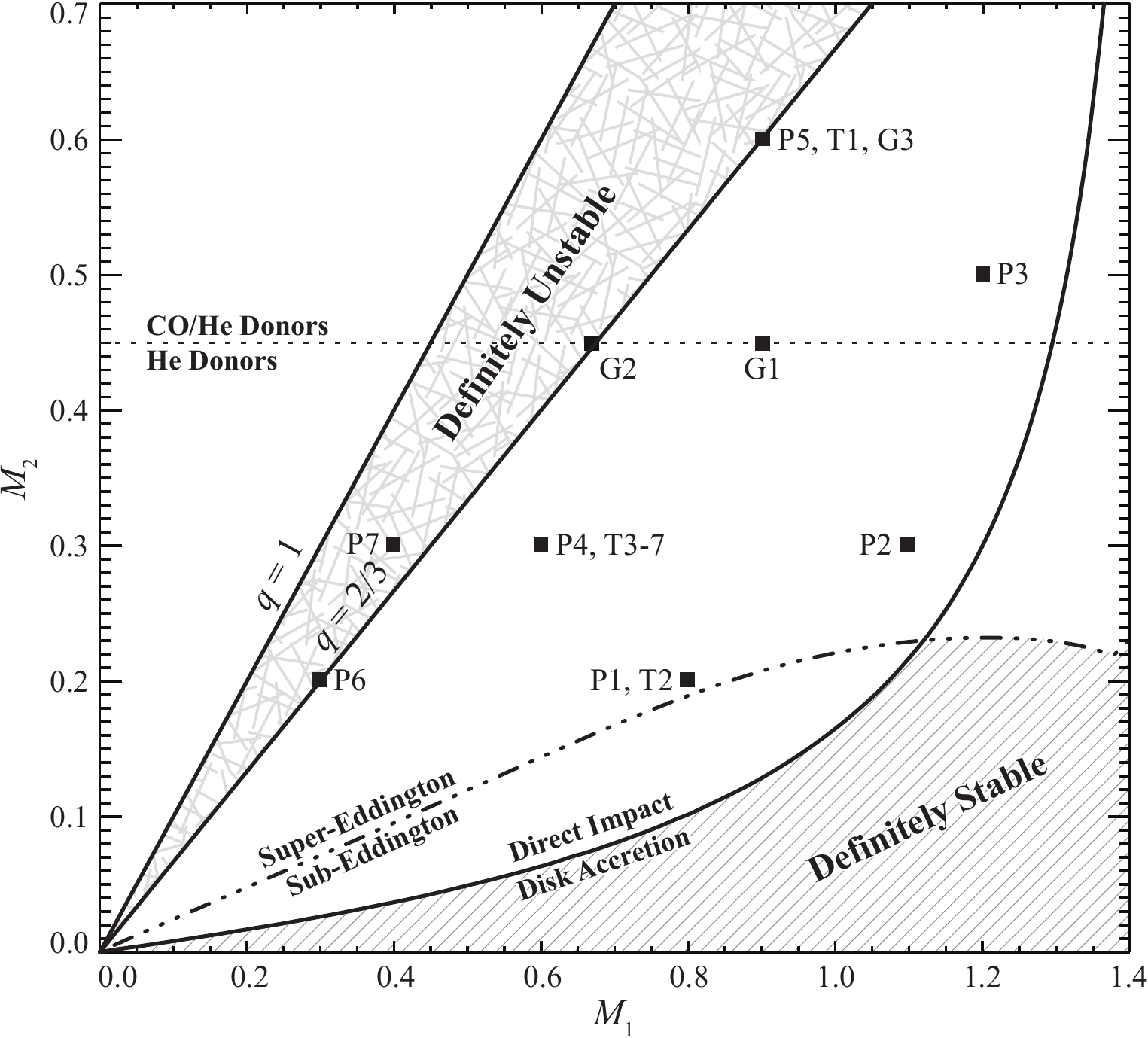}
}
\caption{Analytical estimates of mass transfer stability in double WD
  binaries. The region below the lower solid line indicates the regime where a
  disk forms. Above the line, the accretion stream directly impacts on the
  accretor. The dot-dashed line separates systems that undergo sub-
  vs. super-Eddington accretion, with super-Eddington accretion possibly
  leading to an unstable binary. The hatched region of the plot indicates
  systems for which the accretion is both sub-Eddington and through a disk,
  which are almost certainly stable. For mass ratios $q$ larger than $2/3$,
  the mass transfer is guaranteed to be unstable. The area between the regions
  of guaranteed stability and instability corresponds to the systems that can
  be either stable or unstable, depending on the spin-orbit coupling. The
  filled squares indicate the masses of the systems simulated in this
  work, see \ref{tab:runs}. Adapted from \cite{marsh04}.} 
\label{fig:marsh04}
\end{figure}

Of course, the so far discussed case is highly idealized. A point mass binary
was assumed where the mass lost by the donor was completely incorporated into
the accretor, while in reality mass can be lost from the system. Depending on
whether or not the circularization radius exceeds the accretor radius, the
transferred mass can either form a disk or it can impact directly onto the
accretor surface. If a disk is formed, it would be tidally perturbed  
by the companion star and can thus feed back some fraction of the angular
momentum into the orbit. In the direct impact case most of the angular
momentum is invested into spinning up the accretor, which removes orbital
angular momentum from and is therefore expected to destabilize the orbit.
It should be noted, however, that a deformed donor can also feed back some angular
momentum into the orbit. The orbit and the different binary components can exchange angular momentum both
by advection of mass and by tides. \cite{marsh04} have generalized the simple treatment leading to Equation (\ref{eq:a_dot}) 
for direct impact and tidal spin evolution
\be
\frac{\dot{a}}{2 a}= \frac{\dot{J}}{J_{\rm orb}} - \left( \frac{\dot{J}_1 + \dot{J}_2}
{J_{\rm orb}}\right)_{\rm tid} - \frac{\dot{M_2}}{M_2} \left[ 1 - q - \sqrt{(1+q) r_{\rm c}}\right],
\label{eq:a_dot_marsh}
\ee
where $r_{\rm c}$ is the circularization radius in units of the orbital separation. The second term accounts 
for the exchange of angular momentum due to tides, the third term accounts for the spin of the accretor. If 
tides and accretor spin are ignored, Equation (\ref{eq:a_dot_marsh}) reduces to (\ref{eq:a_dot}). In 
Figure \ref{fig:marsh04} 
we show the stability criteria for binaries of different donor and accretor
masses, superimposed with the initial masses of our simulations (Table
\ref{tab:runs}). \cite{marsh04} find the upper left of the diagram to be
guaranteed unstable, the lower  
right guaranteed stable and an uncertain region along the diagonal of the
diagram. In this work we mainly investigate the ``uncertain'' region
in the diagram.

\section{Numerical modeling of mass transfer}
\label{sec:numerics}
The main tool of this investigation is a 3D smoothed particle hydrodynamics (SPH)
code. For recent
reviews of the method see e.g. \cite{monaghan05} and \cite{rosswog09b}. The orbital
dynamics of a binary system is very sensitive to the (re-)distribution of angular momentum and 
therefore a numerical simulation has to ensure very accurate angular momentum conservation.
Its purely Lagrangian nature and the built-in conservation of mass, energy, linear 
and angular momentum \citep[e.g. Section 2.4 in][]{rosswog09b} make SPH a very powerful tool 
for this type of study. The conservation is exact up to effects that result from finite 
accuracy in gravitational forces due to the use of a tree, see below, and in the numerical 
integration of the ODE set. Both issues, however, are fully under control and 
can be improved to a desired level of accuracy by choosing the corresponding parameters 
for the tree and the ODE integration in an appropriate way, 
at the expense of a larger computational effort, though. For our parameters,
even in the 84-orbit 
P1, corresponding to as many as 16,930 dynamical time scales of the 0.8 \Msun star, energy and angular 
momentum are conserved to better than 1\%. It has to be stated, however, that
the advantage of {\it exact} mass  
conservation via constant SPH particle masses turns into a disadvantage if we are interested
in the resolution of low-density portions of the flow. That is why we turned
to the AMR code FLASH in \cite{guillochon10} to investigate the settling of
the transferred material onto the accretor. 

Details of our SPH code can be found in \cite{rosswog08b}. It uses an artificial viscosity scheme 
\citep{morris97} that reduces the dissipation terms to a very low level away from discontinuities, 
together with a switch \citep{balsara95} to suppress the spurious viscous 
forces in pure shear flows. 
The system of fluid equations is closed by the HELMHOLTZ equation of
state \citep{timmes00a}. It accepts an externally calculated nuclear
composition and allows a convenient coupling to a nuclear reaction network.
We use a minimal nuclear reaction network \citep{hix98} to determine the evolution 
of the nuclear composition and to include the energetic feedback 
onto the gas from the nuclear reactions. A set of only seven abundance groups 
greatly reduces the computational burden, but still reproduces the energy generation of 
all burning stages from He burning to NSE accurately. We use a binary tree \citep{benz90b} 
to search for the neighbor particles and to calculate the gravitational forces.

\begin{table}[!h]
\centerline{
\scriptsize\begin{tabular}{|c|c|c|c|c|c|c|c|c|c|c|c|c|c|}
\hline\hline
Run & Masses & \multirow{2}{*}{Comp.} & $N_{\rm p}$ & \multirow{2}{*}{$a_{\rm 0,9}$} & $P_0$ &
\multirow{2}{*}{$T_{\rm max,8}$} & \multirow{2}{*}{$\rho_{\rm max,7}$} &
$L_{\rm unb}$ & $M_{\rm unb}$ &
\multirow{2}{*}{$N_{\rm orb}$} & \multirow{2}{*}{$a_{\rm dis,9}$} & $M_{\rm trans}$ & \multirow{2}{*}{Comment} \\ 
No. & [$M_{\odot}$] & & [$10^5$] & & [s] & & & $[\%L_{\rm
 tot}]$ & [$10^{-3}M_{\odot}$] & & & [$M_{\odot}$] & \\ 
\hline\hline
\multicolumn{14}{|c|}{Production runs} \\
\hline\hline
P1 & $0.2+0.8$ & He-CO & 2 & 5.77 & 239 & 4.89 & 0.95
& 5.7 & 8.6 & 84 & 6.15 & 0.08 & See \ref{sec:dependence_IC}, \ref{sec:binary1}\\
\hline
P2 & $0.3+1.1$ &He-CO & 2 & 4.75 &151 & 12.23
& 6.15 & 7.11 & 17.6 & 59 & 5.01& 0.13 & \\
\hline
P3 & $0.5+1.2$ & He-ONeMg & 2 & 3.35 & 81 &18.85 & 15.44 & 4.86 
& 17.4 & 31 & 3.27 & 0.09 & \\
\hline
P4 & $0.3+0.6$ & He-CO & 2 & 4.04 & 148 &
4.92 & 0.32 & 2.5 & 3.7 & 45 & 4.07 & 0.07 & \\
\hline
P5 & \multirow{2}{*}{$0.6+0.9$} & \multirow{2}{*}{CO-CO} & \multirow{2}{*}{2}
& \multirow{2}{*}{2.68} & \multirow{2}{*}{62} & \multirow{2}{*}{10.6} &
\multirow{2}{*}{1.75} 
& \multirow{2}{*}{0.8} & \multirow{2}{*}{1.85} & \multirow{2}{*}{29} &
\multirow{2}{*}{2.58} & \multirow{2}{*}{0.10} & See \ref{sec:dependence_IC},
\ref{sec:binary2}\\
G3 &  &  &  &  &  &  & 
&  &  &  &  &  & \citeauthor{guillochon10}\\
\hline
P6 & $0.2+0.3$ & He-He & 2 & 4.39 & 224 & 2.37 & 0.05 & 
1.03 & 0.05 & 22 & 4.29 & 0.05 & \\
\hline
P7 & $0.3+0.4$ & He-He & 2 & 3.60 & 141& 2.72 & 0.11 & 0.6 & 0.65 &
14 & 3.41 & 0.06 & \\ 
\hline
P8 & $0.9+1.2$ & CO-CO & 2 & 1.91 & 31& 71.15 & 16.1 & 2.3 & 34.1 & 29 & 1.82 & 0.07 & \\
\hline\hline
\multicolumn{14}{|c|}{Test runs} \\
\hline\hline
G1 & $0.45+0.9$ & He-CO & 1 & 3.36 & 91 & 16.11 & 1.59 & 2.2 &
4.6 & 30 & 3.35 & 0.12 & \citeauthor{guillochon10} \\
\hline
G2 & $0.45+0.67$ & He-CO & 1 & 3.12 & 90 & 12.3 & 0.48
& 0.55 & 0.9 & 21 & 3.01 & 0.10 & \citeauthor{guillochon10} \\
\hline
T1 & $0.6+0.9$ & CO-CO & 2 & 2.31 & 49 & 16.4 & 1.83
& 0.48 & 2.01 & 1 & 2.02 & 0.15 & app. IC\\
\hline
T2 & $0.2+0.8$ & He-CO & 2 & 4.95 & 190 & 5.24 & 0.96
& 1.91 & 1.8 & 2 & 4.43 & 0.08 & app. IC\\
\hline
T3 & $0.3+0.6$ & He-CO & 2 & 3.89 & 139 & 5.59 & 0.33
& 0.4 & 0.5 & 6 & 3.64 & 0.08 & $a_0=1.13 a_0^{\rm Egg}$ \\
\hline
T4 & $0.3+0.6$ & He-CO & 0.4 & 4.00 & 145 & 4.99 & 0.35
& 2.1 & 2.6 & 20 & 4.0 & 0.07 & $AV_1$\\
\hline
T5 & $0.3+0.6$ & He-CO & 0.4 & 4.00 & 145 & 6.10 & 0.36
& 1.91 & 2.5 & 23 & 3.99 & 0.08 & $AV_2$\\
\hline
T6 & $0.3+0.6$ & He-CO & 0.4 & 4.00 & 145 & 4.70 & 0.33
& 2.09 & 2.6 & 18 & 4.00 & 0.08 & $AV_3$\\
\hline
T7 & $0.3+0.6$ & He-CO & 0.4 & 4.00 & 145 & 3.99 & 0.31
& 2.13 & 2.5 & 19 & 4.12 & 0.07 & $AV_4$\\
\hline\hline
\end{tabular}
}
\caption{Summary of the runs performed for this paper. $a_{\rm 0,9}$ is the
  initial separation in units of $10^9$ cm, $T_{{\rm max},8}$ is the peak
  temperature in units of $10^8$ K, $\rho_{{\rm max},7}$ the peak density in
  units of $10^7$ \gcc and $L_{\rm unb}$ is the fraction of angular momentum
  contained in material that is unbound at the end of the simulation. $a_{\rm
    dis,9}$ is the separation (in $10^9$ cm) at which tidal disruption of the
  donor occurs and $M_{\rm trans}$ is the amount of material (in \msun) lost
  by the donor prior to disruption.}
\label{tab:runs}
\end{table}

\subsection{Initial conditions}
In what follows we will summarize 
how initial conditions are constructed for double WD binary systems, both in this and in previous work.

\subsubsection{Previous work}
In their pioneering work, \cite{benz90b} followed the dynamical evolution of a $0.9 + 1.2\ M_\odot$
WD binary system. They constructed cold, isolated WDs in hydrostatic equilibrium and placed
them without spin-rotation at an initial separation such that the secondary's Roche radius was smaller than
its unperturbed initial radius by 8\%. One of the major conclusions of \citeauthor{benz90b} was that within about 
two orbital periods the secondary is completely destroyed and transformed into a thick accretion disk around 
the accretor. As we will show below, this short time to merger is an artifact of the initial 
separation being too small. We have followed the recipe of \citeauthor{benz90b} for the initial conditions and find exactly 
the same result as presented in their paper. We have further compared the orbital evolution of a 50,000 particle binary system, once
according to the original Benz et al. description and according to our scheme, as described in Section \ref{sec:our_IC}. Again, with their 
initial conditions the merger set in within about one orbital period. With our initial conditions and 
everything else being the same, the binary survived for 14 orbits of continued mass transfer. 

In \cite{segretain97} and \cite{yoon07a} the stars were placed at a mutual separation so that
the secondary fills its Roche lobe. In \citeauthor{segretain97} the WDs had no initial spin, while in \citeauthor{yoon07a}
the WDs begin tidally locked. \cite{guerrero04} and \cite{loren09} initially placed the stars in a detached
configuration and then slowly decreased the separation to the point where mass
transfer began. However, their relaxation procedure did not allow the stars to reach complete hydrostatic
equilibrium. If the stars are brought together too quickly, this can cause undesired (and unrealistic) oscillations in the WDs
and may again lead to an abbreviated mass transfer phase. For example, the mass transfer phase in their $0.4 + 1.2\ M_\odot$ 
system (Figure 10 in \cite{guerrero04}) only lasts for a single orbit.
This is also the case for the simulation of a $0.6 + 0.8\ M_\odot$ system 
\citep[see Figures 1 and 2 in][]{loren09}.

Our initial conditions are constructed in a way that is similar to
\cite{rasio95} (Section \ref{sec:our_IC}). \citeauthor{rasio95} constructed a
synchronized binary system with mass ratio  
of $q=0.5$, and also followed the dynamical evolution once a few particles had crossed the $L_1$ 
point. They found a merger after five orbital periods and also questioned the earlier 
use of non-equilibrium initial conditions. We have tried to reproduce their calculations as closely as
possible by adopting the same particle number, equation of state, and the same prescription for evolving the entropic 
function $K(s)$ (Equation (\ref{eq:entropic_function})), but still we only found a merger after fifteen orbital periods. 
\cite{fryer08} place the two stars close enough such that the donor is already
overfilling its Roche radius, which leads to a premature merger. The recent
work of \cite{pakmor10} focuses primarily on the coalescence of the binary,
but their initial conditions are not appropriate for studies of long-term
stability. 

\subsubsection{Our approach}
\label{sec:our_IC}
\begin{figure}[t]
\centerline{
 \includegraphics[width=\linewidth]{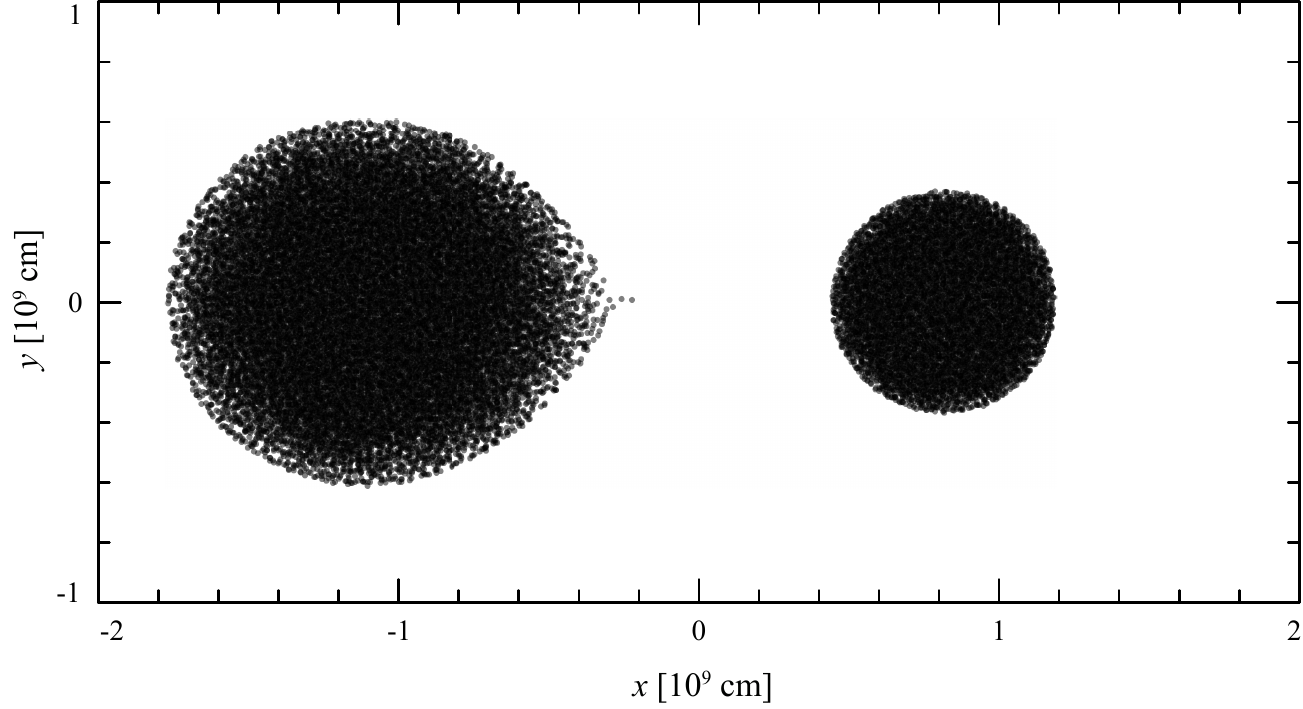}
}
\caption{Equilibrium configuration ($0.9 + 1.2\ M_\odot$) at the onset
of mass transfer. The figure shows a projection of the SPH particles onto the orbital
plane. Each particle's location within this plane is shown as a translucent grey circle.} 
\label{fig:IC1}
\end{figure}

\begin{figure}[t]
\centerline{
 \includegraphics[height=2.15in]{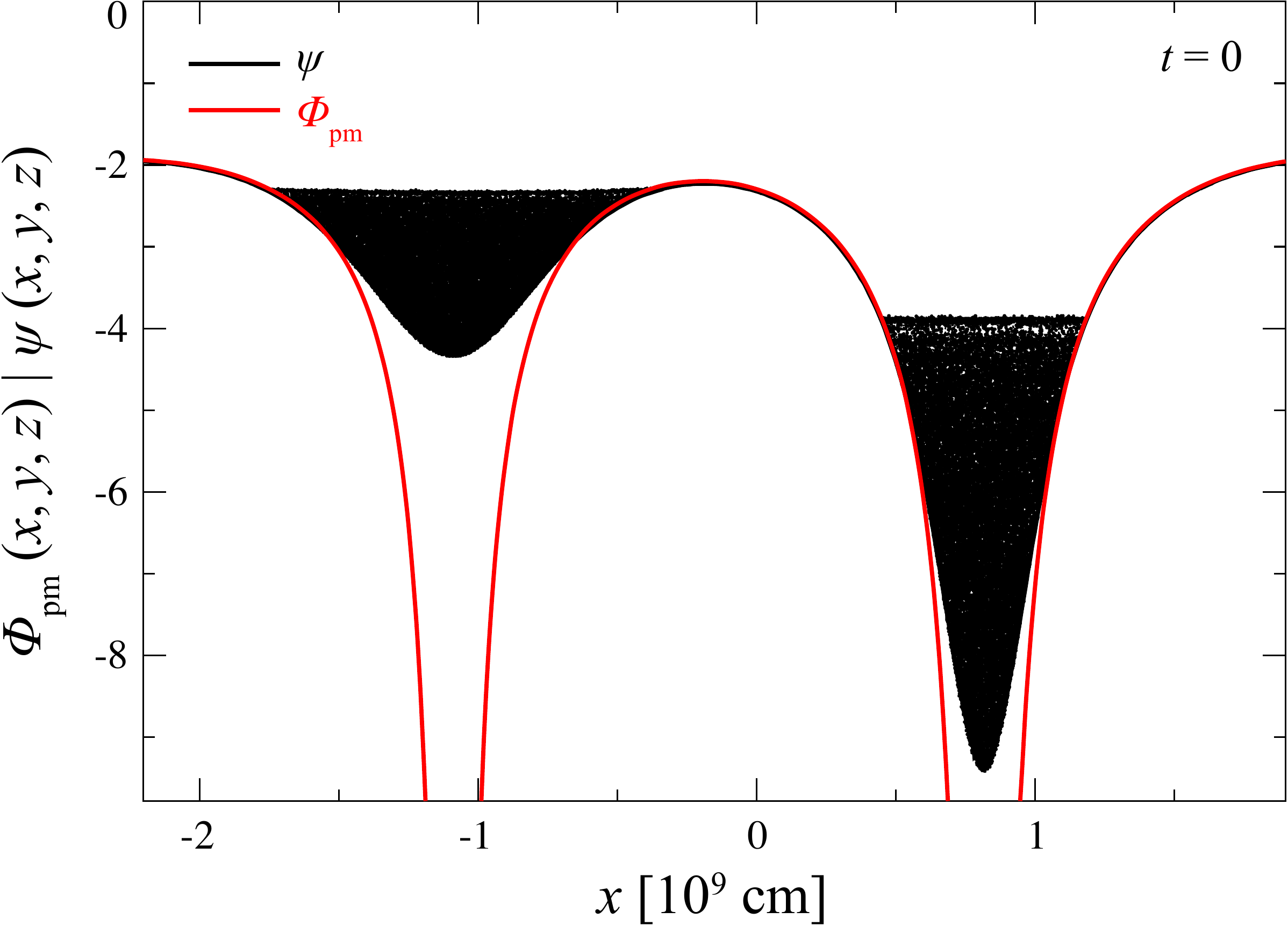}\hspace{0.5cm}
 \includegraphics[height=2.15in]{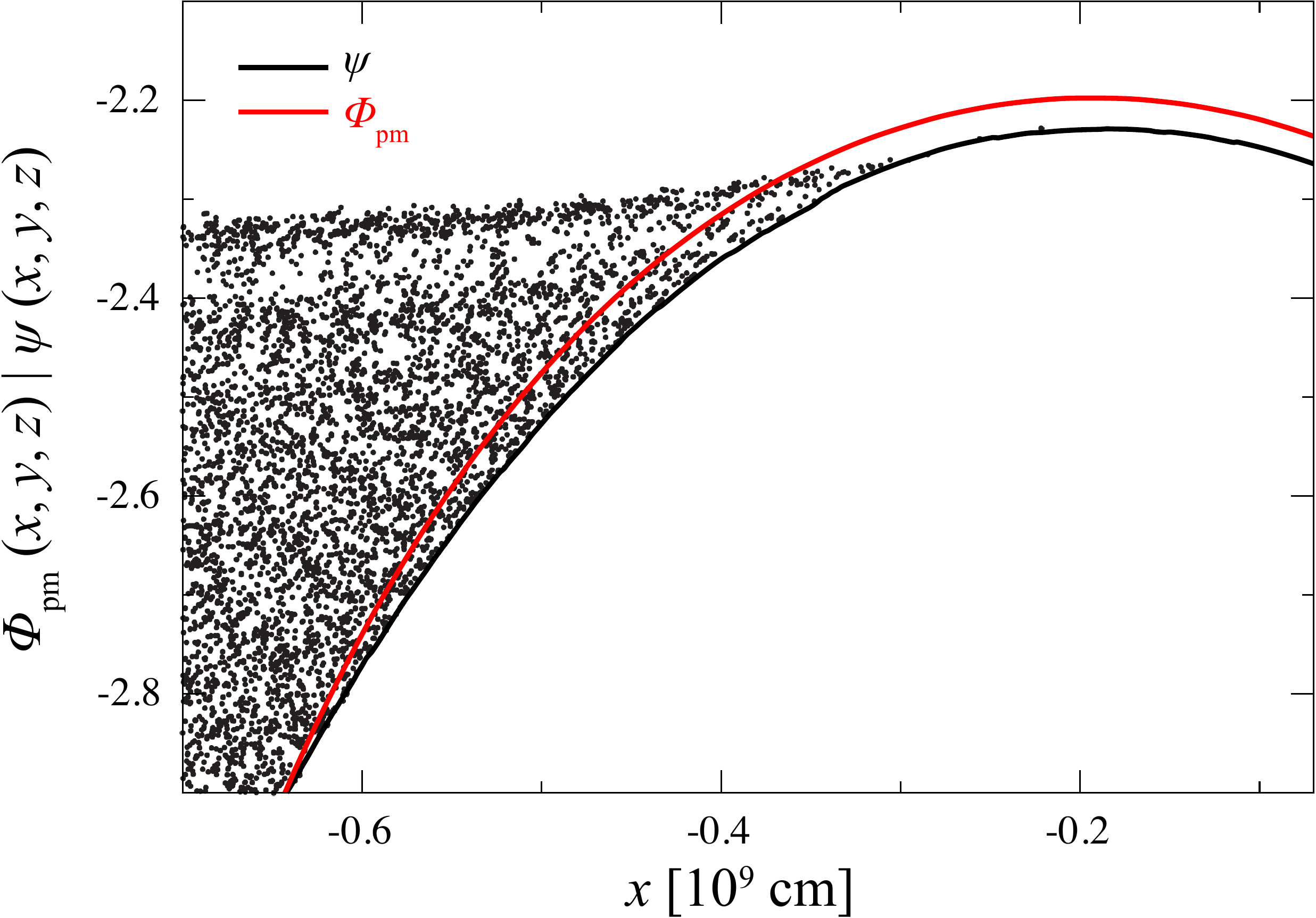}
}
\caption{Projection of the SPH particles on the $(x,\Phi_{\rm pm}/\Psi)$
  plane. The dotted red line is the point mass Roche potential, $\Phi_{{\rm
      pm}}(x,y=0,z=0)$. The solid black line is the value of the potential
  $\Psi$ along the $x-$axis, the SPH particle values are shown as filled black
  circles. {\it Left}: Wide view showing both stars. {\it Right}: Zoom into
  the neighborhood of the Lagrange point $L_1$. The fluid potential $\Psi$
  lies below the point mass value $\Phi_{\rm pm}$ and therefore the point mass
  approximation underestimates the distance where mass transfer sets in.} 
\label{fig:IC2}
\end{figure}

We primarily focus on synchronized binary systems, but we also perform some tests with initially non-spinning WDs. 
Our procedure is similar to the one described in \cite{rosswog04b}. We begin our initialization procedure with 
stars that are set up according to one-dimensional, cold ($T=10^5$ K) equilibrium
models. These stars are then mapped into three dimensions and are then driven
to an accurate numerical equilibrium by applying a velocity-dependent 
acceleration $\vec{f}_{\rm damp}$ while keeping the temperature constant. 
The stars are then placed at an initial separation that is large enough to
avoid any immediate mass transfer. The full SPH code is then used to relax the
initial conditions to a tidally deformed binary system at the brink of
numerically resolvable mass transfer. 
This phase of the initialization is performed in a corotating frame, i.e. centrifugal and Coriolis forces must be accounted for\footnote{As
the velocities vanish for synchronized systems, Coriolis forces can (and should) be ignored in this case
to reach equilibrium faster.}. We calculate the orbital frequency $\omega$ in a way 
that the total forces on the individual stellar centers of mass vanish exactly \citep{rosswog04b} and
thus account for finite size deviations from the Kepler frequency. The acceleration applied to a
particle $i$ then reads
\be
{\bm f}_i= {\bm f}_{{\rm grav}, i} + {\bm f}_{{\rm hyd}, i} - {\bm \omega} \times \left({\bm \omega} \times {\bm r}_i\right)
- 2 {\bm \omega} \times {\bm v}_i - \frac{{\bm v}_i - {\bm \Omega}_{1,2} \times ({\bm r}_{1,2} - {\bm r}_i)}{\tau_i}.
\ee
Here, ${\bm f}_{{\rm grav}, i}$ and ${\bm f}_{{\rm hyd}, i}$ are the gravitational and hydrodynamic acceleration, ${\bm r}_i$ and
${\bm v}_i$ position and velocity in the corotating frame, ${\bm
  \Omega}_{1,2}$ is either the primary or the secondary star's angular
velocity, 
and $\tau_i$ a damping time scale. In this frame, ${\bm \Omega}_{1,2} = 0$ for
synchronized systems and ${\bm \Omega}_{1,2} = -{\bm \omega}$ for non-spinning
stars. During this 
relaxation process, we adiabatically decrease the orbital separation $a$ in a way that the orbital 
shrinkage time scale, $\tau_{\rm shrink}=a/\dot{a}$, is substantially longer than dynamical timescale of the secondary $\tau_{\rm dyn, 2}$,
\be
\tau_{\rm shrink} = \frac{a}{\dot{a}} = \frac{1}{\epsilon \sqrt{G \bar{\rho}_2}} \gg \tau_{\rm dyn, 2},
\ee
where $\bar{\rho}_2$ is the average density of the secondary. We found that
spurious oscillations are minimized for values of $\epsilon \lesssim 0.05$. As
soon as the first SPH particle crosses the  
Lagrange point $L_1$, we stop the relaxation process, transform to a space-fixed frame with the same coordinate
origin and start the dynamical simulation. This moment is adopted as the time origin ($t=0$) of the simulation,
shown in Figure \ref{fig:IC1} for a $0.9 + 1.2$ \Msun system. In Figure \ref{fig:IC2} we show the potential of the initial conditions, where 
$\Phi_{\rm pm}$ is the usual Roche potential as calculated for a point mass binary,
\be
\Phi_{\rm pm}({\bm r})= - \frac{GM_1}{|{\bm r} - {\bm r}_1|} 
     - \frac{GM_2}{\left|{\bm r} - {\bm r}_2\right|}
     - \frac{1}{2}\left({\bm \omega}_{\rm k} \times {\bm r}\right)^2
\ee
with $M_1/M_2$ and ${\bm r}_1/{\bm r}_2$ the stellar masses and centre of mass positions and ${\bm \omega}_{\rm k}$
the point mass Kepler frequency. The quantity $\Psi$ is the corresponding quantity as calculated for the
actual fluid configuration
\be
\Psi({\bm r})= \Phi\left({\bm r}\right) - \frac{1}{2}\left({\bm \omega} \times {\bm r}\right)^2,
\label{eq:psi_potential}
\ee
where ${\bm \omega}$ is the orbital frequency of our (fluid) binary and $\Phi$ the gravitational potential due to the fluid.

As we will show below in detail, the accuracy of the initial conditions
has a strong impact on all major aspects of the simulations, prolonging the
mass transfer stage by dozens of orbits in all investigated cases. 

\subsection{Numerical treatment}
In this section we explore the sensitivity of the binary evolution on physical
and numerical factors, including numerical resolution, shock heating
treatment, the artificial viscosity prescription, and the initial conditions. 

\subsubsection{Dependence on numerical resolution}
The duration of mass transfer and the separation at which it begins depends on
the extent to which we can resolve the outer layers of the WD donor. 
In SPH, where usually the particle masses are not evolved in time, this means that even in a decent 
numerical effort with a million SPH particles the initial resolvable mass transfer rate is 
already a substantial fraction of the total system mass.
The numerically resolvable mass transfer $\dot M_{\rm lim}$ can be estimated as 
\begin{equation}\label{eq:mmin}
\dot M_{\rm lim}\approx \frac{1\ {\rm particle \; mass}}{{\rm orbital \; period}}
\approx 2 \times 10^{-8}
\left(\frac{N_{\rm p}}{10^6}\right)^{-1}
\left(\frac{M_{\rm tot}}{1\ M_\odot}\right)^{3/2}
\left(\frac{a_0}{2 \times 10^9\ {\rm cm}}\right)^{3/2}\frac{M_\odot}{\rm s},
\end{equation}
where $a_{0}$ is the separation between the stars
at the onset of mass transfer. In order to better resolve regions of lower density,
SPH particles of different masses could be used, but we refrain from such a possibility as it can
introduce a substantial level of numerical noise.
\begin{figure}[!t]
\centerline{
 \includegraphics[height=1.9in]{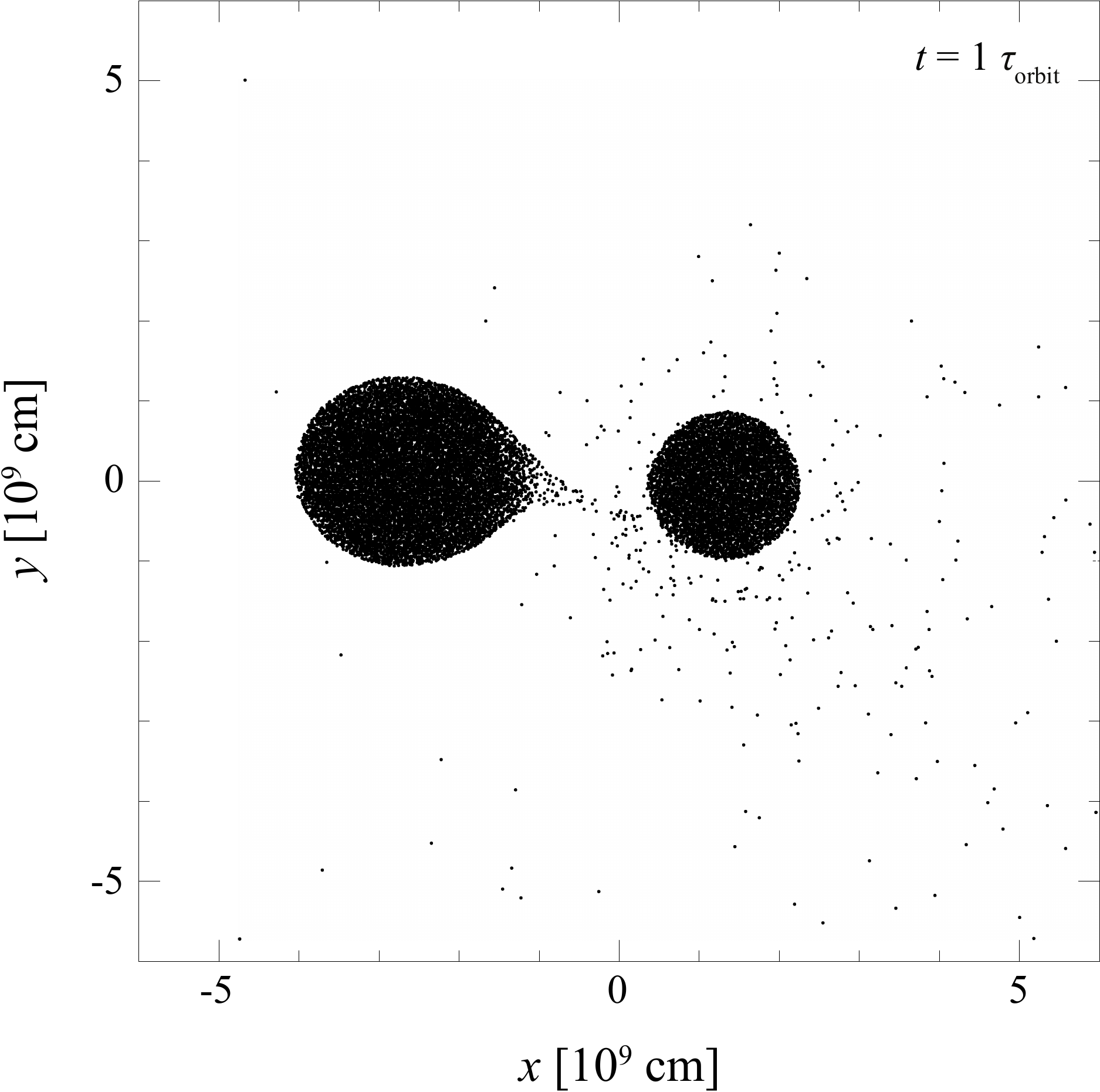}\hspace{0.5cm}
 \includegraphics[height=1.9in]{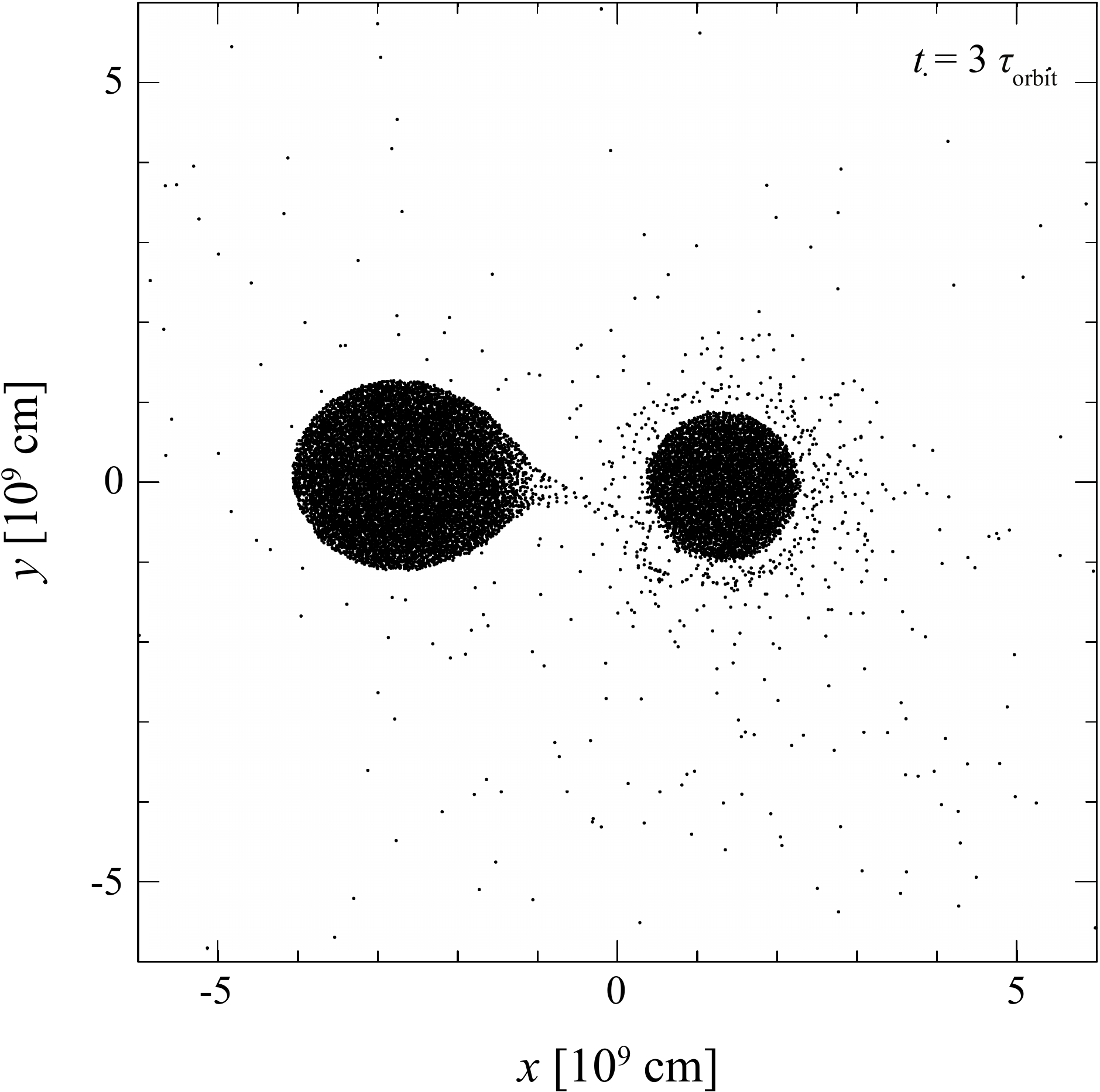}\hspace{0.5cm}
 \includegraphics[height=1.9in]{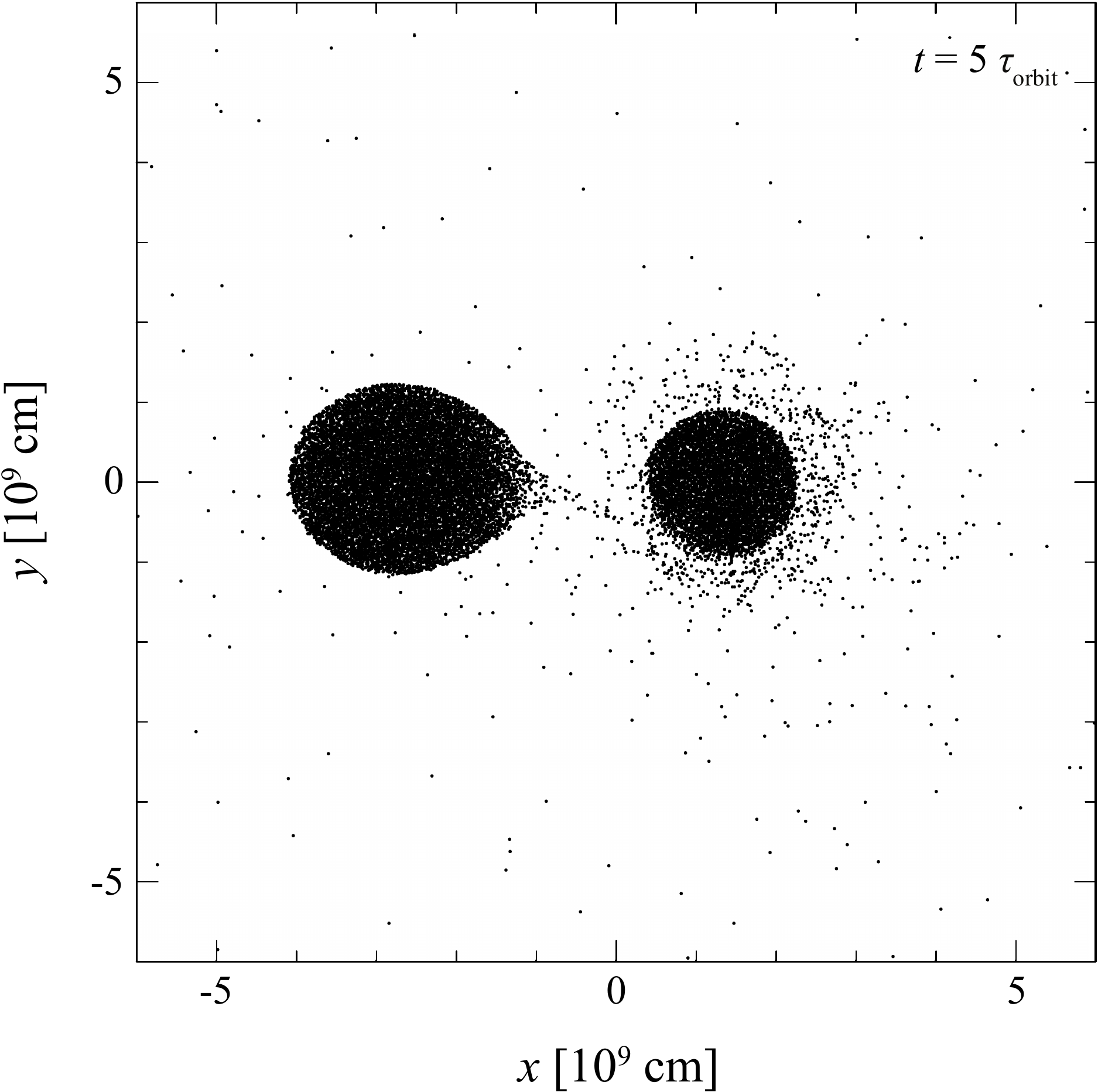}
}
\vspace{0.5cm}
\centerline{
 \includegraphics[height=1.9in]{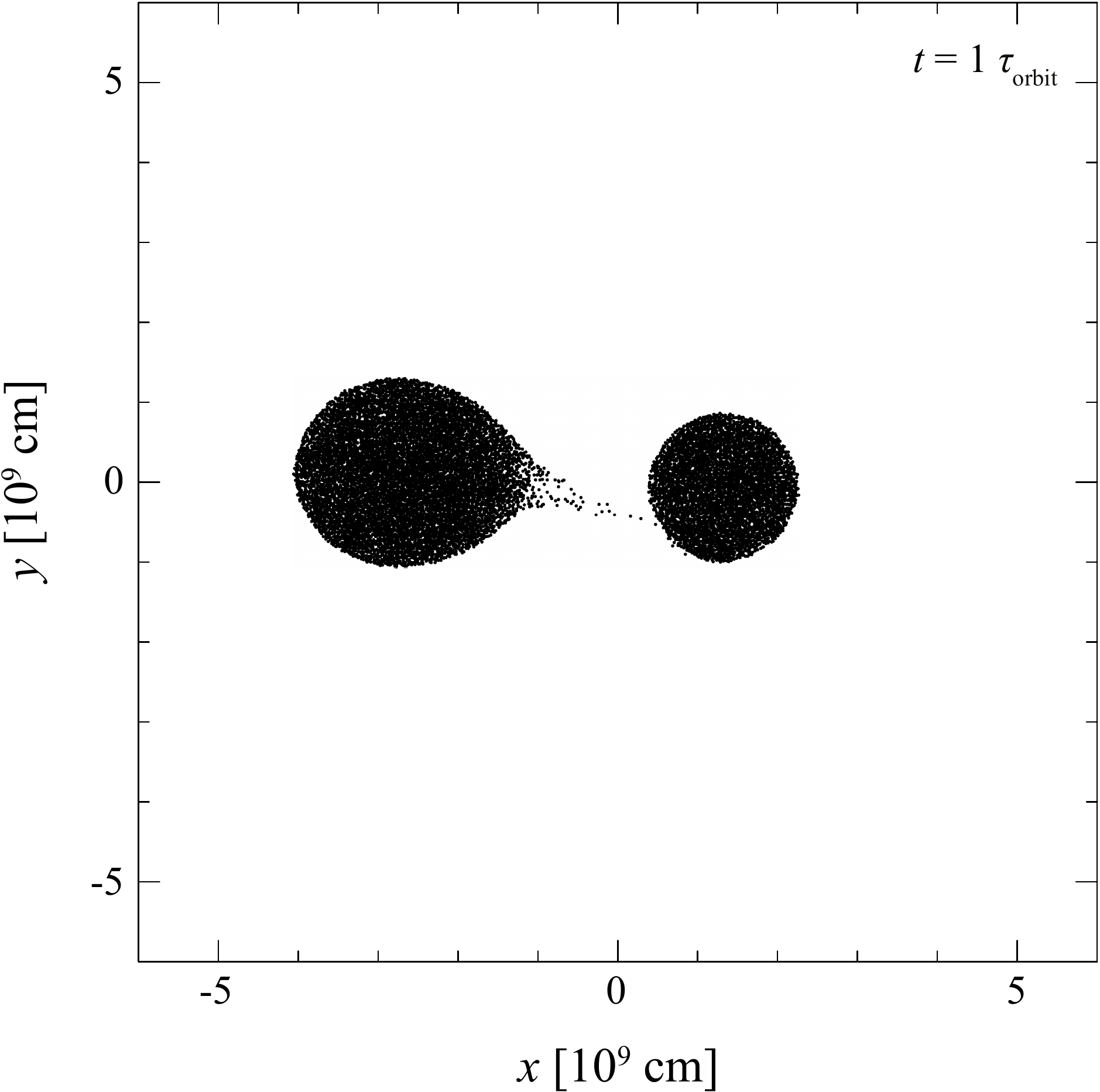}\hspace{0.5cm}
 \includegraphics[height=1.9in]{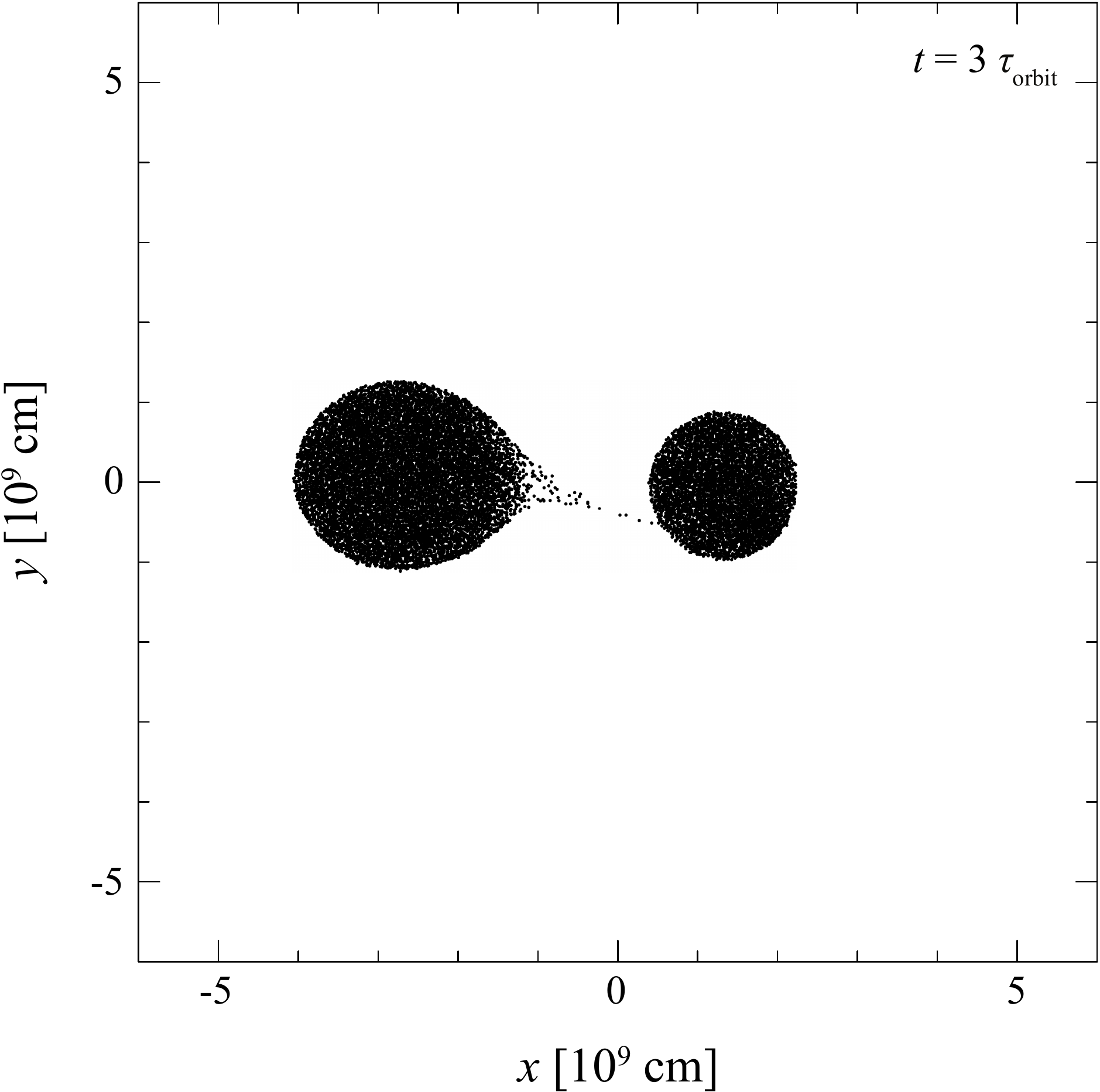}\hspace{0.5cm}
 \includegraphics[height=1.9in]{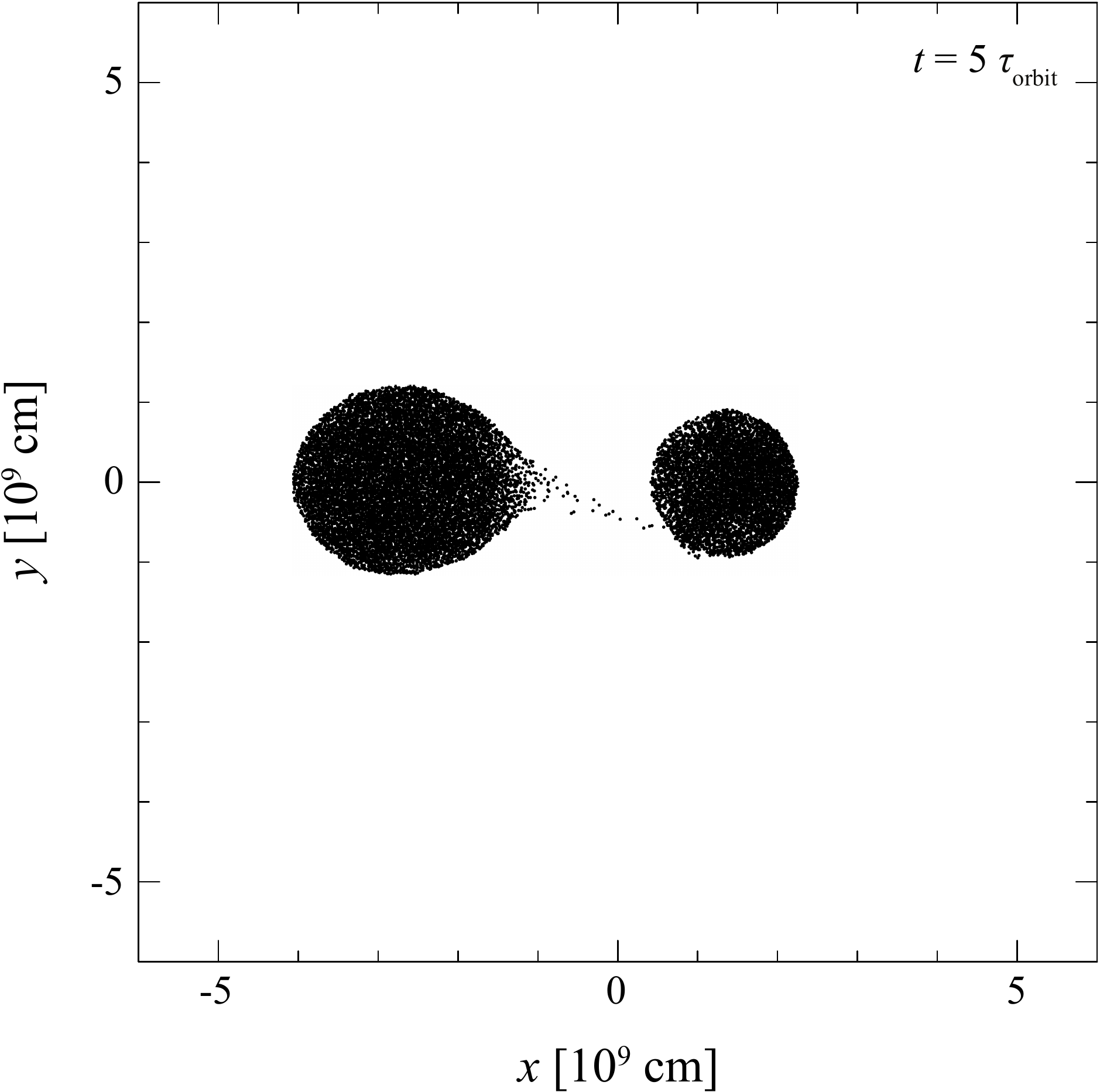}
}
\caption{Particle distribution for a test run with (upper row) and without
(bottom row) shock heating. Shown are snapshots after 1, 3 and 5 orbital periods.} 
\label{fig:part_shock_no_shock}
\end{figure}

\subsubsection{Comparison with the StarCrash code}
As a validation of our code we have also performed a comparison with the results obtained with
the StarCrash code. This code was developed originally by Rasio and Shapiro to 
study merging binaries \citep{rasio92,rasio94,rasio95}. The major conceptual difference in their
code is the use of a fast Fourier transform instead of a tree to calculate the 
gravitational forces.

To enable a valid comparison, for this test we also use a polytropic EOS, $P= K \rho^{5/3}$, 
with $\Gamma=5/3$, we evolve the entropic function according to
\be
\frac{dK_i}{dt}=\frac{\Gamma-1}{2\rho^{\Gamma-1}_i}\sum_j m_j\Pi_{ij}
{\bm v}_{ij}\cdot \nabla_i W_{ij},\label{eq:entropic_function}
\ee
where $m_j$ is the mass of particle $j$, $\Pi_{ij}$ the artificial viscosity tensor depending 
on parameters $\alpha$ and $\beta$, ${\bm v}_{ij} \equiv {\bm v}_i-{\bm v}_j$ the velocity difference between particle $i$
and $j$, and $W$ is the smoothing kernel. 
We fix our artificial viscosity parameters in this comparison to the values recommended in their 
StarCrash documentation ($\alpha=0.5$, $\beta=1$).
We use the subroutines of the StarCrash code to set up a synchronized 
binary system of mass ratio $q=0.5$ with 40,000 SPH particles, both simulations start from 
these initial conditions. The codes yield nearly indistinguishable results, after about 15 orbital 
periods the donor WD becomes tidally disrupted. Although \cite{rasio95} also utilizes the StarCrash code, they start with slightly different initial conditions, with the two stars being closer to one another at $t = 0$. As a result, they find that the donor only survives 5 orbits, despite using the same EOS and SPH particle number. We suspect that they started from a slightly too small initial separation, but the very good agreement of the two codes for the same initial 
conditions is encouraging. In passing we note that these results are very similar to those 
\cite{dsouza06}, which will be discussed in more detail in Section \ref{sec:dsouza}.

\subsubsection{Effect of shock heating treatment}
As mentioned previously, recent grid-based simulations \citep{motl02,dsouza06,motl07} showed
long-lived mass transfer phases, in contrast to the earlier SPH
simulations. In these calculations artificial viscosity was only used in the
momentum equation, but not in the energy equation, which means that entropy
production in shocks was neglected. It was suspected that at least part of the
difference in the mass transfer duration may be due to the suppression or
emergence of shocks/ignoring the change of entropy \citep{fryer08}. 

\begin{figure}[!t]
\centerline{
\includegraphics[height=2.5in]{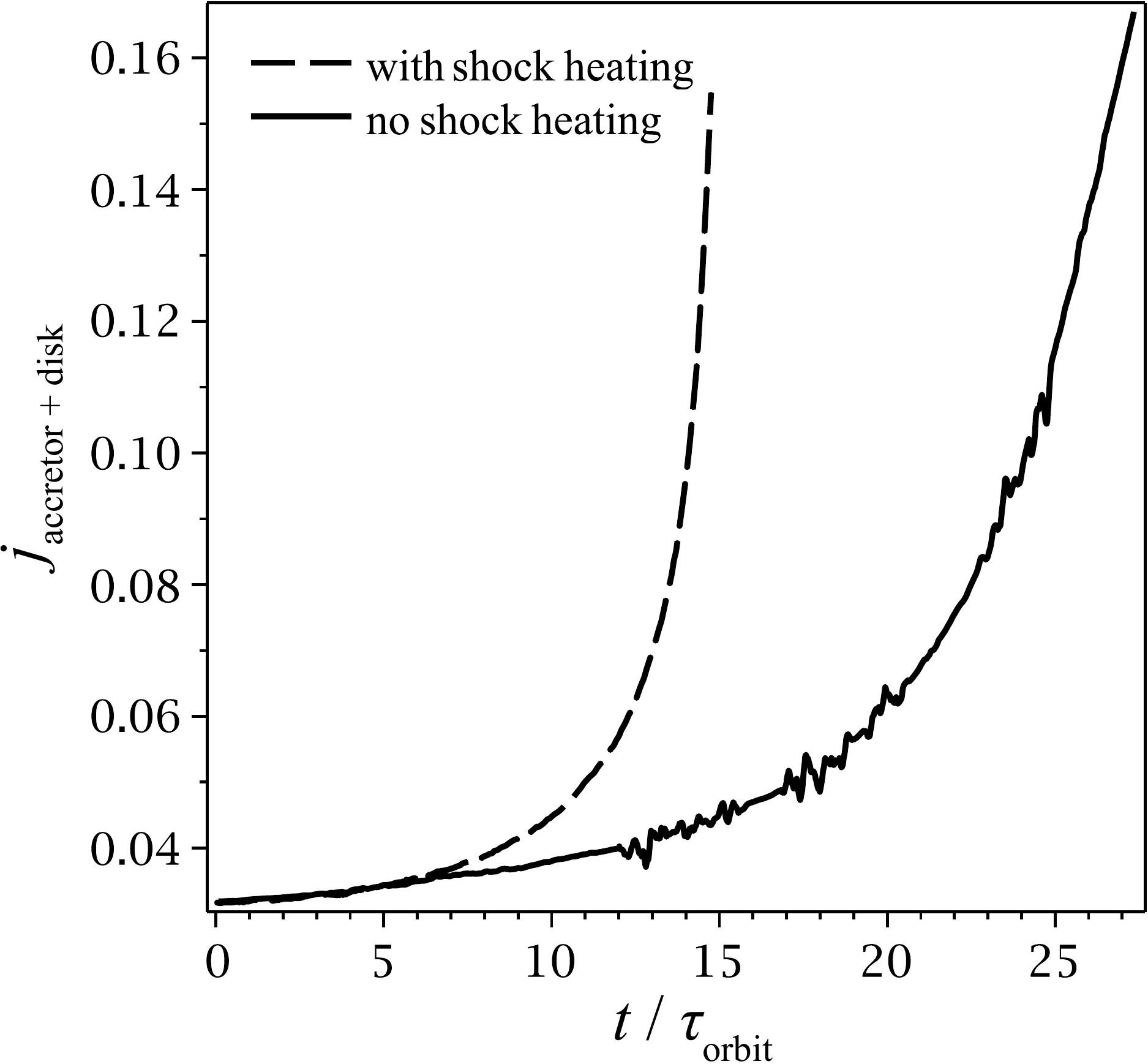} \hspace{0.5cm}
\includegraphics[height=2.5in]{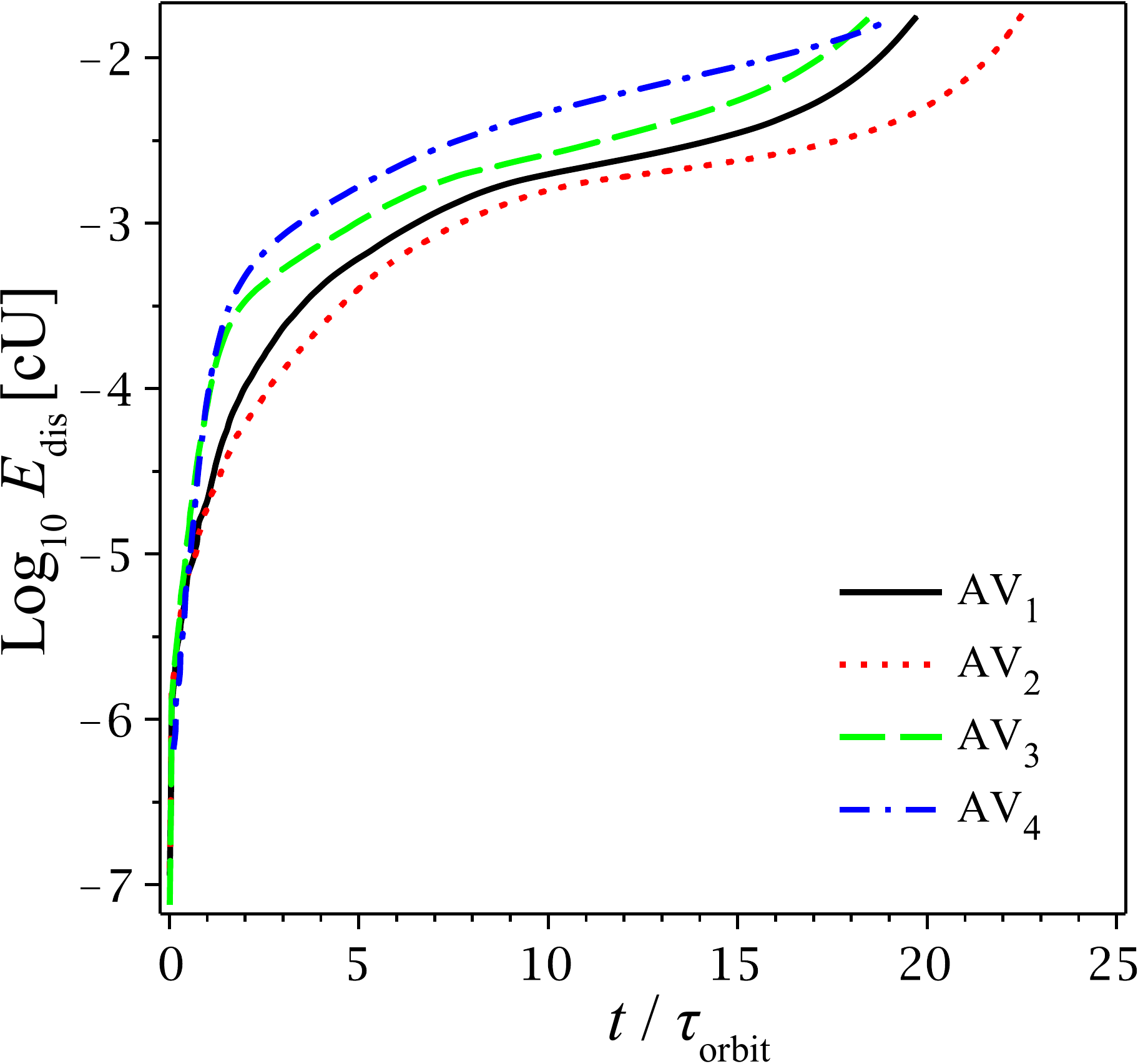}
}
\caption{Left: specific angular momentum evolution (of accretor plus disk) for the test case where we explore
  the effect of the (non-)inclusion of shock heating onto the orbital evolution. Right: evolution of 
  the total energy dissipated, $E_{\rm diss}$, by the different artificial viscosity
  prescriptions. See text for details. Time is measured in units of
  the initial orbital period, $\tau_{\rm orbit}$.}
\label{fig:angmom_shock_no_shock}
\end{figure}

We perform two test calculations at the example of a $q=0.5$ binary, similar to \cite{dsouza06}. 
We follow their approach and use a $\Gamma=5/3$-polytrope, but in one case we evolve the entropy 
function according to Equation (\ref{eq:entropic_function}) and in the other we keep $K$ constant. 
A mass ratio of 0.5 results in a direct impact of the accretion stream onto the accretor surface. In the setup that does not include shocks, the accreted matter is smoothly incorporated into the accretor (Figure \ref{fig:part_shock_no_shock}, lower row), while the setup that includes shocks produces a high-entropy halo around the accretor (Figure \ref{fig:part_shock_no_shock}, upper row). This high-entropy halo applies a greater torque on the binary as it has a larger effective lever arm than the low-entropy halo that is produced when shocks are neglected. As a result, the matter within the high-entropy halo is able to remove more angular momentum from the orbit over the same number of orbital periods (Figure \ref{fig:angmom_shock_no_shock}), leading to a more rapid evolution of the binary separation. This supports the findings of \cite{fryer08} that the large increase in orbital separation reported by \cite{dsouza06} is in part an artifact of shock suppression.

\subsubsection{Impact of the artificial viscosity prescription}
As we have shown that the entropy generation associated with shocks can affect the stability of double degenerate systems, we need to consider the importance of other entropy-generating mechanisms. To this end we compare 
different implementations of artificial viscosity. All have the ``standard'' form originally suggested
by \cite{monaghan83}, but differ in terms of constant or non-constant parameter values and additional 
``limiter switches'':
\begin{itemize}
\item $AV_1$: constant parameters $\alpha=1$ and $\beta=2$
\item $AV_2$: constant parameters $\alpha=2$ and $\beta=4$
\item $AV_3$: time-dependent parameters \citep{morris97}, controlled as described in \cite{rosswog08b}, but
without ``Balsara switch'' \citep{balsara95}
\item $AV_4$: time-dependent parameters plus Balsara switch as described in \cite{rosswog08b}.
\end{itemize}
As test case we run simulations with a 0.3 He and a 0.6 CO WD, each one
modeled by 20,000 particles (T4-T7, Table \ref{tab:runs}). 
Synchronized initial conditions are constructed as described in Section \ref{sec:our_IC} and all tests used
the Helmholtz EOS. Artificial viscosity does indeed have an impact: in the lowest viscosity case, $AV_4$, 
the system merges after 19, in highest-viscosity case, $AV_2$, after 22 orbital periods. To quantify
the dissipation we calculate the thermal energy produced by artificial viscosity before merger
\be
E_{\rm diss}= \sum_i m_i \int_0^{t} \left(\frac{du_i}{dt}\right)_{\rm diss} dt',
\ee
where
\begin{equation}\label{eq:udiss}
\left(\frac{du_i}{dt}\right)_{\rm diss}=\frac
12\sum_j m_j\Pi_{ij}{\bm v}_{ij}\nabla_i W_{ij}
\end{equation}
is the artificial viscosity contribution to the energy equation of particle
$i$. The quantity $E_{\rm diss}$ as a 
function of time (for $t< t_{\rm disr}$ where $t_{\rm disr}$ is the time
when the donor star is tidally disrupted) is shown in Figure
\ref{fig:angmom_shock_no_shock} (right panel), for all four artificial
viscosity prescriptions.  
Interestingly, the prescription with the lowest average artificial dissipation parameters, $AV_4$, dissipates 
the most energy. The reason for this is that the combination of low viscosity parameters and low resolution 
allow the SPH particles to build up a large level of velocity noise (i.e. fluctuations around the local average 
value). This leads to an increase in the typical value of ${\bm v}_{ij}$, and
thus more dissipation, despite the lower viscosity parameters and suppression
of dissipation in shear flows. 

The increased rate of dissipation results in a more efficient spin-up of the
accretor, with the angular momentum for this spin-up coming at the orbit's
expense. This leads to a reduction in the number of orbital periods the binary
can survive prior to merger. We perform our production runs with prescription 
$AV_4$ which yields a lower limit on the duration of the mass transfer phase.

\subsubsection{Dependence on initial conditions}
\label{sec:dependence_IC}
As we have discussed, most of the earlier WD merger simulations were
constructed using approximate initial conditions where the separation was
estimated by comparison of the stellar radius with the Roche lobe size.
To illustrate the importance of
carefully constructed initial conditions, here we  
compare the evolution of a binary initialized using both ``approximate'' and
``accurate'' initial conditions (ICs). 

\begin{figure}[!h]
\centerline{
 \includegraphics[height=2.5in]{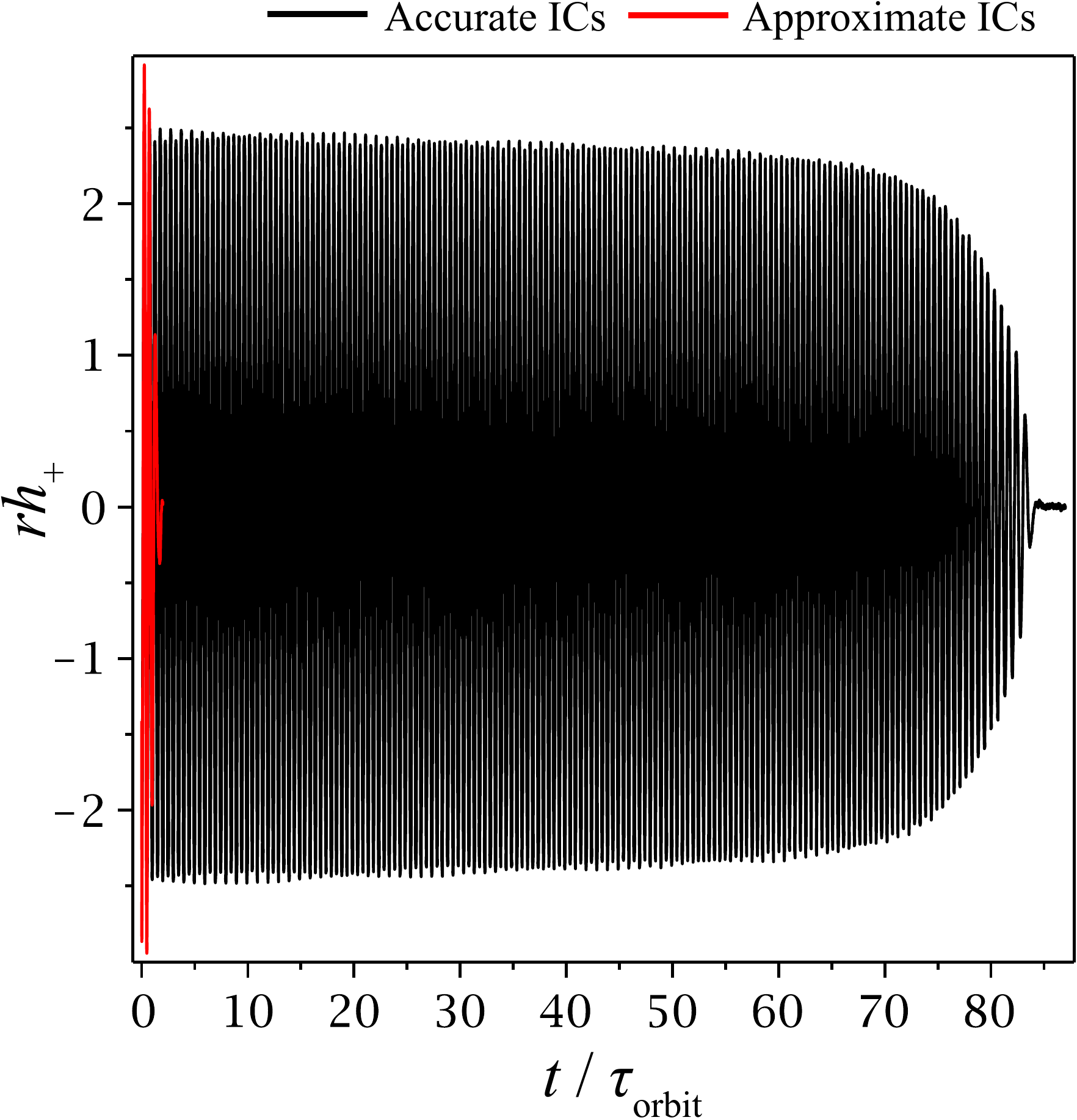}\hspace{0.5cm}
 \includegraphics[height=2.5in]{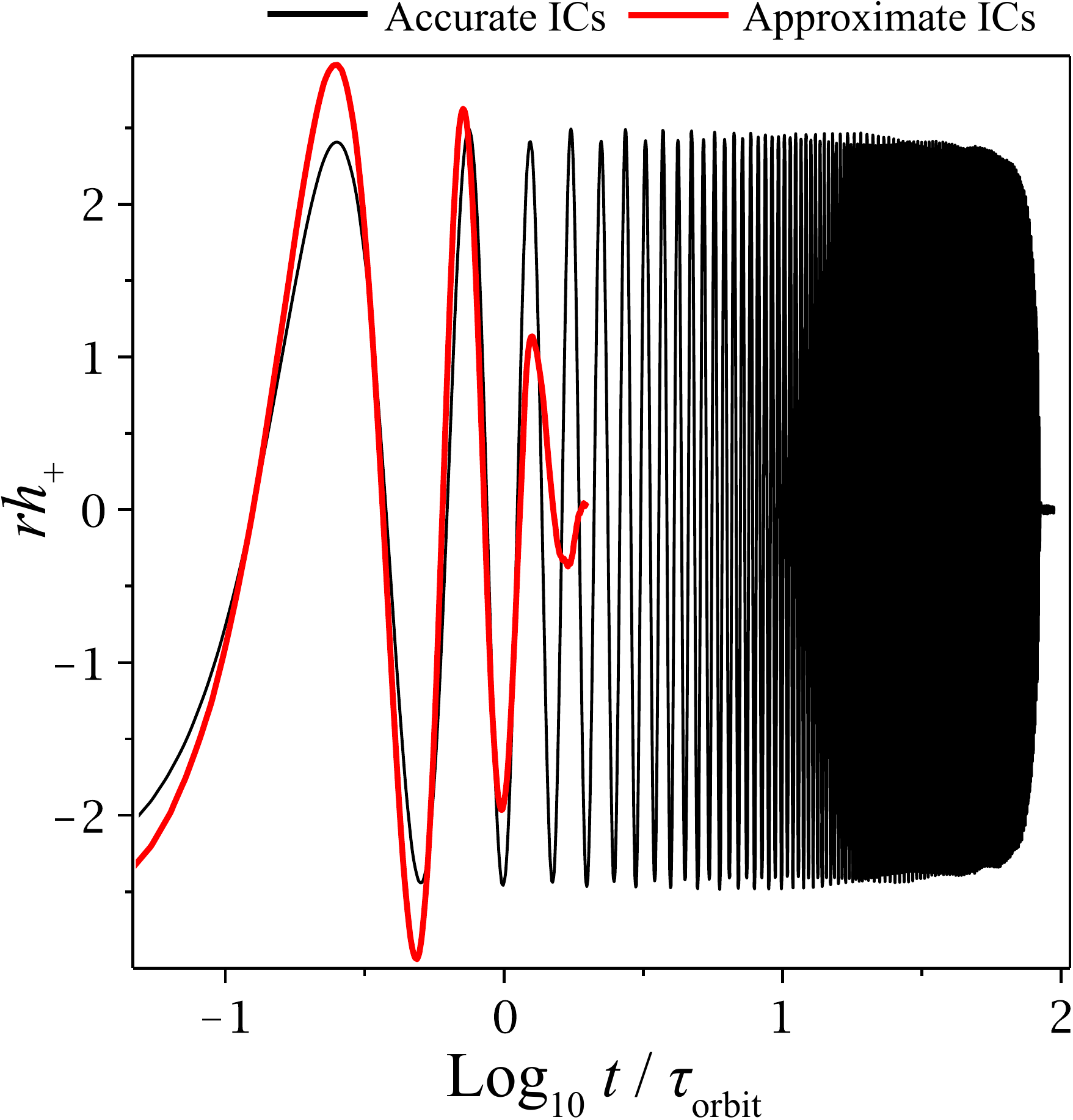}
}
\caption{Gravitational wave amplitude $h_+$ for the $0.2 + 0.8\ M_\odot$ system ($r$ is the distance to the observer) as a function of
time in units of the initial orbital period of the system, $\tau_{\rm orbit}$, both with
``approximate'' (red) and ``accurate'' ICs (black). } 
\label{fig:GWs_02_08}
\end{figure}
\begin{figure}[!h]
\centerline{
 \includegraphics[height=2.5in]{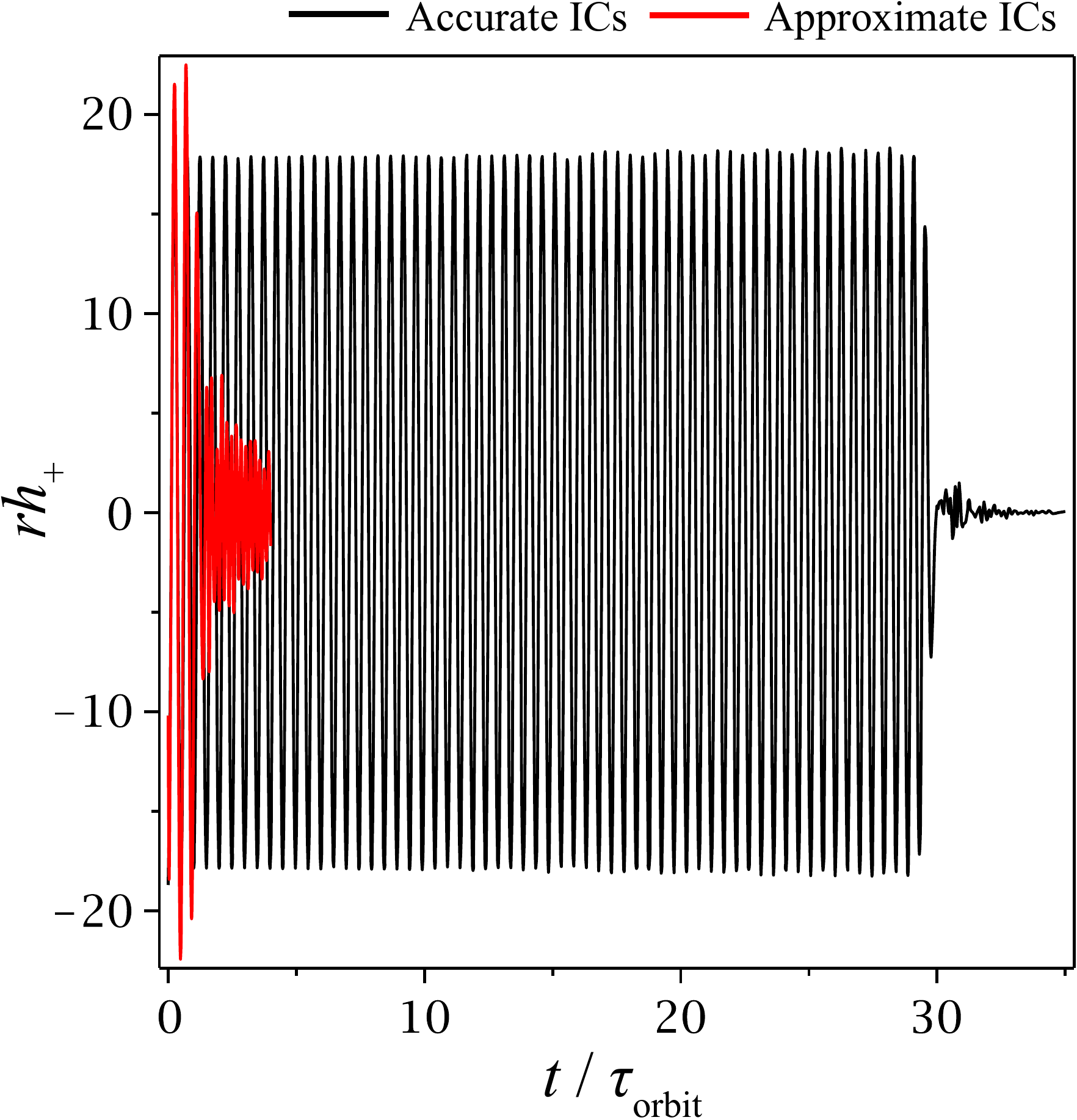}\hspace{0.5cm}
 \includegraphics[height=2.5in]{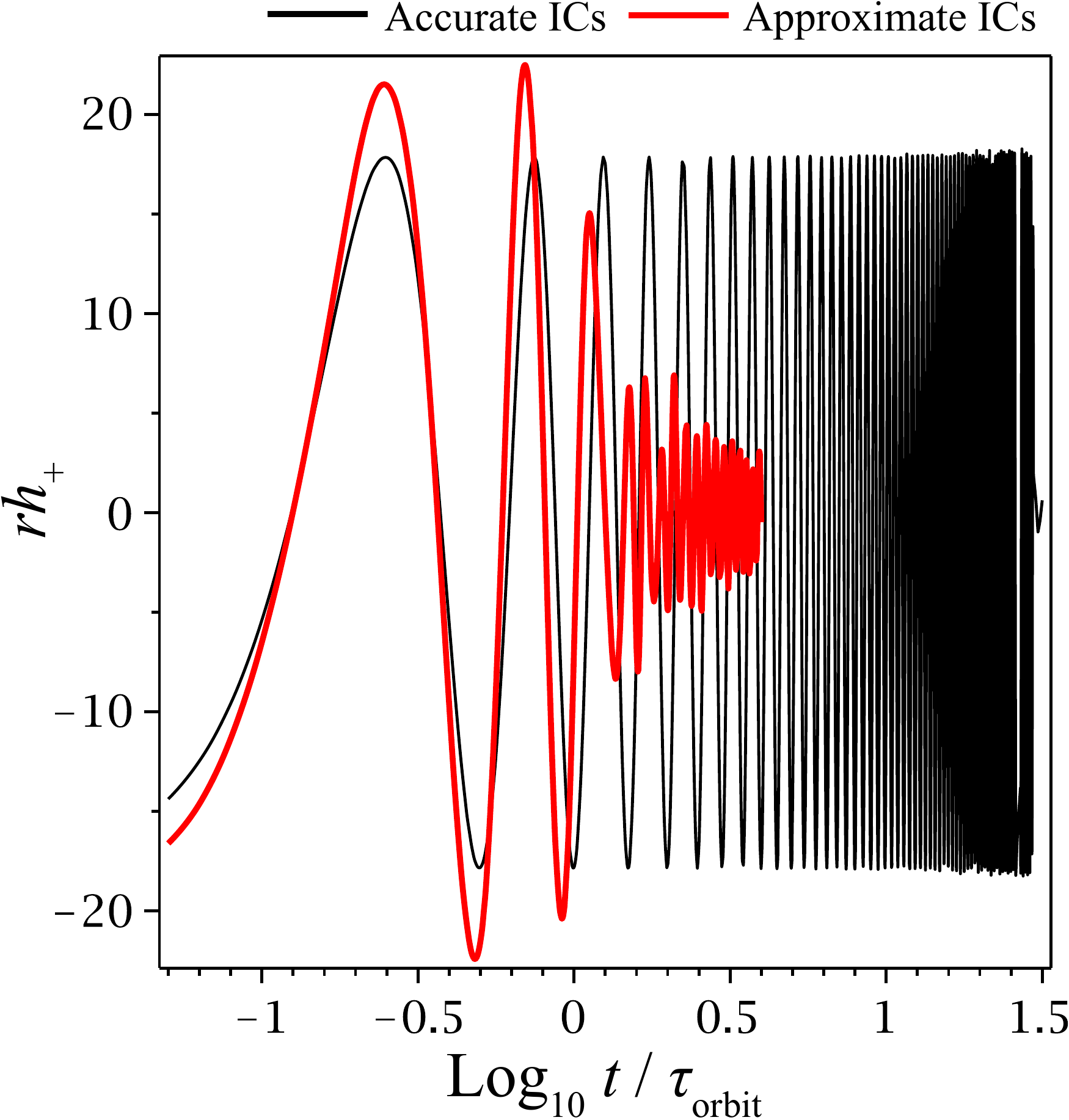}
}
\caption{Same as Figure \ref{fig:GWs_02_08}, but for the $0.6 + 0.9\ M_\odot$ system.}
\label{fig:GWs_06_09}
\end{figure}

To construct the approximate ICs, we equate Eggleton's formula \citep{eggleton83} for the Roche radius
with the radius $R_2$ of the secondary as found from the SPH particle positions. The initial separation $a_0$ is thus given by
\be
a_0^{\rm Egg}= R_2 \; \frac{0.6 q^{2/3}+\ln(1+q^{1/3})}{0.49q^{2/3}}\label{eq:a0_Eggleton}.
\ee
The stars, previously relaxed in isolation (i.e. without the companion's tidal field), are subsequently placed 
at this mutual separation with a spin that has the same period as the orbit of the binary. We refer to these ICs
as ``approximate'' and compare them to the carefully constructed ICs described
in Section \ref{sec:our_IC}. The described procedure is plausible, as we will
see below, its approximate nature introduces a slew of artifacts.

Two representative examples of binaries are compared, a 0.2 \Msun He + 0.8 \Msun CO binary and a 0.6 \Msun CO + 0.9 \Msun CO binary.
Each system is simulated using both the approximate (corresponding to runs T1 and T2 in Table \ref{tab:runs} and Figure \ref{fig:marsh04}) and accurate (P1 and P5) ICs. 
Some marked differences are seen between the simulations initialized using these two different prescriptions. 
The approximate prescription systematically underestimates the 
separation at which mass transfer sets in by about 14\% in both cases. This is mainly because it neglects the tidal 
deformation of the donor star.
As a result, binaries initialized using approximate ICs carry a deficit of angular momentum when compared to those evolved using the more accurate prescription. The difference in angular momentum between the two IC prescriptions is 11\% and 13\% for the $0.2+0.8$ \Msun and 
$0.6+0.9$ \Msun cases, respectively. This slight discrepancy severely impacts their subsequent evolution, with the duration of the mass transfer phase being underestimated by about two order of magnitudes. In the $0.2 + 0.8$ \Msun case, the approximate 
ICs yield two orbits (478 s) of mass transfer while the accurate ICs show a 
mass transfer for as long as 84 orbital periods (20,328 s). The difference in the $0.6 + 0.9$ \Msun case is 
similarly dramatic (1 vs 29 orbital periods).

\begin{figure}[!t]
\centerline{
\begin{tabular}{@{}ccc@{}}
 \includegraphics[height=2.15in]{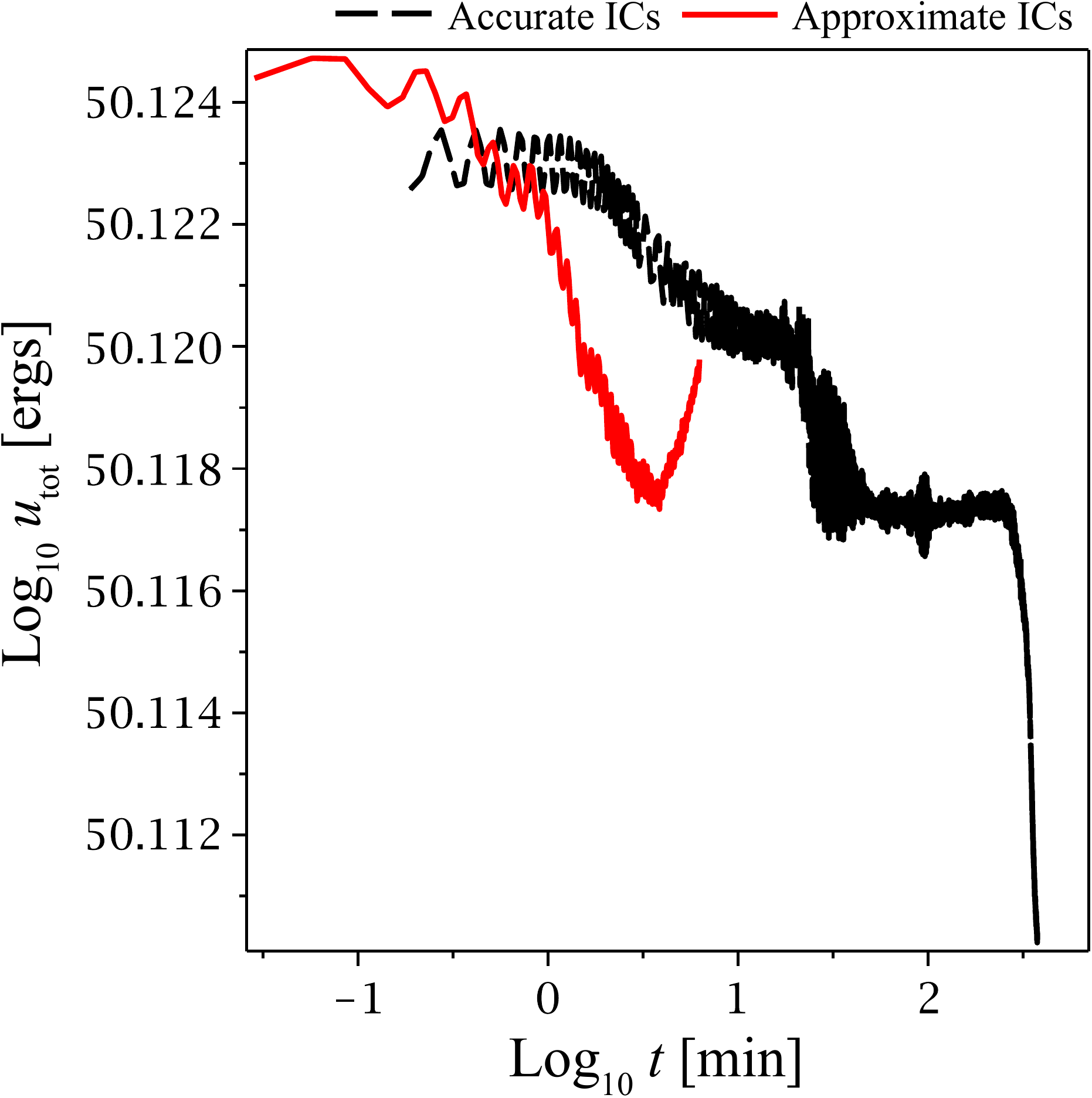}&
 \includegraphics[height=2.15in]{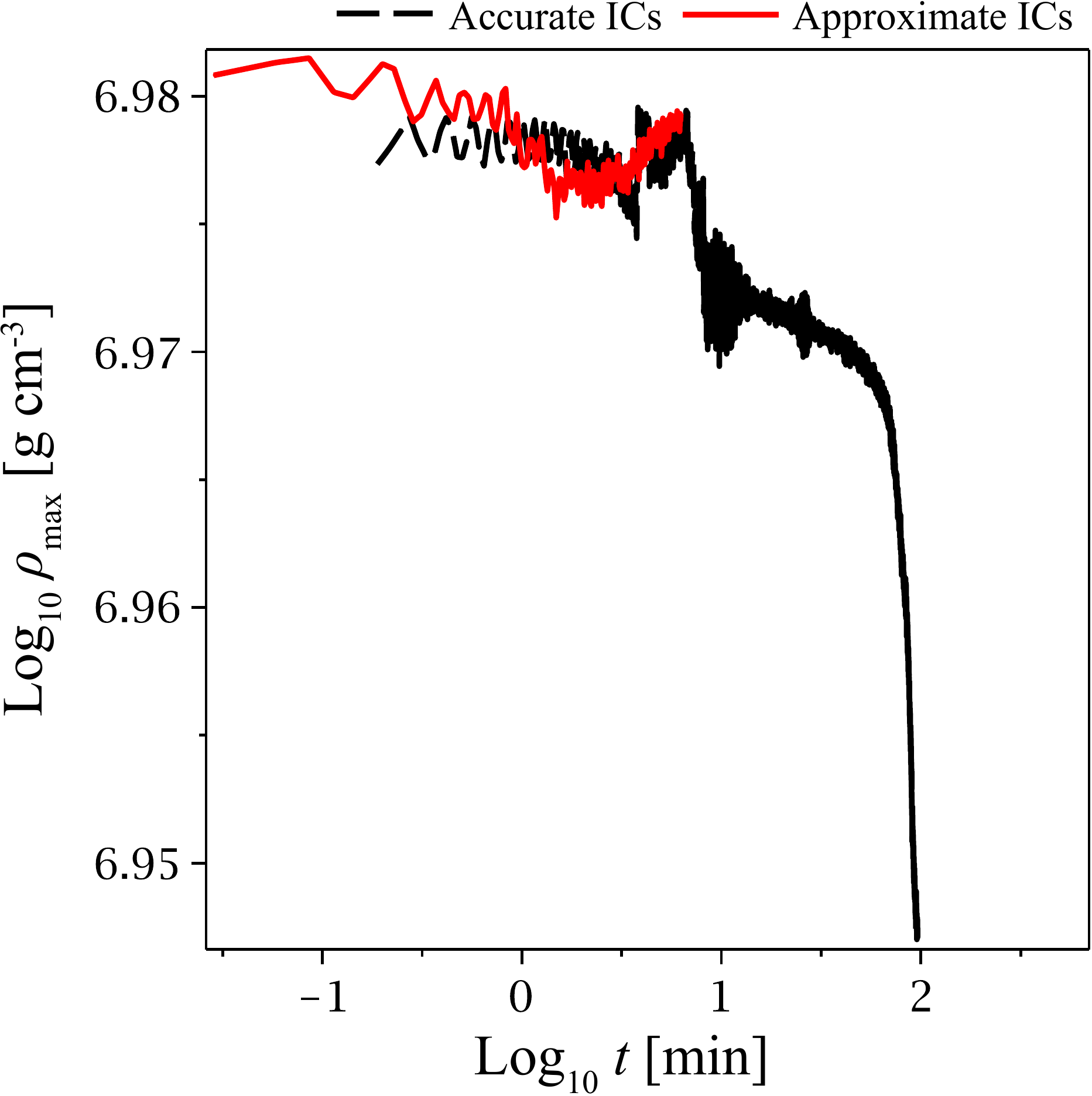}&
 \includegraphics[height=2.15in]{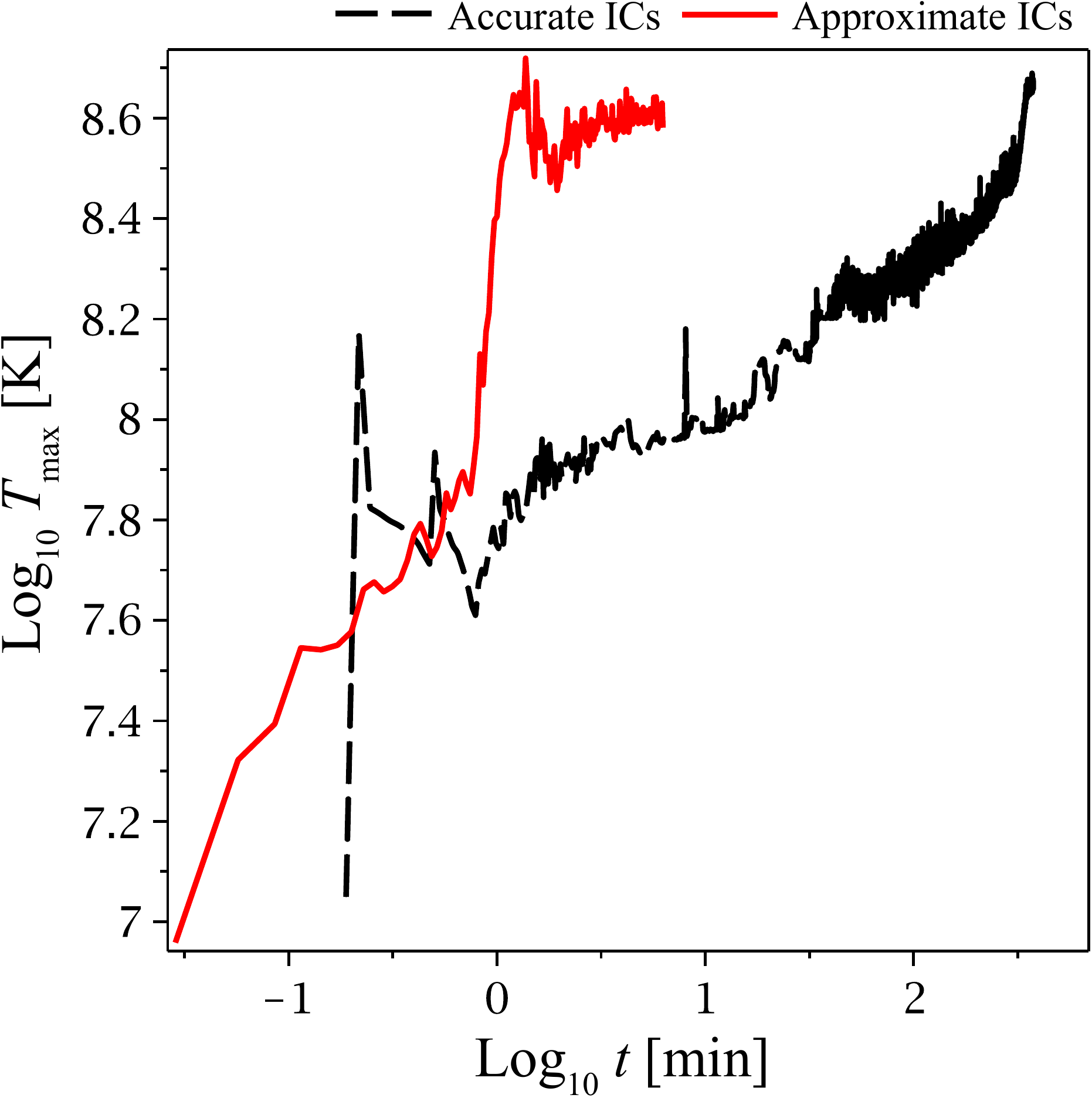}
\end{tabular}
}
\caption{Evolution of thermodynamic quantities for a $0.2 + 0.8\ M_\odot$ system both with approximate (red)
and accurate ICs (black). Shown are: total internal energy
(left), maximum density (center) and temperature (right) as a function of time $t$ in units of minutes.} 
\label{fig:thermal_02_08}
\end{figure}
\begin{figure}[!t]
\centerline{
\begin{tabular}{@{}ccc@{}}
 \includegraphics[height=2.15in]{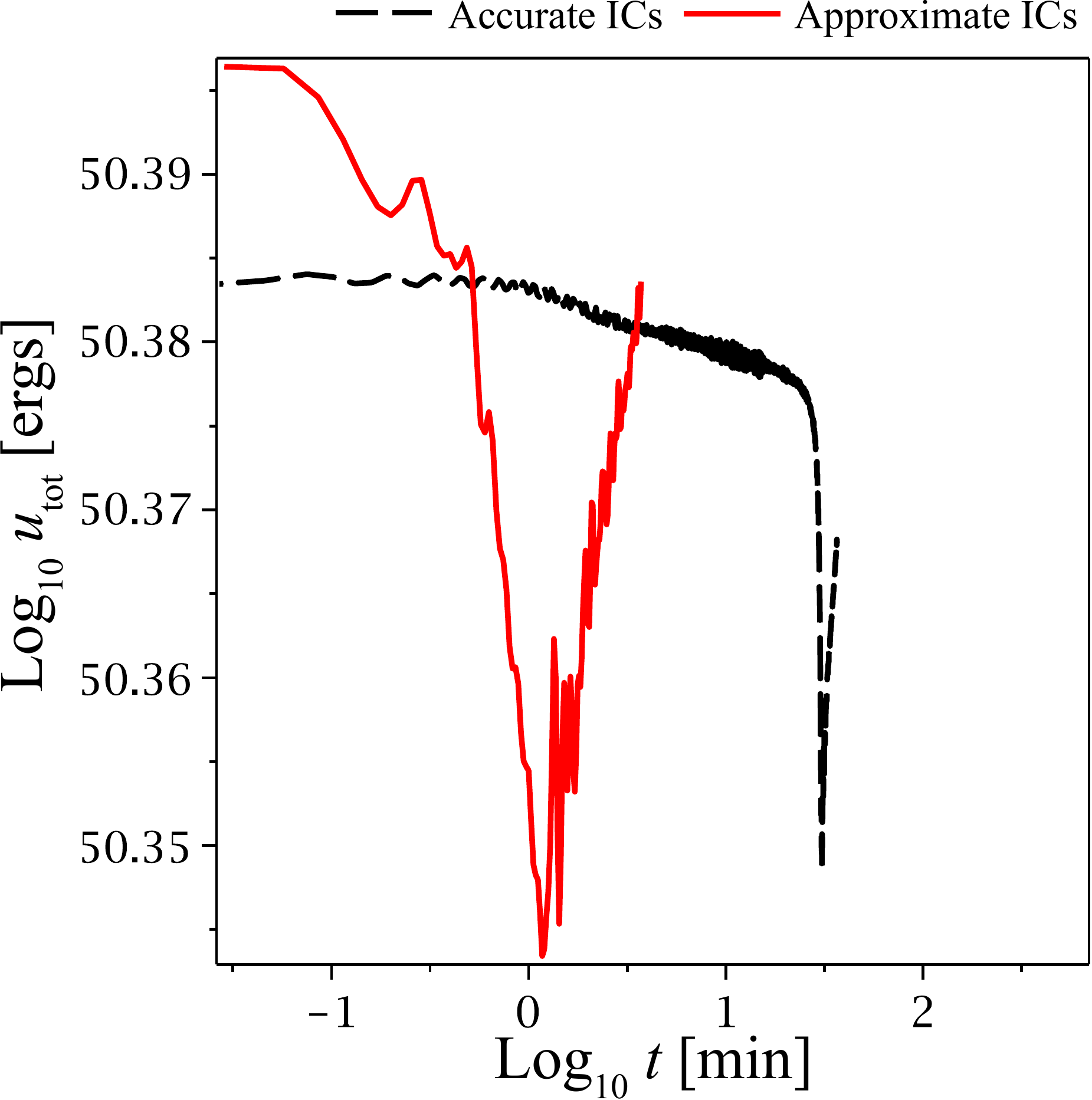}&
 \includegraphics[height=2.15in]{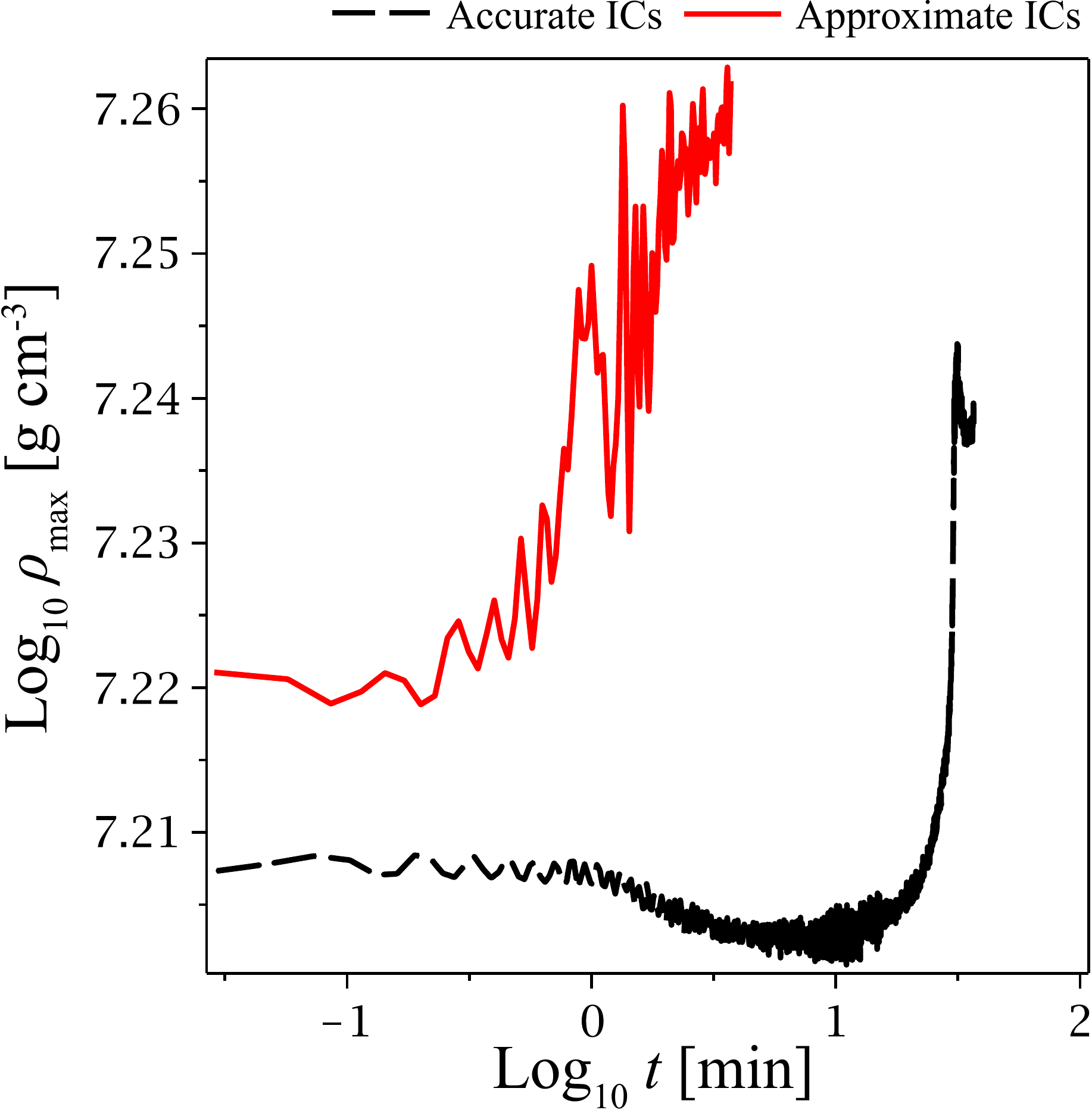}&
 \includegraphics[height=2.15in]{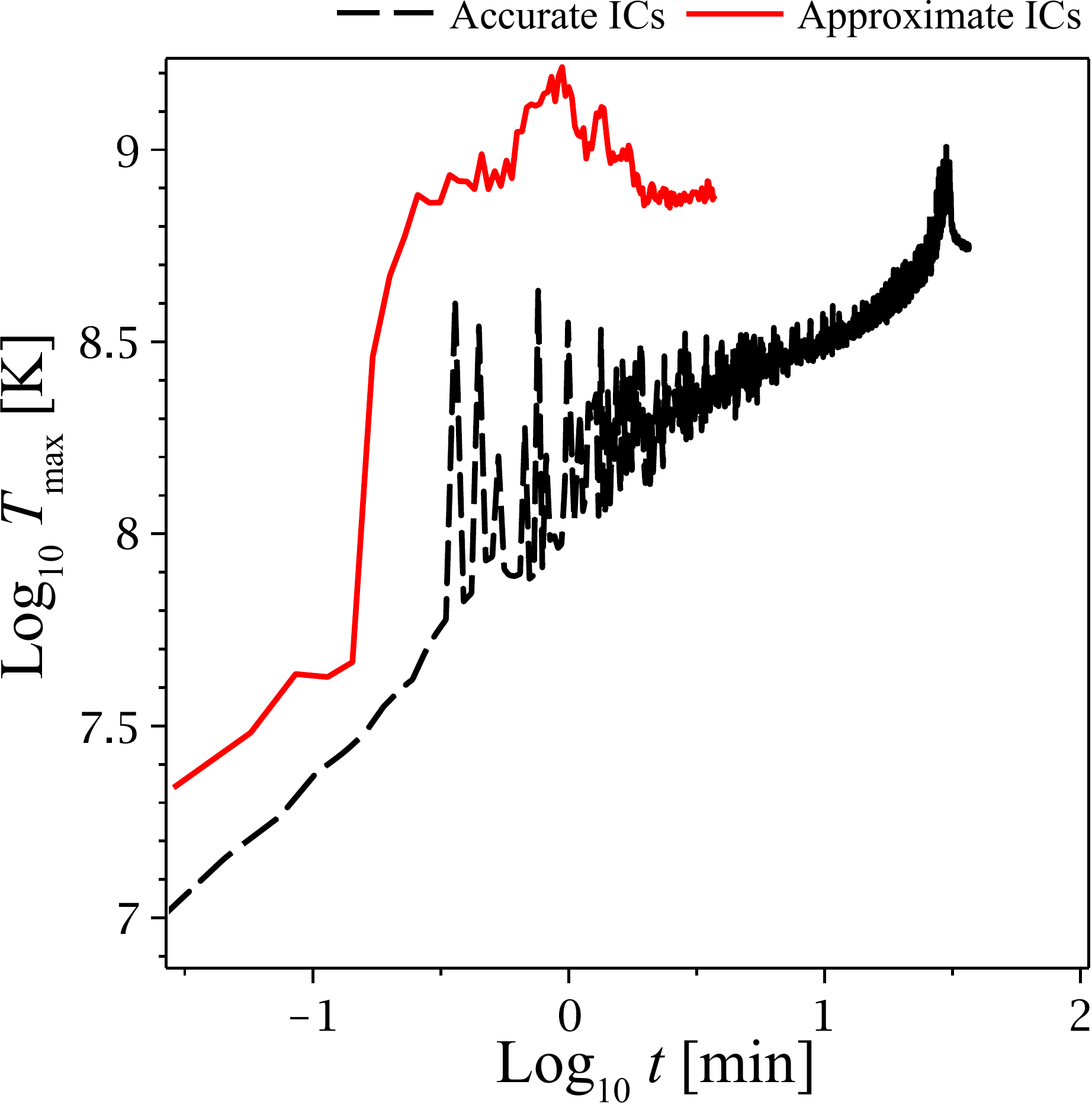}
\end{tabular}
}
\caption{Similar to Figure \ref{fig:thermal_02_08} but for the $0.6 + 0.9$ \Msun system.}
\label{fig:thermal_06_09}
\end{figure}

The effect of varying ICs on the orbital dynamics is most clearly seen when
comparing the resulting gravitational wave signals (Figures
\ref{fig:GWs_02_08} 
and \ref{fig:GWs_06_09}). The approximate ICs give rise to a rapidly fading
signal, while in the more accurate case the amplitude remains essentially
constant up to the point of tidal disruption. The $0.6+0.9$ \Msun case, which
is less stable than the $0.2+0.8$ \Msun system (Figure
\ref{fig:marsh04}), shows only a slightly  
noticeable increase in the gravitational wave amplitude up to disruption. In
both cases, the approximate ICs overestimate the peak amplitude 
and frequency. The approximate case gives an orbital frequency of 46 mHz in
the $0.6+0.9$ \Msun case and 10 mHz in the $0.2+0.8$ \Msun case,
while the accurate ICs yield 38 and 8.4 mHz, respectively (a difference of
about 20 \%). This is mainly due to the difference in separation at final
plunging (40 \% difference for the $0.2+0.8$ \Msun case and 30 \% for the
$0.6+0.9$ \Msun case).  
\begin{figure}[!h]
\centerline{
 \includegraphics[height=2.7in]{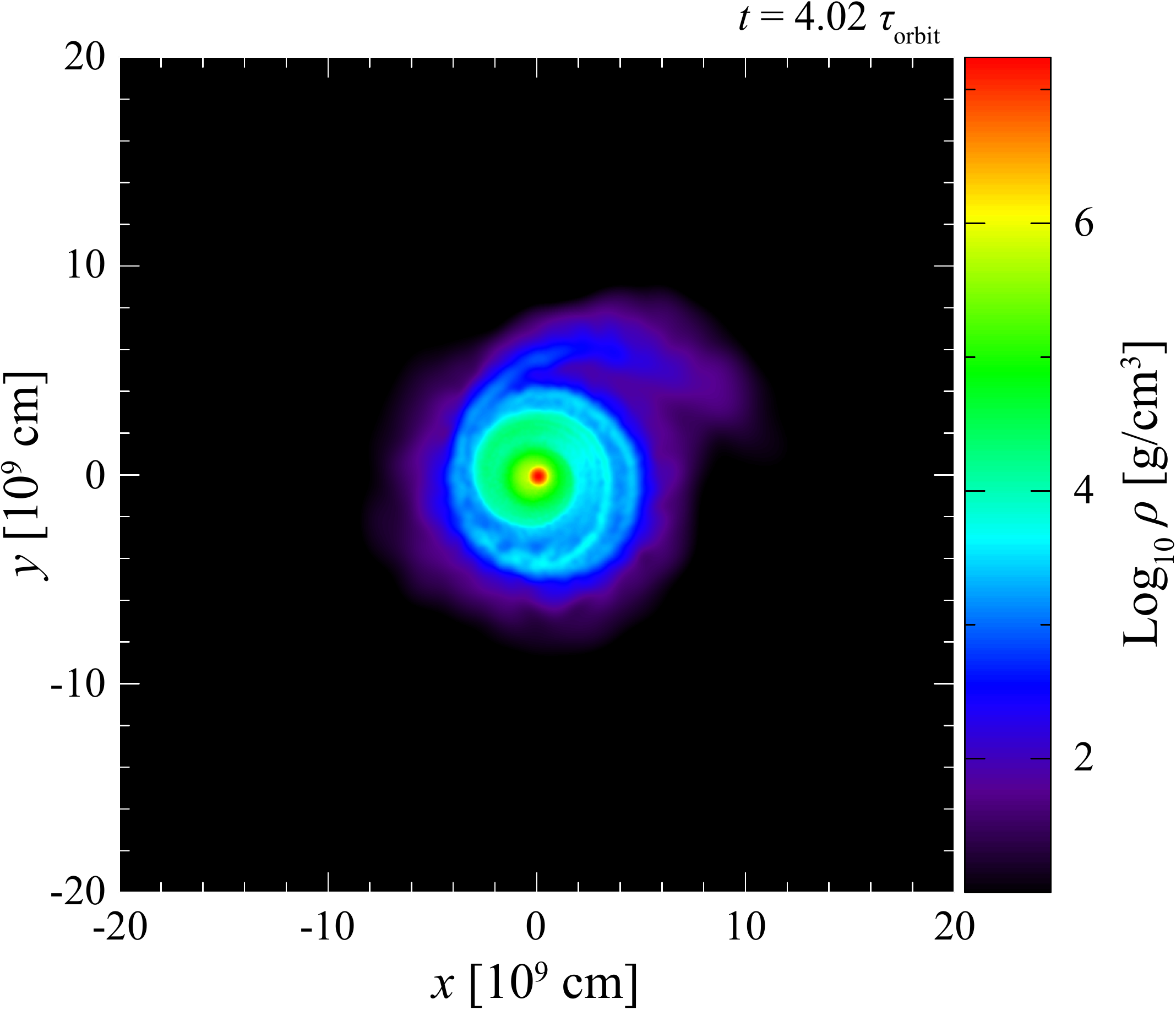}\hspace{0.5cm}
 \includegraphics[height=2.7in]{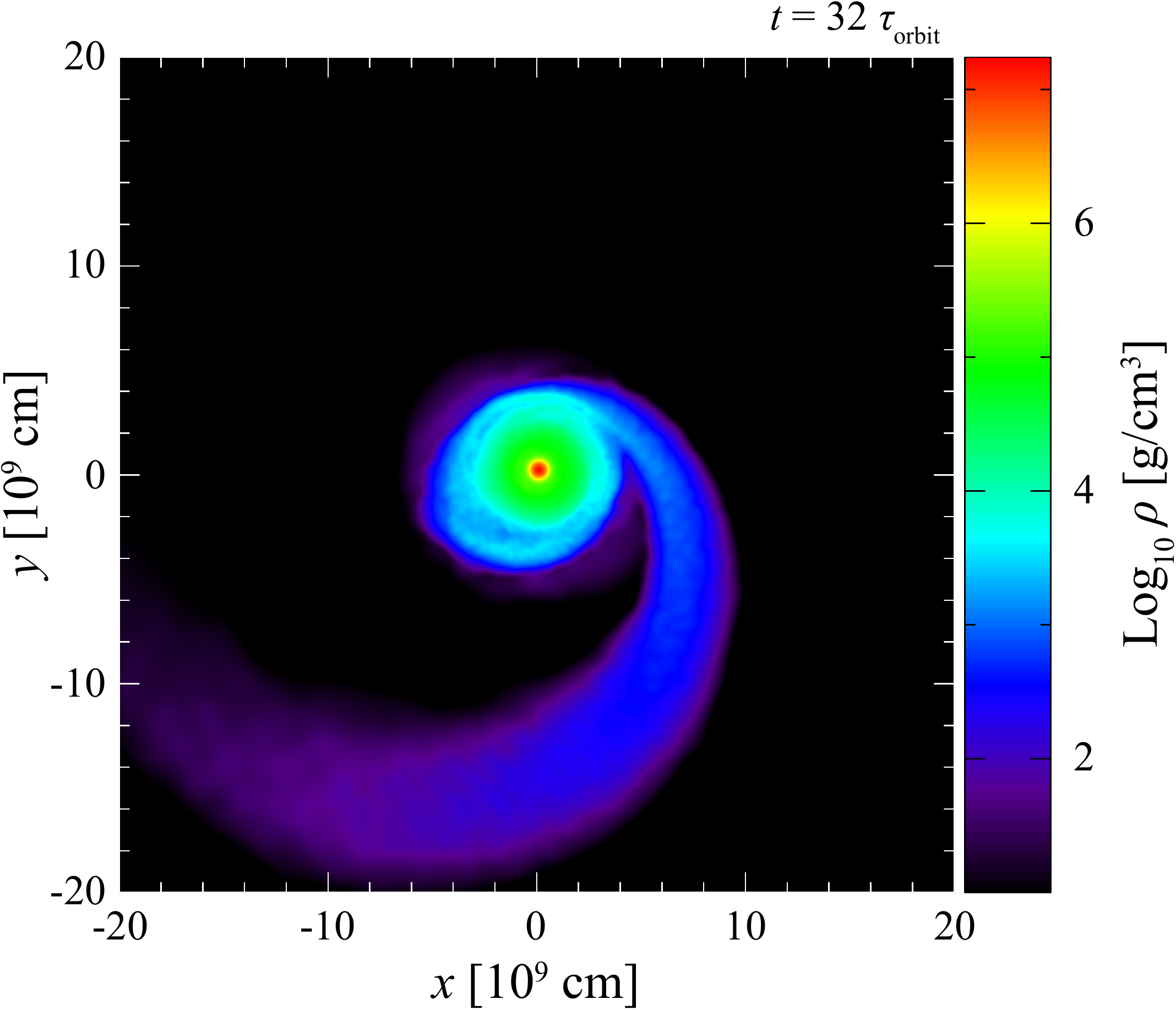}
}
\caption{Remnant structure resulting from the merger of a $0.6 + 0.9\ M_\odot$ binary 
system, captured 3 orbits after plunging. The left (right) panel shows the result
for approximate (accurate) ICs. The initial angular momentum
content in the system in this case differs by 13\% between the two different
IC prescriptions. This slight discrepancy severely impacts their subsequent
evolution, with the duration of the mass transfer phase being severely
underestimated by the approximate prescription. The approximate  
ICs yield one orbit before plunging while the accurate ICs show a 
mass transfer for as long as 29 orbital periods.} 
\label{fig:WD06WD09_remnant}
\end{figure}

As a result of the more abrupt artificial decrease in separation, the
approximate initial conditions also give rise to large spurious oscillations
of the central 
remnant core. Such oscillations are imprinted
not only on the gravitational wave signal (see Figure \ref{fig:GWs_06_09})
but are also present on the 
thermodynamical trajectories of the merger products. As shown in
Figures~\ref{fig:thermal_02_08} and \ref{fig:thermal_06_09}, both peak
densities and temperatures 
are overestimated when using the approximate ICs. The use of accurate ICs
helps to better characterize the time evolution of the temperature, density,
and angular momentum during mass transfer as well as the final morphology of
the merger product, as seen in Figure \ref{fig:WD06WD09_remnant}. Due to the
larger 
initial separation for the more accurate ICs, the matter possess more total
angular momentum, which results in more mass in 
the trailing arm and less mass in the disk than when starting from the
approximate IC. The material ejected on eccentric orbits 
will fall back later onto the remnant \citep[similar to the case of neutron
  star mergers; see e.g.][]{rosswog07a,lee07}, the timescale of this fallback
matter being seriously underestimated when using the approximate ICs.

\begin{figure}[t]
\centerline{
 \includegraphics[height=2.0in]{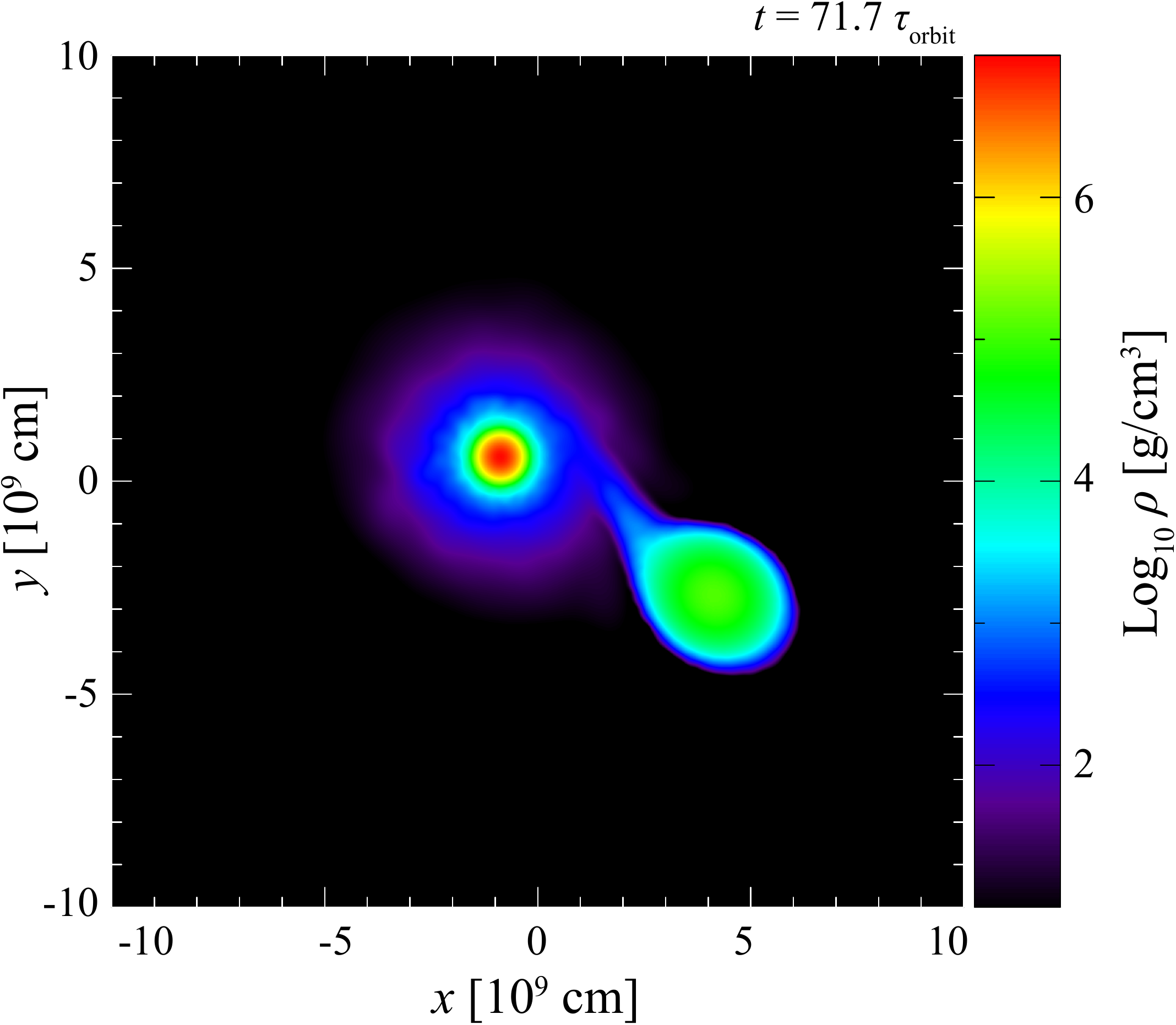}\hspace{0.25cm}
 \includegraphics[height=2.0in]{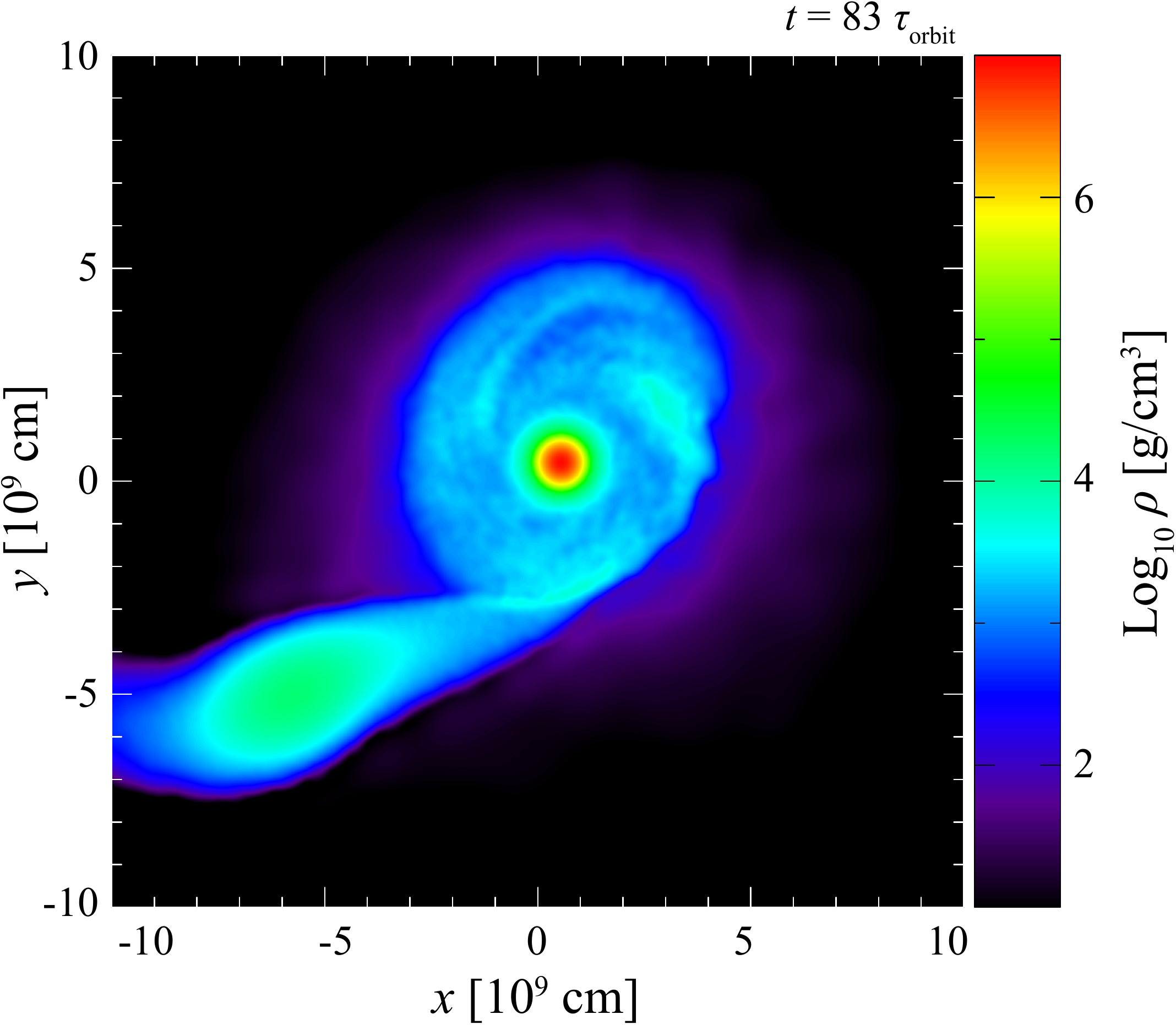}\hspace{0.25cm}
 \includegraphics[height=2.0in]{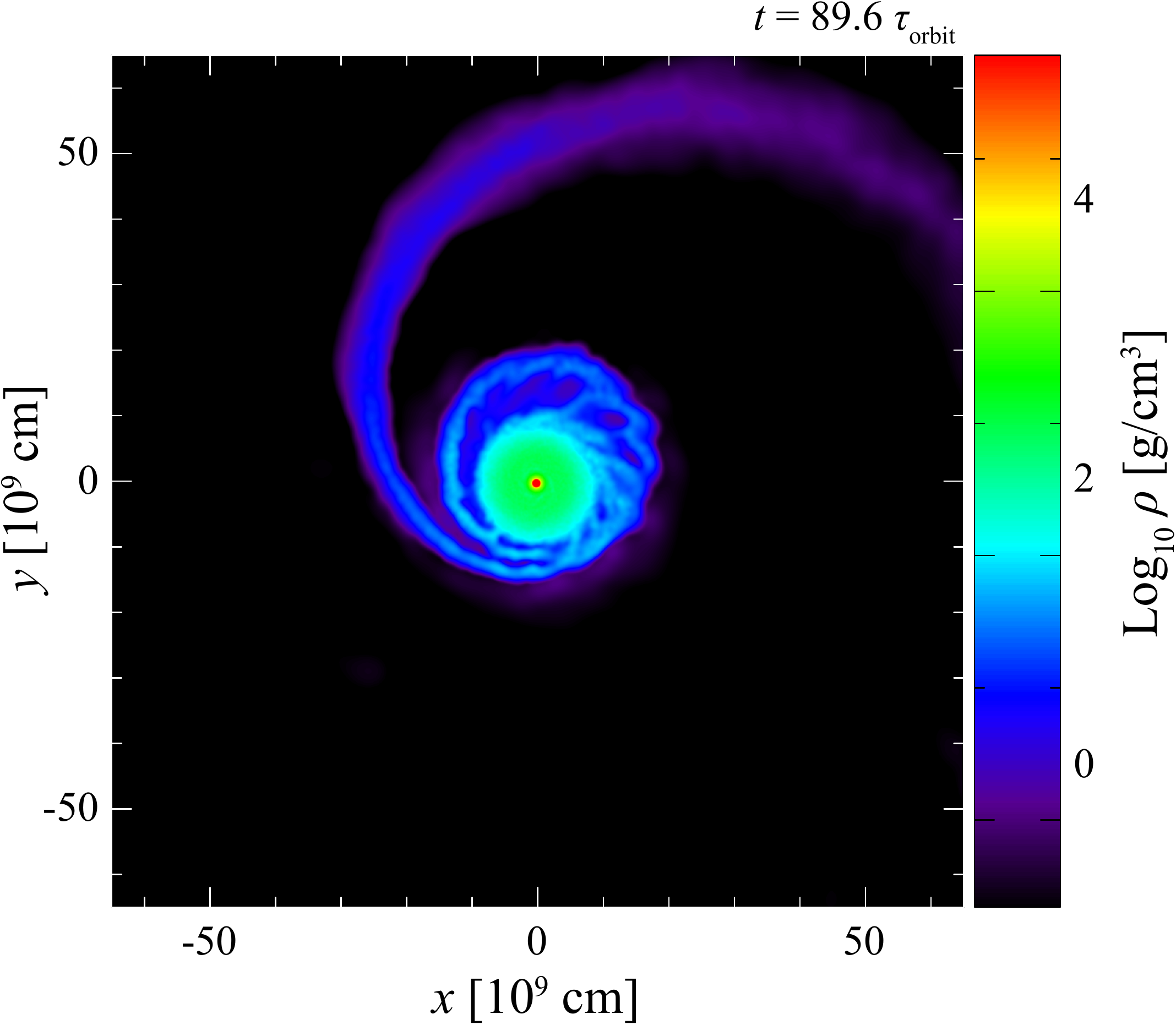}
}
\caption{Dynamical evolution of a 0.2 He and a 0.8 \Msun CO
 WD (at 71, 83 and 89 times the initial orbital
 period). 
 The mass transfer in this system continues for as long as 84 orbital periods until
 the He WD is finally destroyed. Times are shown in
 the upper right corner of each frame in units of the initial
 orbital period.} 
\label{fig:WD02WD08}
\end{figure}

\section{White Dwarf Merger Simulations with Accurate Initial Conditions}
\label{sec:results}
The use of accurate initial conditions and a correct treatment of shock
heating is necessary for a reliably temporal evolution of the
temperature, density, and angular momentum during the mass transfer phase. The
evolution of two representative binary systems is described here in detail:
the merger of a synchronized binary system with a 0.2 He WD and a 
0.8 CO WD (P1, Table \ref{tab:runs}) and a tidally locked binary with $0.6 +
0.9$ \Msun CO WDs (P5). Their evolution is then subsequently compared with the
wide range of binary systems that we have simulated as well as with the
results of earlier work. 

\subsection{0.2 + 0.8 {\msun}: A borderline unstable system near the disk
  stability limit}\label{sec:binary1} 
\begin{figure}[t]
\centerline{
\begin{tabular}{@{}cc@{}}
\includegraphics[height=2.5in]{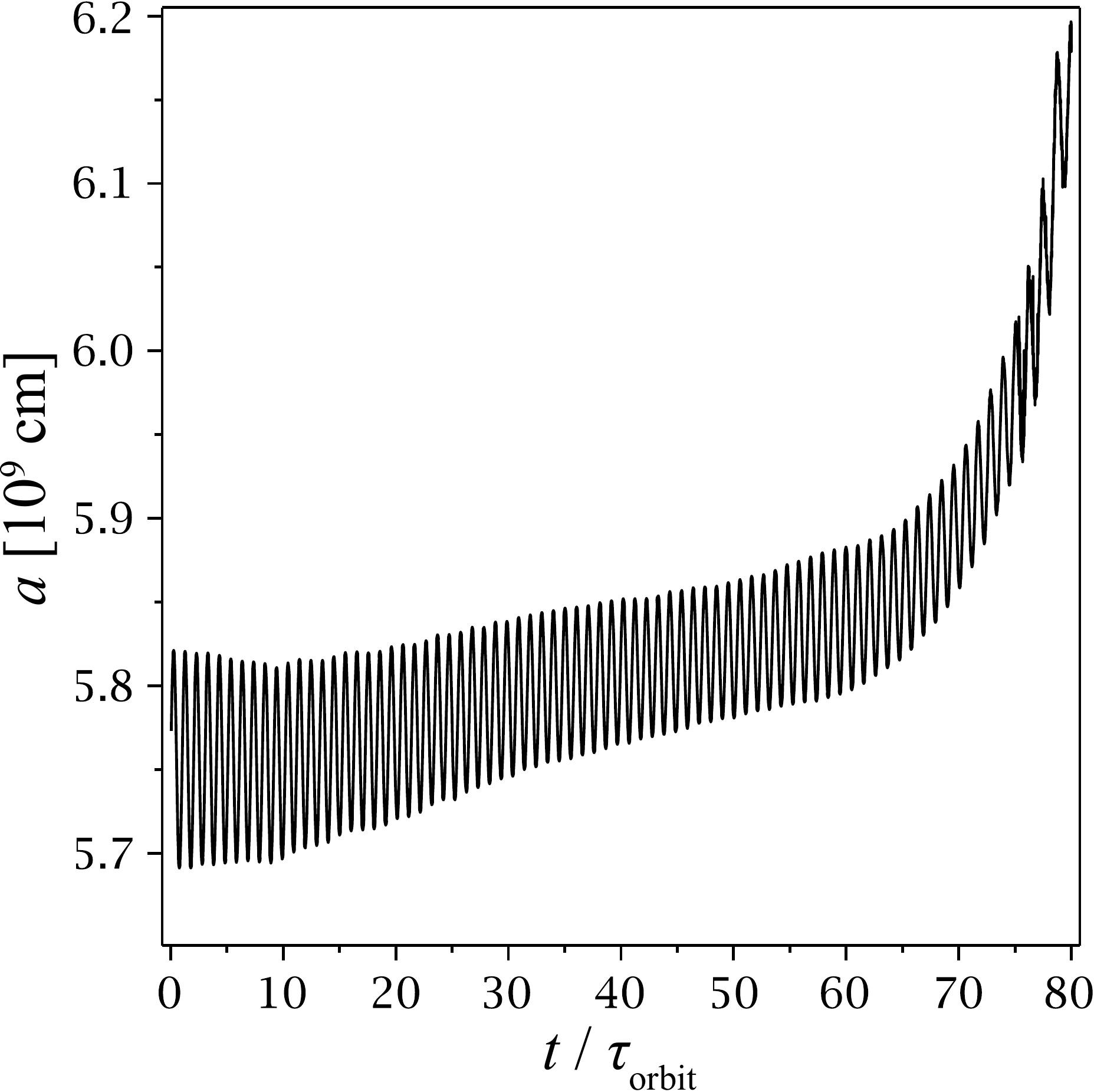}&
\includegraphics[height=2.5in]{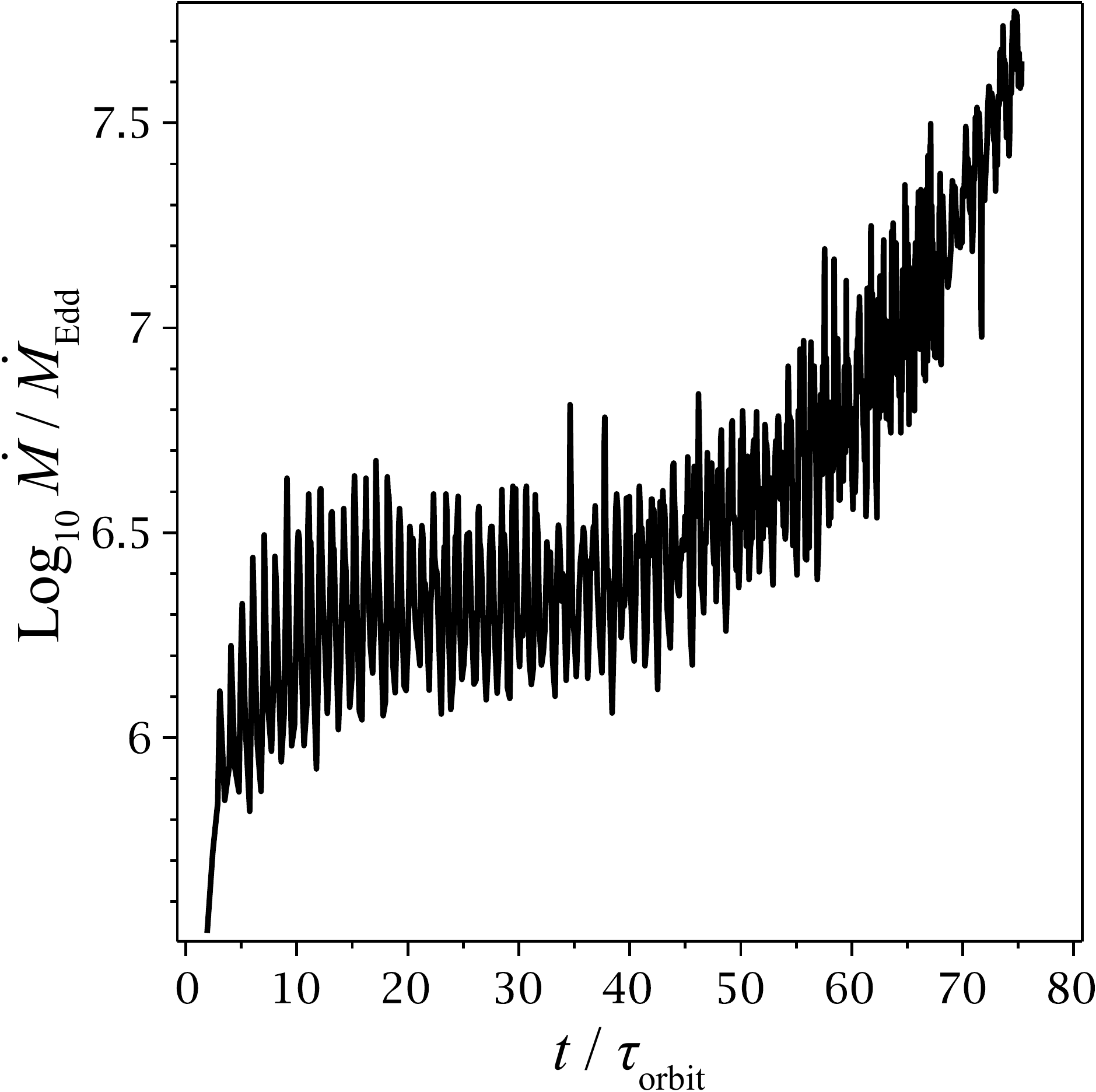}
\end{tabular}
}
\caption{Evolution of orbital separation (in $10^9\ {\rm cm}$; left)
and mass transfer rate (in units of the Eddington value; right)
of the $0.2 + 0.8 M_\odot$ binary, P1 in Table \ref{tab:runs}.}
\label{fig:WD02WD08_200Kam}
\end{figure}

According to the stability analysis of \cite{marsh04} a 0.2 He + 0.8 CO \Msun
binary system (marked as ``P1'' in  
Figure \ref{fig:marsh04}) should still be in the ``direct impact'' regime where the 
pericentre separation of the transferred mass is smaller than the accretor radius, 
but near the stable, disk forming regime. This is clearly seen in simulation
snapshots of the matter distribution plotted in Figure \ref{fig:WD02WD08}. In
this case, we find that the rotation of the 
donor star remains synchronized with the orbital motion throughout the mass transfer 
phase up to the point of tidal disruption. More importantly, as soon as mass
transfer sets in, the 
orbit is observed to become rapidly eccentric and the mutual stellar
separation begins to oscillate (with about 5\% amplitude changes), see  
Figure \ref{fig:WD02WD08_200Kam}. This oscillation, as expected, also modulates the mass 
transfer rate (Figure \ref{fig:WD02WD08_200Kam}).

The average orbital separation in this system is observed to secularly
increase and, as a result, the mass transfer is stabilized, similar to what  
is expected for AM CVn stars. The transfer rate stays nearly 
constant at a level of $\approx 10^{6.2}$ $\dot{M}_{\rm Edd}$ for about 50 orbital periods and continues until 
the donor star is completely disrupted. This occurs after as many as 84 orbital or 16,930 dynamical time scales 
of the accretor. Before being disrupted, the donor star looses about 0.08
\msun. Our results are in agreement with the Marsh et al. analysis, which
suggests the mass transfer phase to be long-lived but still unstable. 

\begin{figure}[t]
\centerline{
\begin{tabular}{@{}ccc@{}}
\includegraphics[width=2.25in]{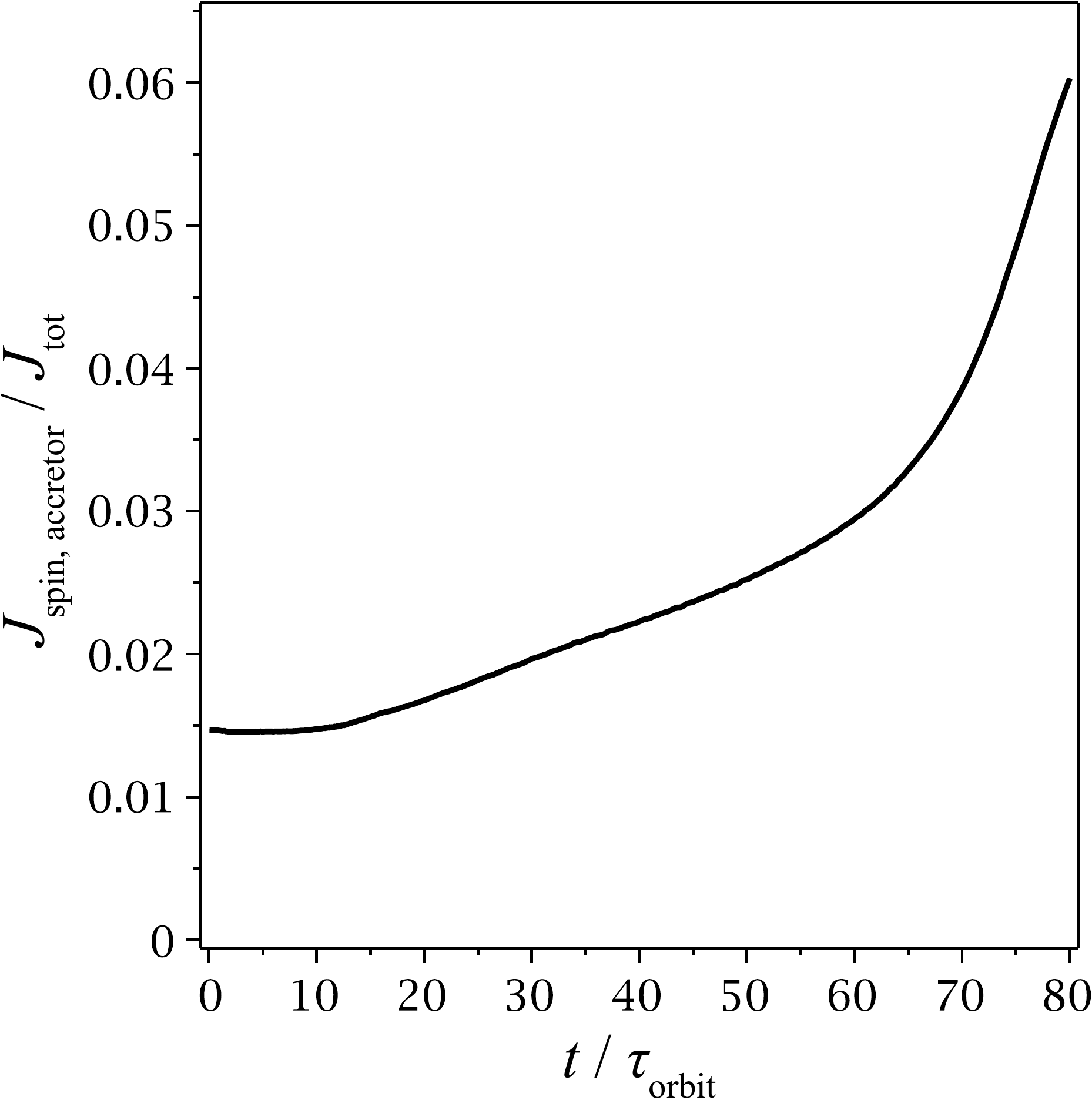}&
\includegraphics[width=2.25in]{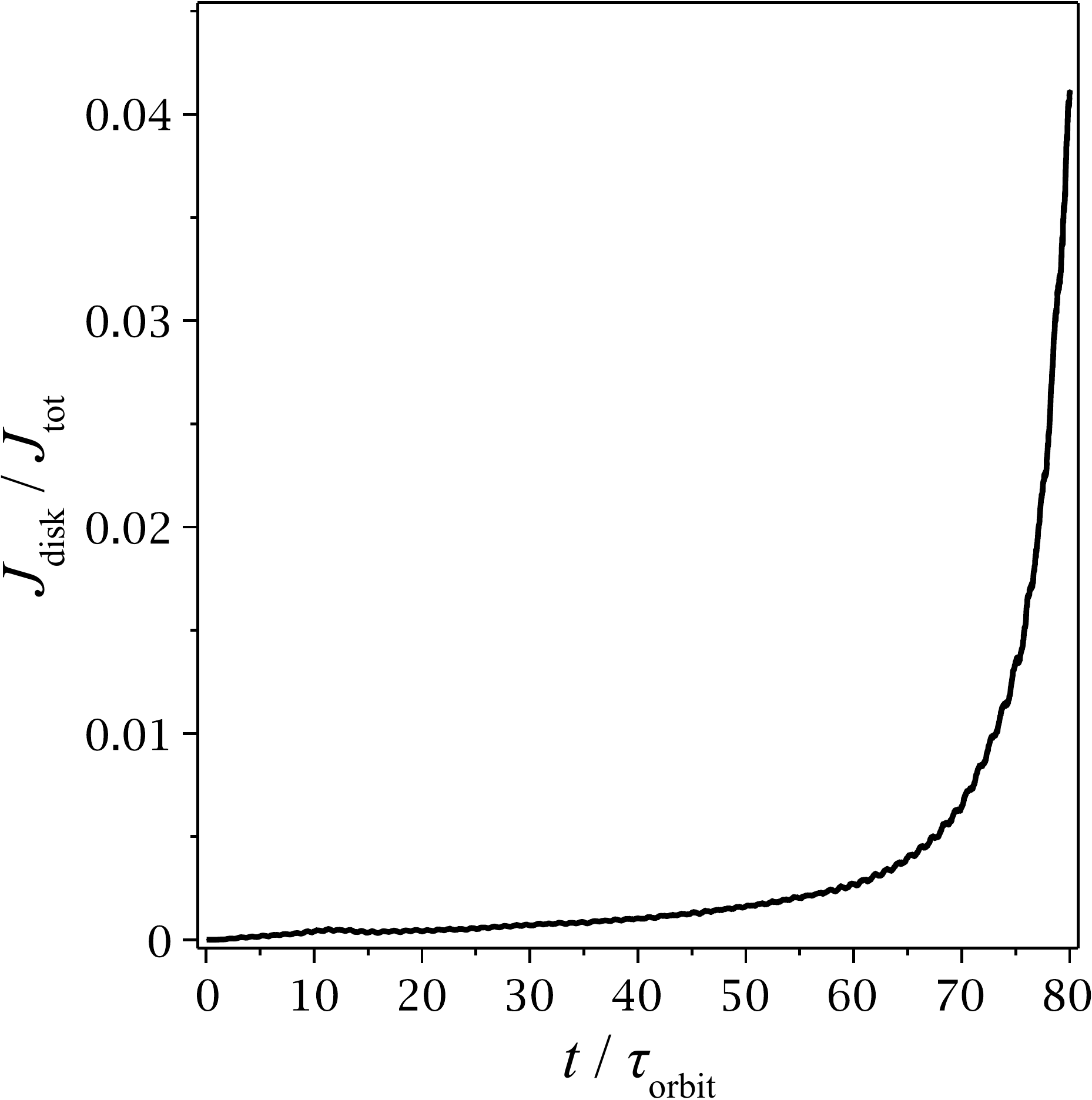}&
\includegraphics[width=2.25in]{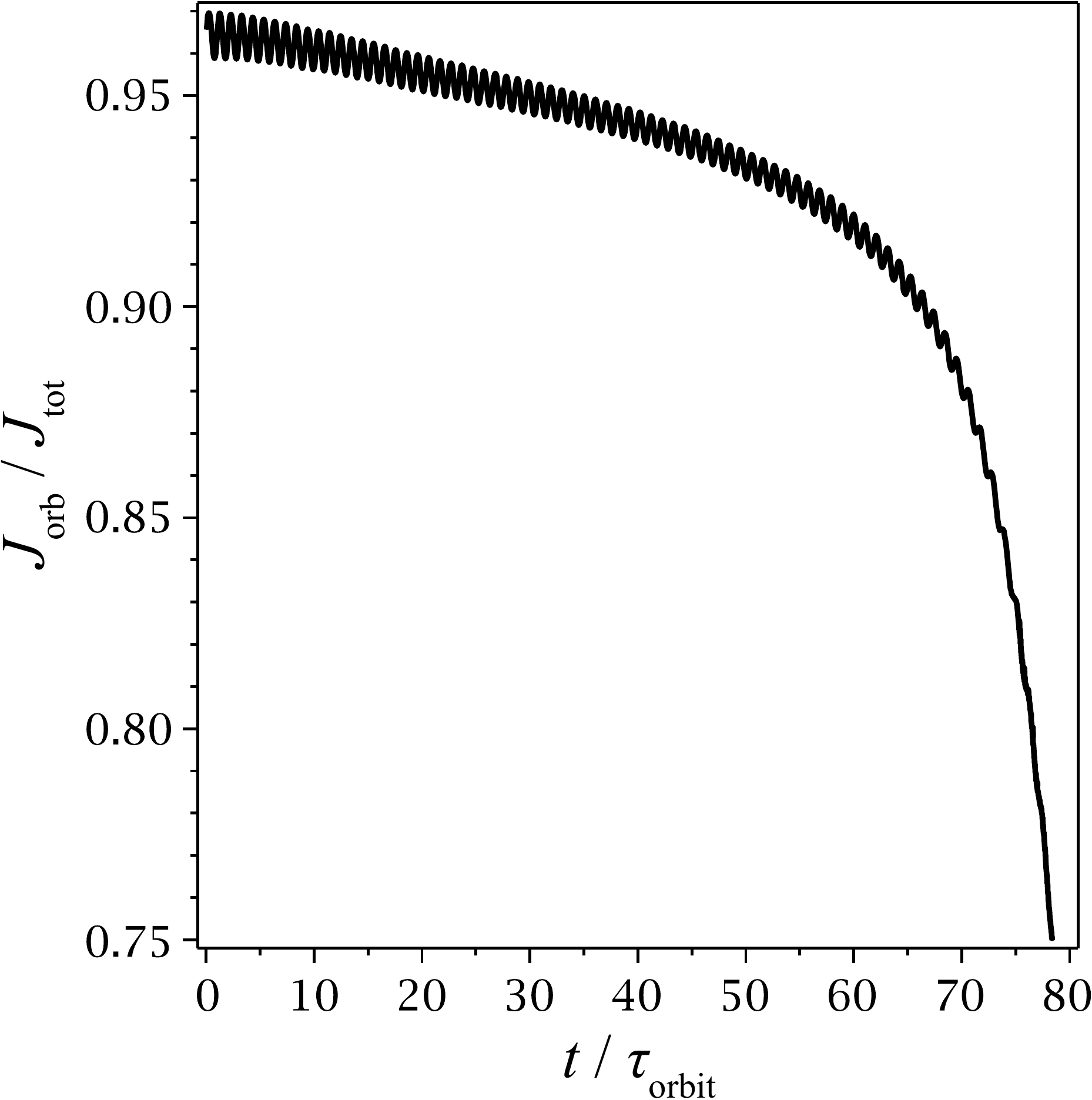}
\end{tabular}
}
\caption{Evolution of spin, disk and orbital angular momentum of the 
  $0.2 + 0.8\ M_\odot$ case (P1 in Table \ref{tab:runs}). $J_{\rm tot}$ is the initial total
angular momentum of the system.}
\label{fig:WD02WD08_200KJ}
\end{figure}

The binary system during the mass transfer phase sheds matter through the
outer Lagrange points, resulting in the loss of approximately 4\% of the total
angular momentum. As we have argued,  
the orbital dynamics is very sensitive to the loss of angular momentum which, in turn, prevents 
efficient mass transfer stabilization. The orbital angular momentum that is carried away is shared 
between the accreting star and the disk (Figure \ref{fig:WD02WD08_200KJ}),
and, due to the constant change in orbital separation, is observed to
oscillate (Figure \ref{fig:WD02WD08_200KJ}).

\begin{figure}[!h]
\centerline{
 \includegraphics[height=2.5in]{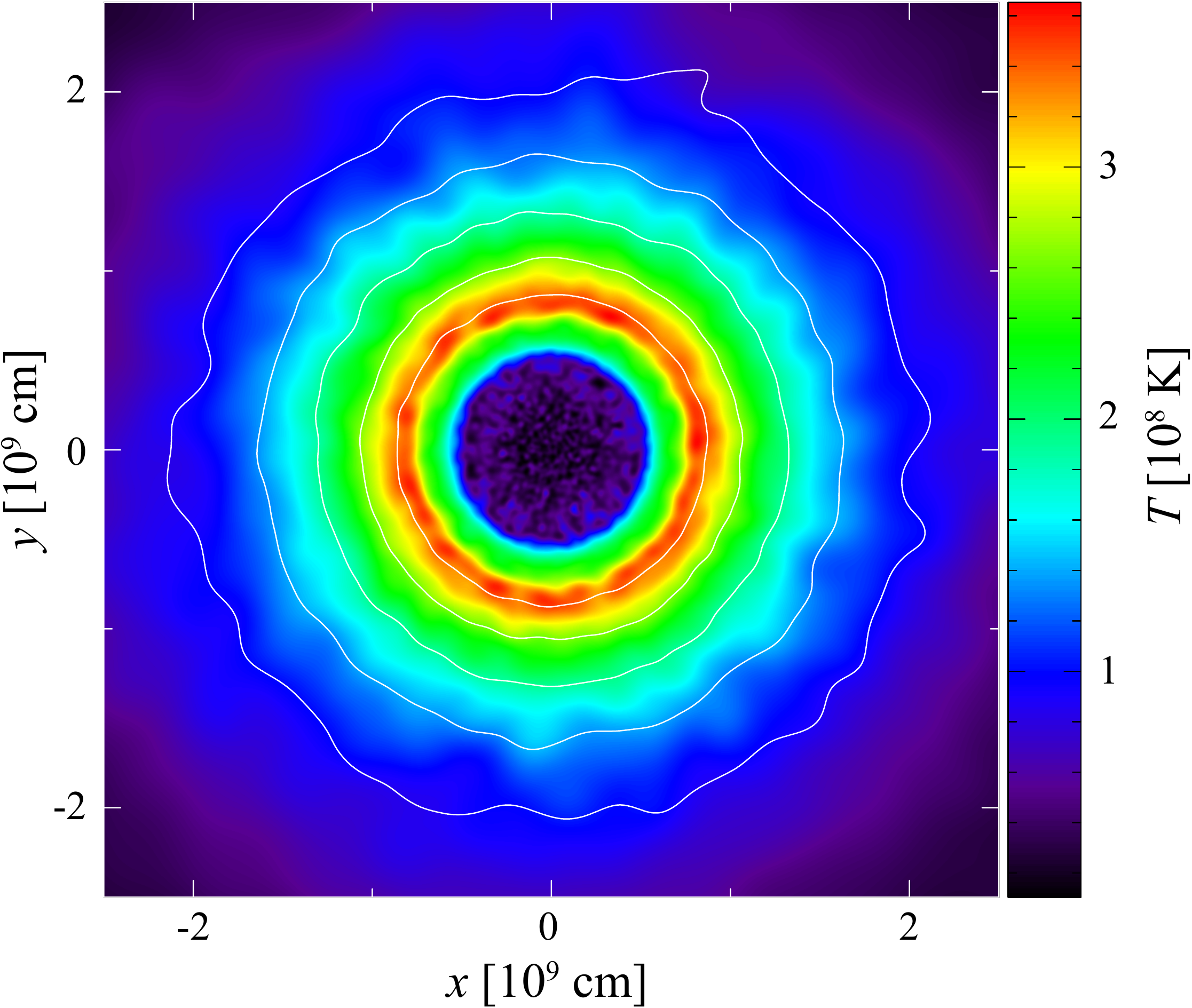}\hspace{0.5cm}
 \includegraphics[height=2.5in]{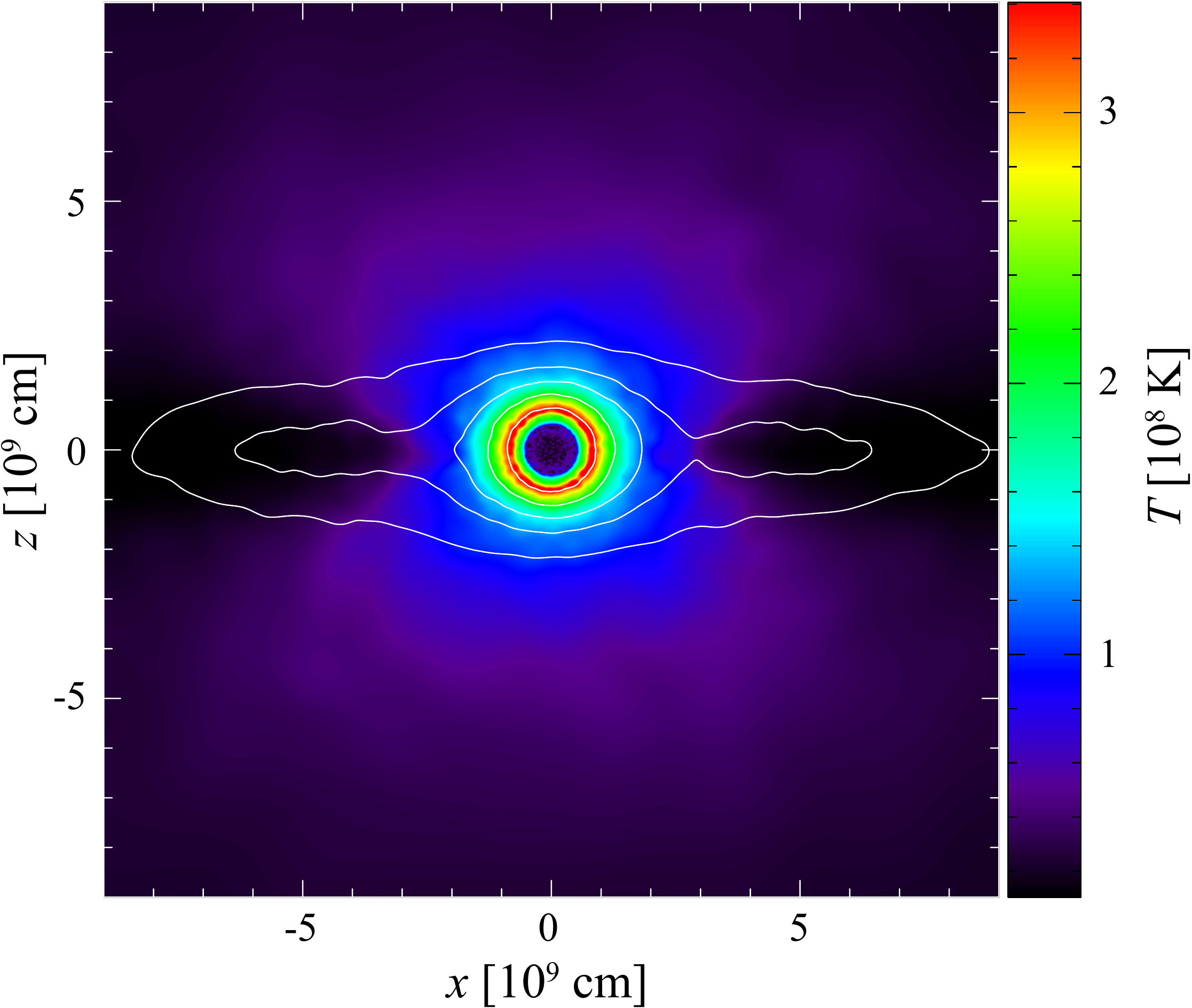}}
\caption{Densities and temperatures of the final remnant resulting from the merger of $0.2+0.8$ \Msun 
  binary (P1): $xy$-plane (left) and $xz$-plane (right). 
  Color-coded is the temperature (in units of $10^8$ K), the 
overlaid white contours refer to $\log{\rho}$ ($\rho$ in \gcc). The contours
in the left panel show densities ranging from $\log_{10} \rho = 4$ (innermost
contour) to $\log_{10} \rho = 2$ (outermost contour) in steps of 0.5, while
the contours in the right panel range from $\log_{10} \rho = 4.1$ to
$\log_{10} \rho = 2.9$ in steps of 0.3.}  
\label{fig:rho_T_remnant0208}
\end{figure}

The densities and temperatures of the final remnant core are shown in Figure \ref{fig:rho_T_remnant0208}
for the $0.2+0.8$ \Msun case. The highest temperatures are reached in the shearing region between the central core and the 
surrounding thick disk. The peak temperatures of around $4 \times 10^8$ K are reached at densities of only about $10^{5}$ \gcc, so
that no efficient carbon burning is expected to occur.

\subsection{0.6 + 0.9 \Msun: An unstable binary system with direct impact mass transfer}\label{sec:binary2}
\begin{figure}[!t]
\centerline{
 \includegraphics[height=2.15in]{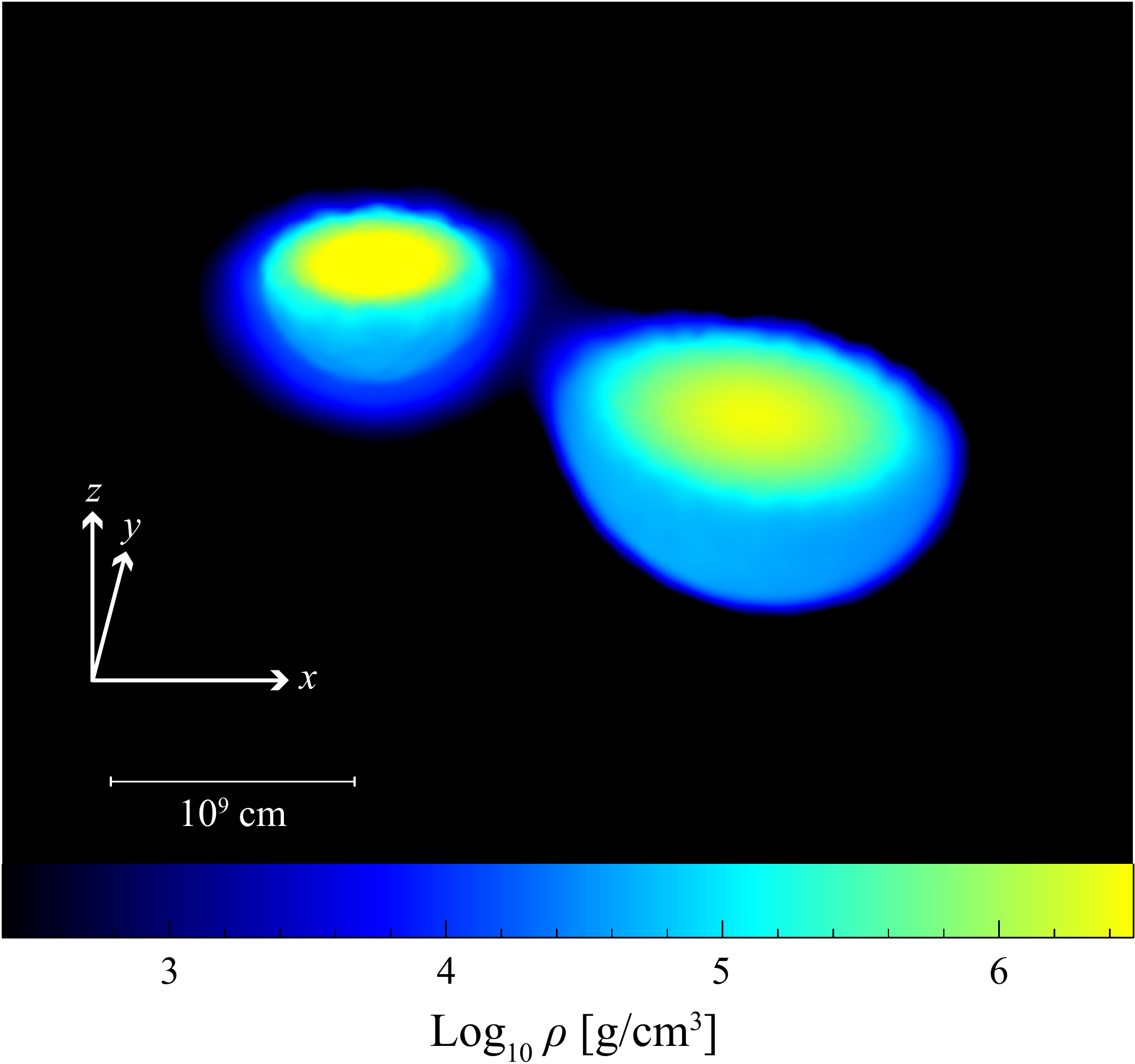}\hspace{0.15cm}
 \includegraphics[height=2.15in]{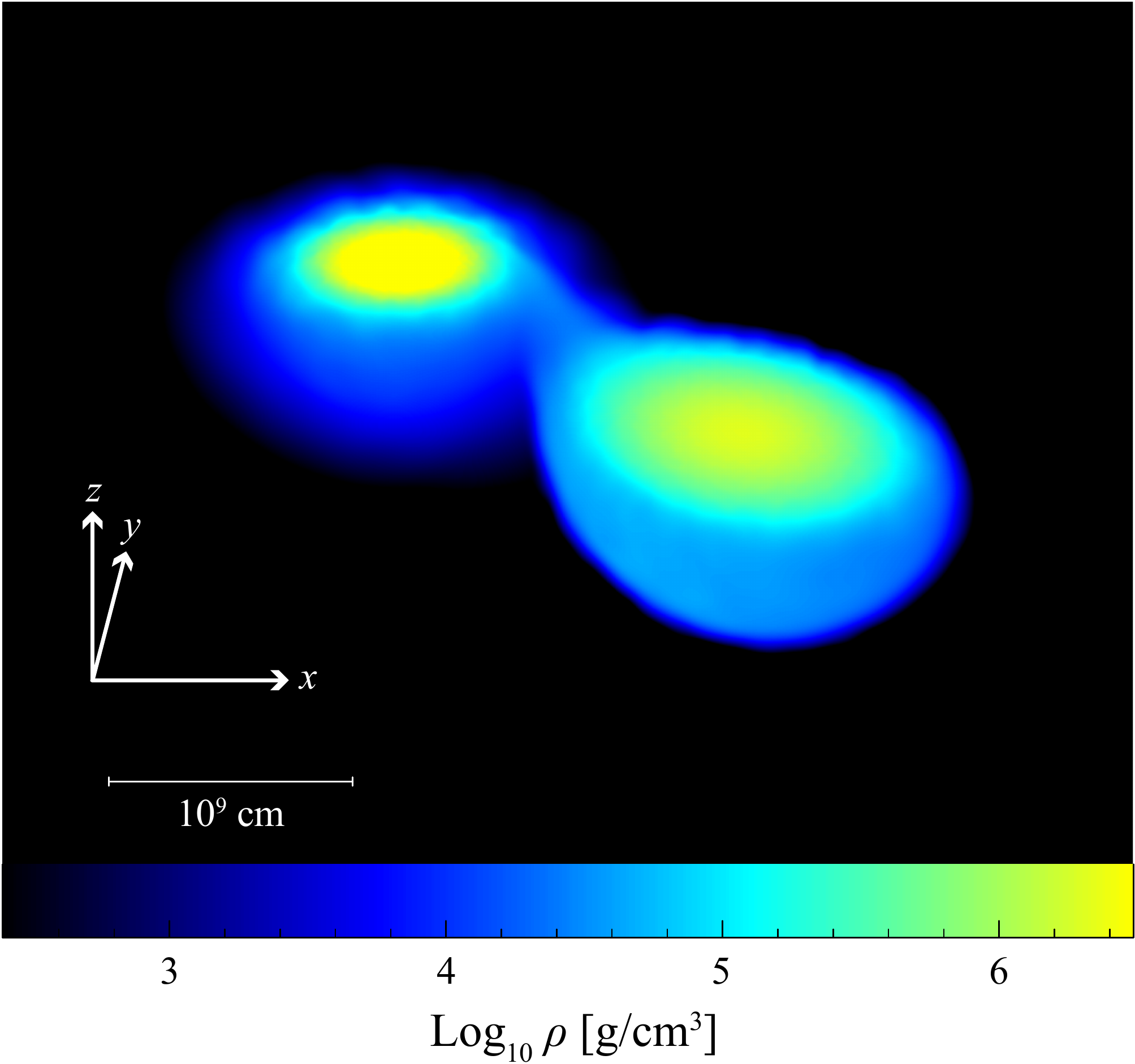}\hspace{0.15cm}
 \includegraphics[height=2.15in]{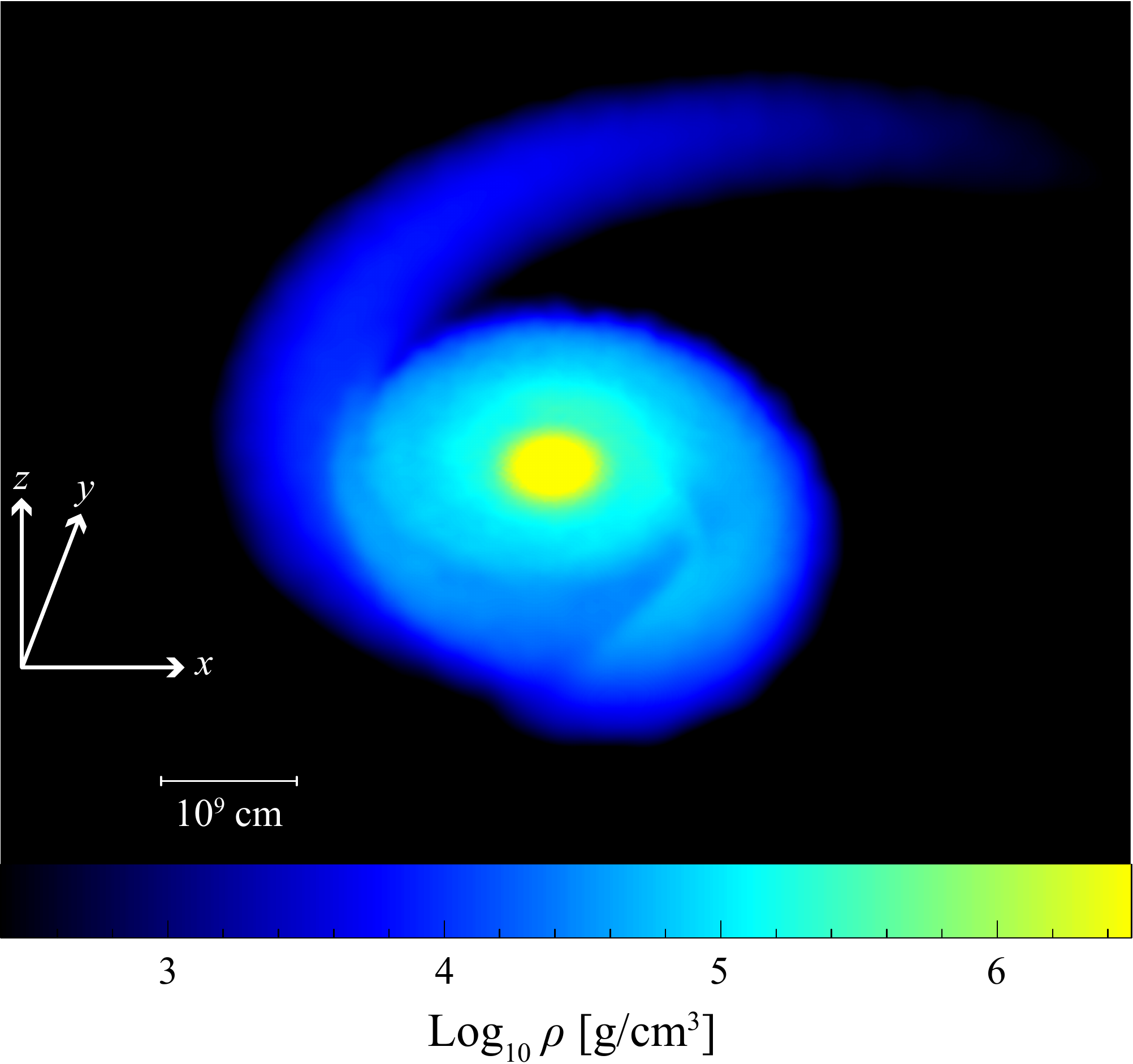}
}
\caption{Evolution of the dynamically unstable system of two CO white
 dwarfs with $0.6 + 0.9$ \msun. The panels show
 three-dimensional renderings of the density at 16, 27 and 30 times the
 initial orbital period.} 
\label{fig:WD06WD093d}
\end{figure}

\begin{figure}[!h]
\centerline{
 \includegraphics[height=2.5in]{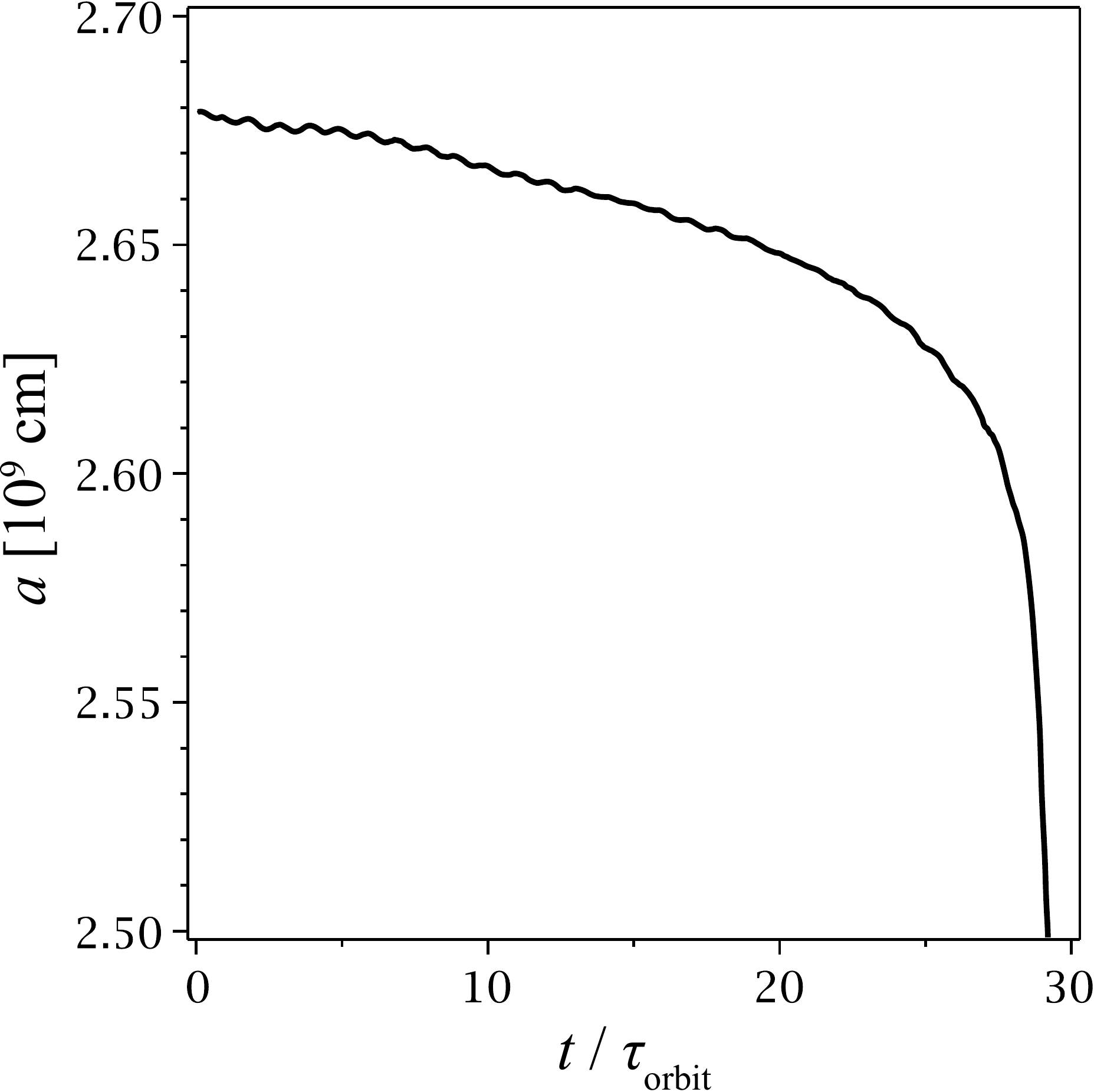}\hspace{0.5cm}
 \includegraphics[height=2.5in]{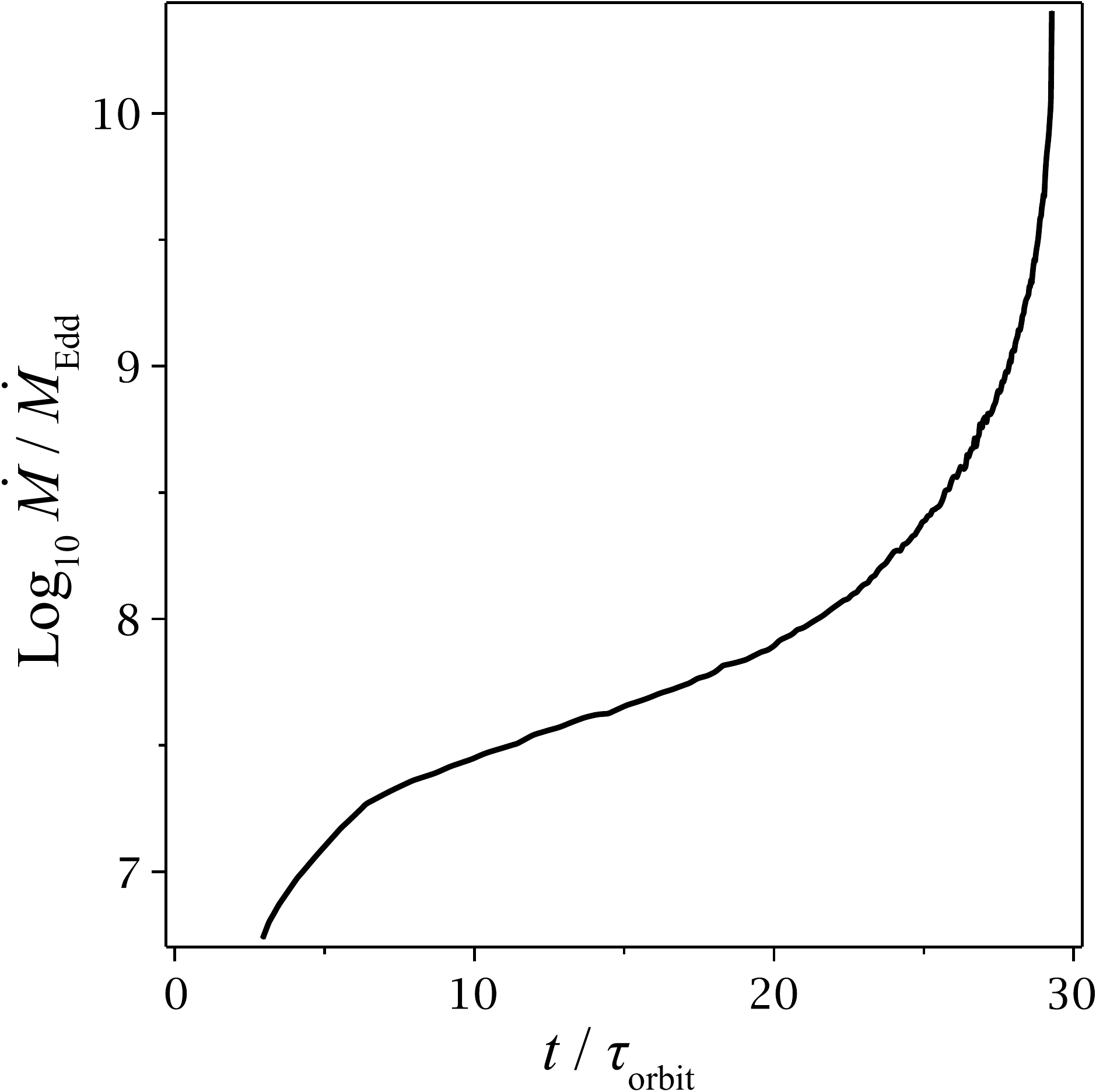}
}
\caption{Orbital separation (in $10^9$ cm; left) and mass transfer rate (in
units of the Eddington value: right) for 
  the unstable $0.6 + 0.9$ \Msun system, P5 in Table \ref{tab:runs}.}
\label{fig:WD06WD09_a_Mot}
\end{figure}

\begin{figure}[!t]
 \center\includegraphics[height=2.5in]{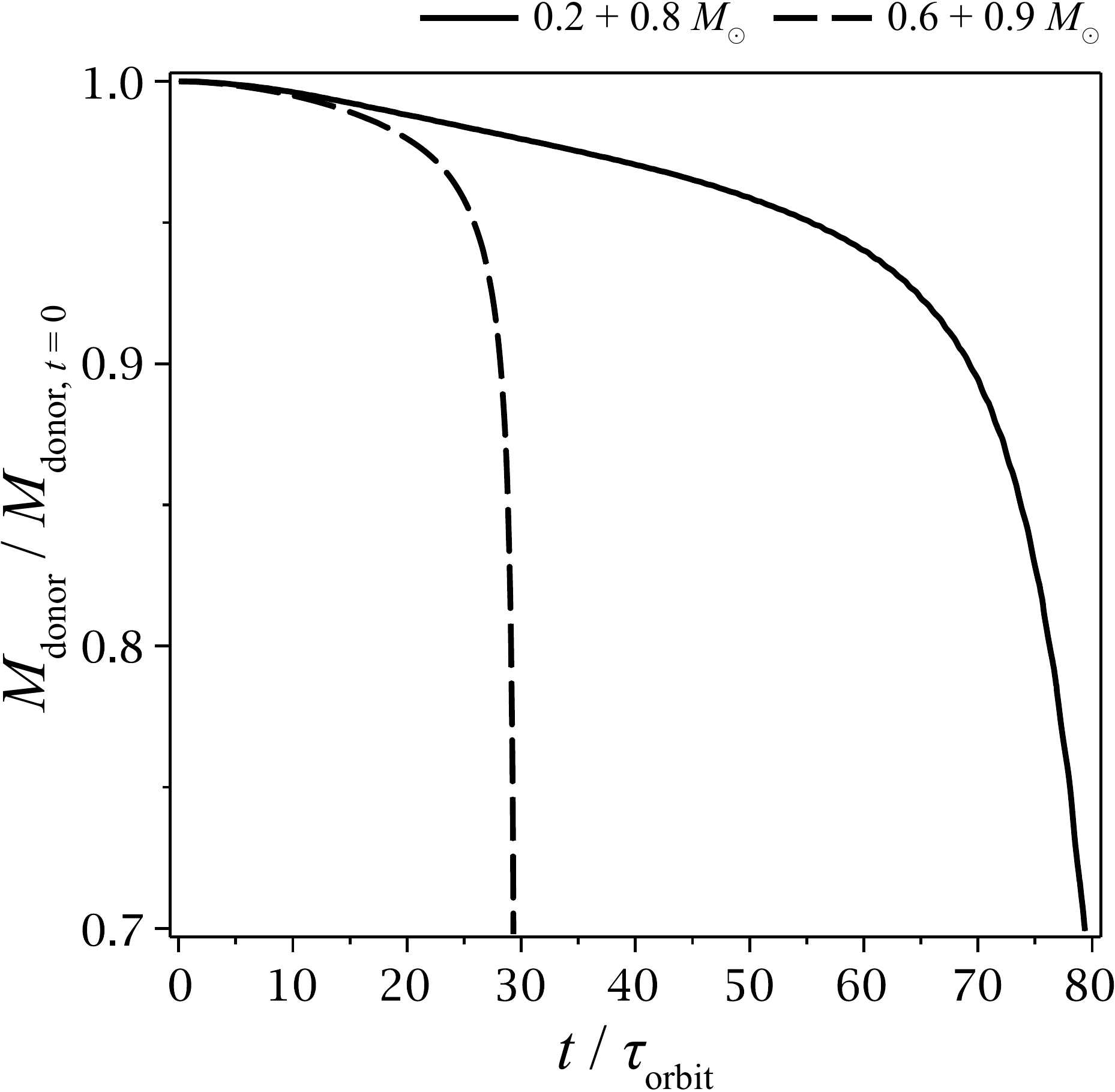}
\caption{Evolution of the donor mass normalized to its initial value for both the $0.2+0.8$ (P1) and
  $0.6+0.9$ \Msun case (P5).} 
\label{fig:M2_evolution}
\end{figure}

A $0.6 + 0.9$ \Msun binary system is predicted to be in the
unstable, direct impact regime (P5 in Figure \ref{fig:marsh04}). This is seen in 
the three dimensional density renderings plotted in Figure
\ref{fig:WD06WD093d}, which show the binary system at three different stages
of its merger evolution. Although clearly unstable, the mass transfer 
continues for about 29 orbital periods before disruption. In contrast to the
$0.2+0.8$ \Msun case, the orbit systematically shrinks from the 
onset of the mass transfer phase (Figure \ref{fig:WD06WD09_a_Mot}) and shows
no noticeable oscillations. 
The depletion of the donor star in the $0.6+0.9$ \Msun and $0.2+0.8$ \Msun
binary systems is shown in Figure \ref{fig:M2_evolution}. In the $0.2+0.8$
\Msun case, the donor mass decreases very regularly to about 90\%, followed by
a phase where the mass loss speeds up dramatically and the donor becomes
subsequently disrupted during the remaining 20 orbits. Once mass transfer
begins in the $0.6 + 0.9$ \Msun case, the rate of mass loss from the donor
star increases exponentially with time, leading to a swift depletion of the
donor. 

\begin{figure}[!h]
\centerline{
 \includegraphics[height=2.35in]{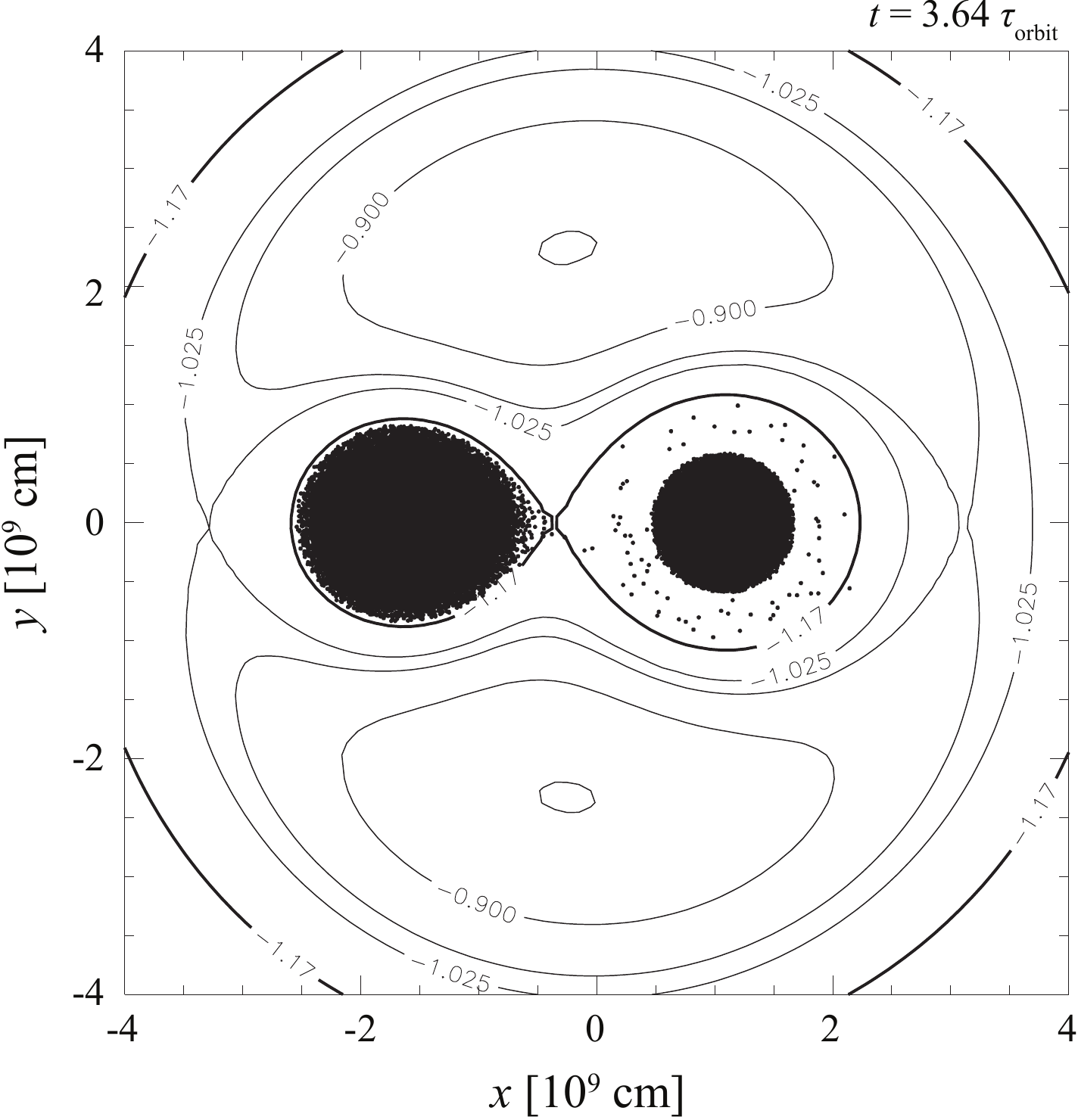}\hspace{0.25cm}
 \includegraphics[height=2.35in]{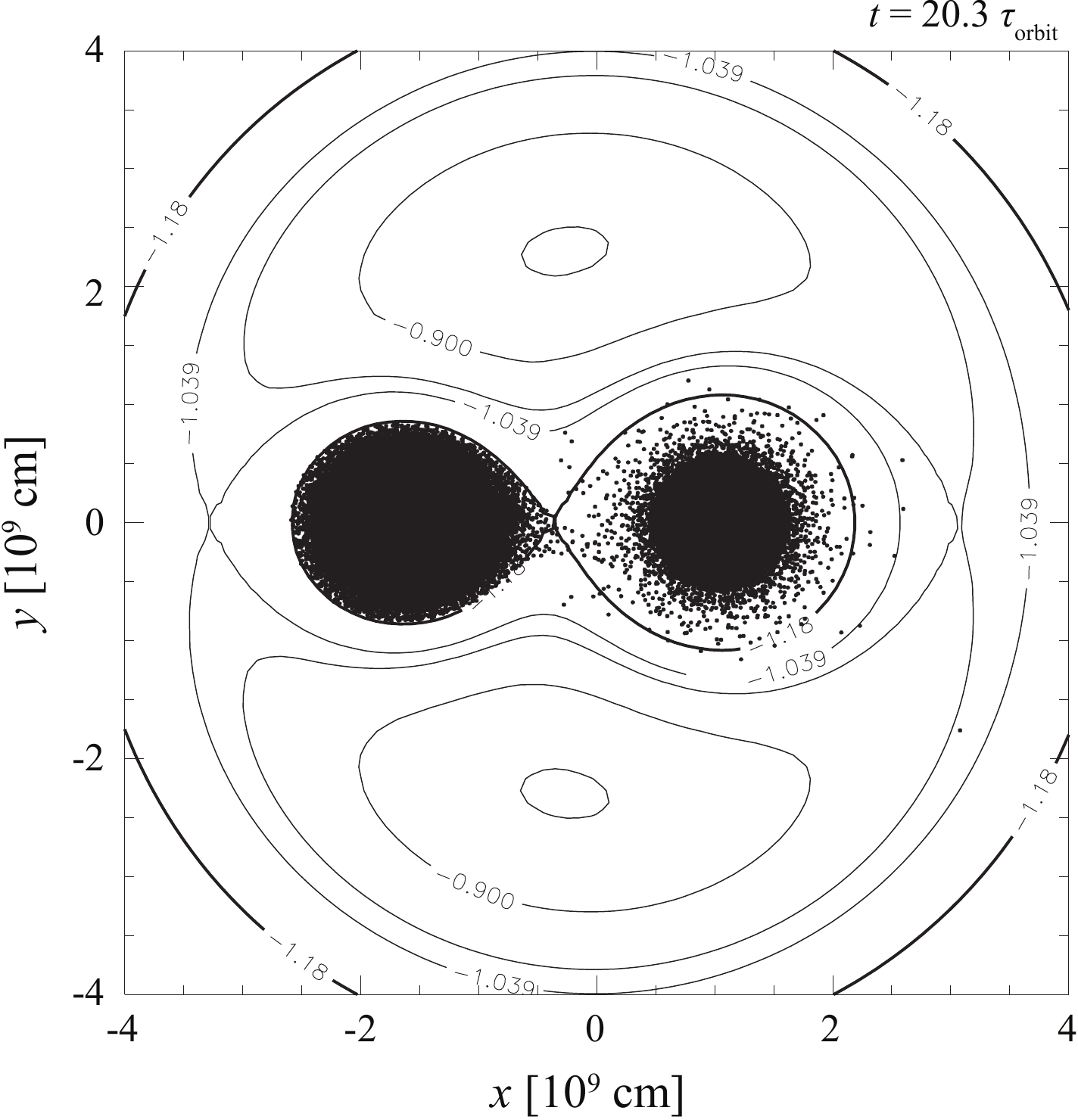}\hspace{0.25cm}
 \includegraphics[height=2.35in]{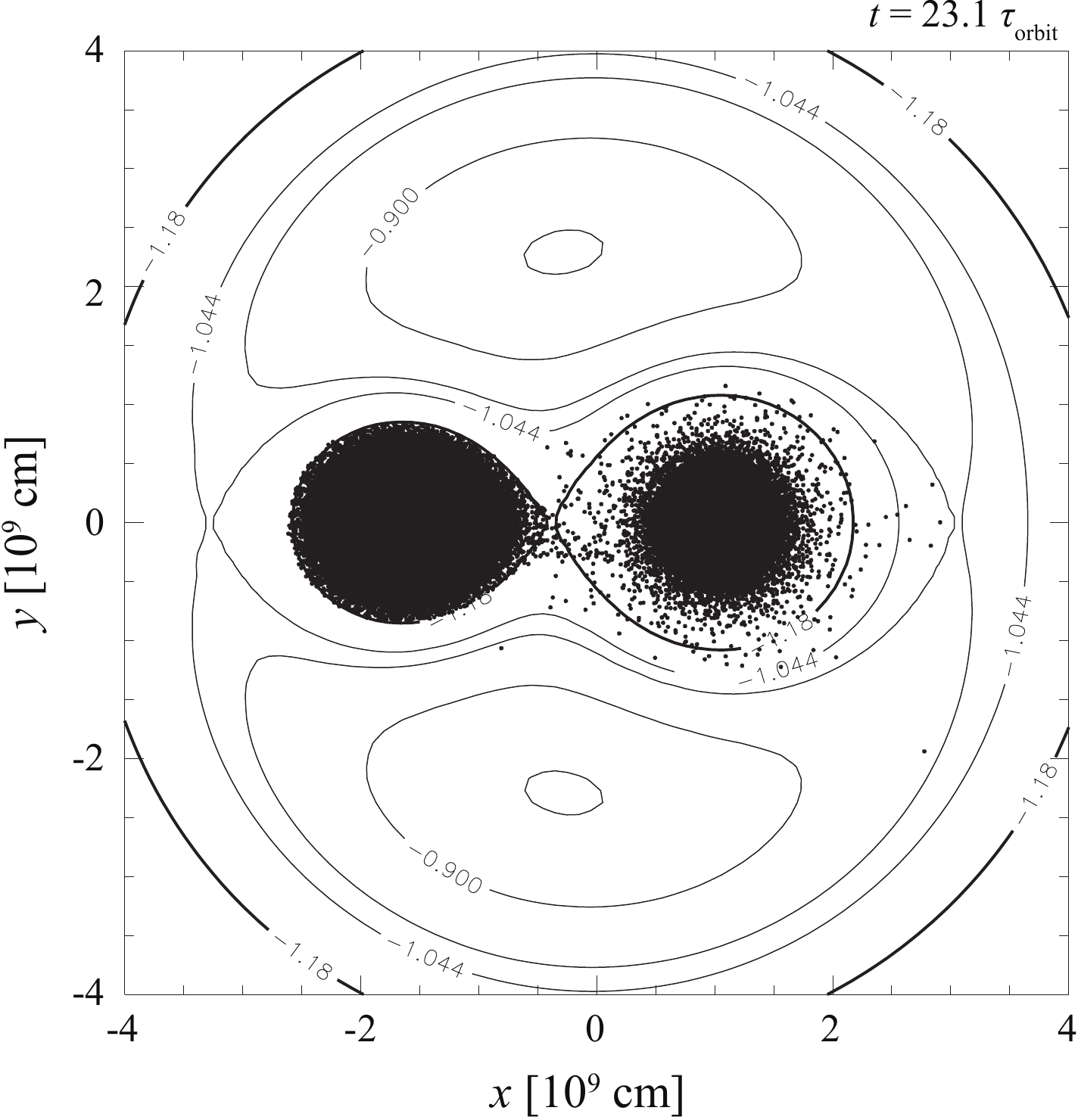}
}
\caption{Roche contours for the system with $0.6 + 0.9$ \Msun
components. Snapshots are taken after 3, 20 and 23 times the initial
orbital period.} 
\label{fig:WD06WD09roche}
\end{figure}

\begin{figure}[!h]
\centerline{
 \includegraphics[height=2.15in]{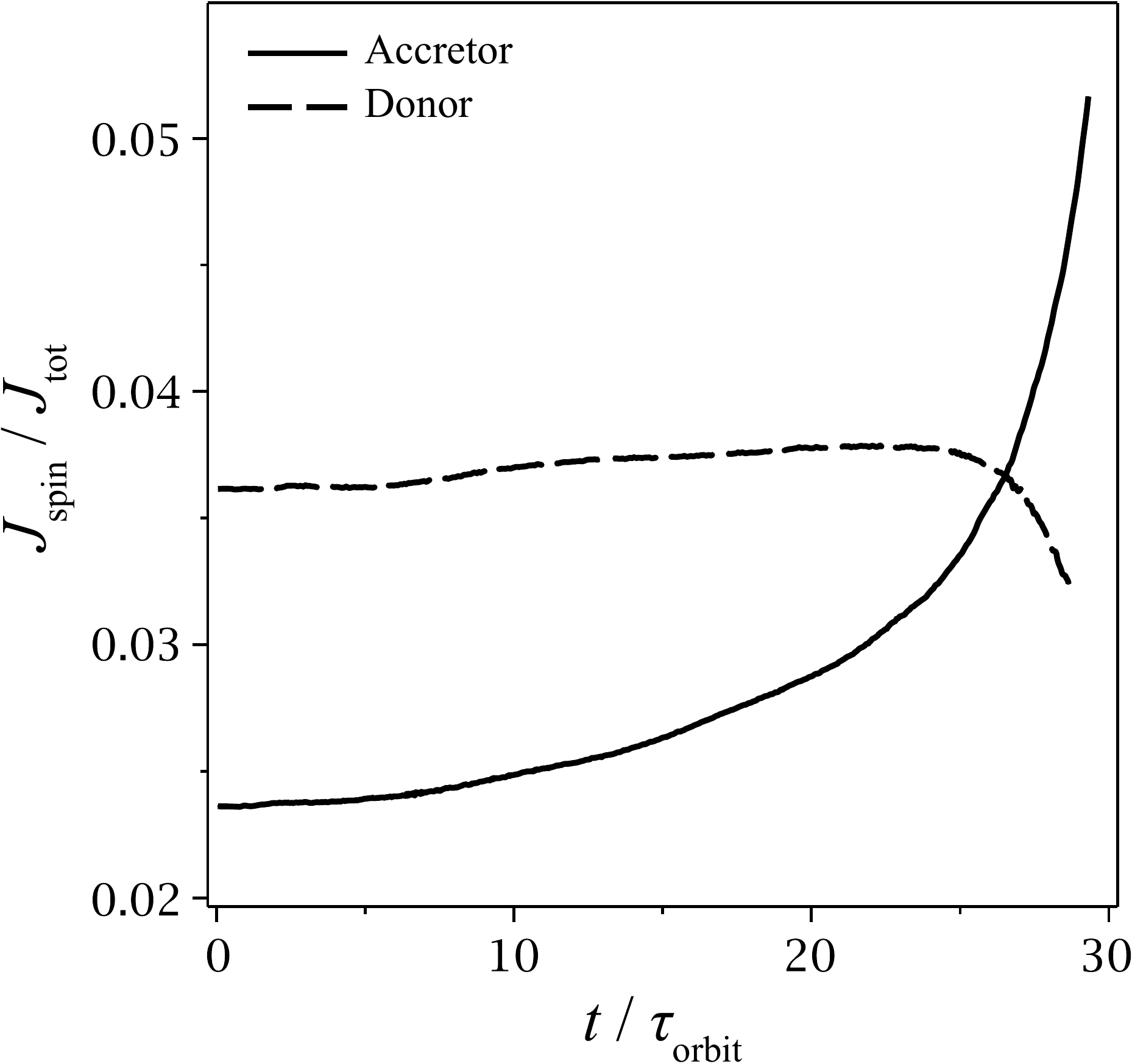}\hspace{0.5cm}
 \includegraphics[height=2.15in]{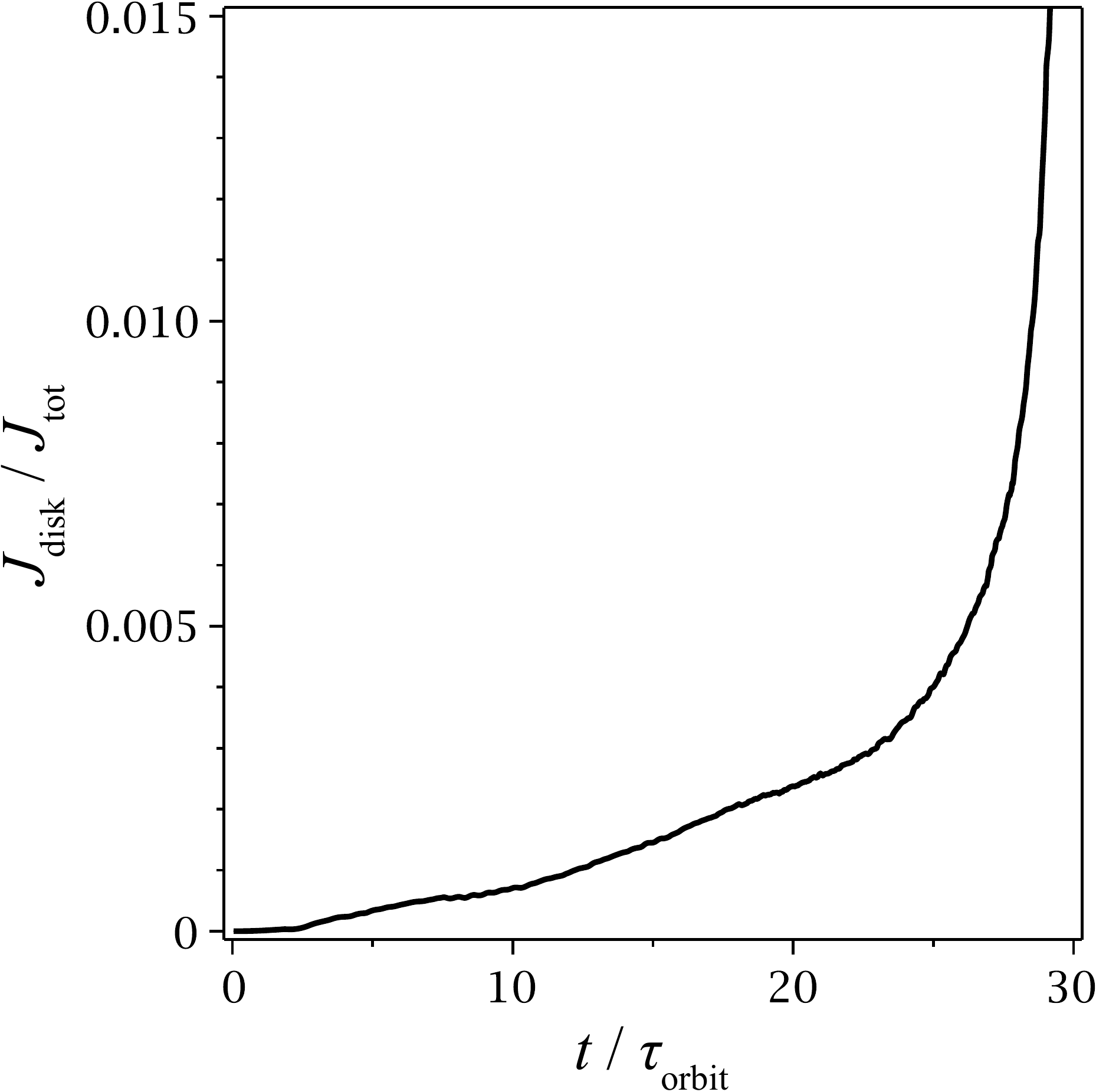}\hspace{0.5cm}
 \includegraphics[height=2.15in]{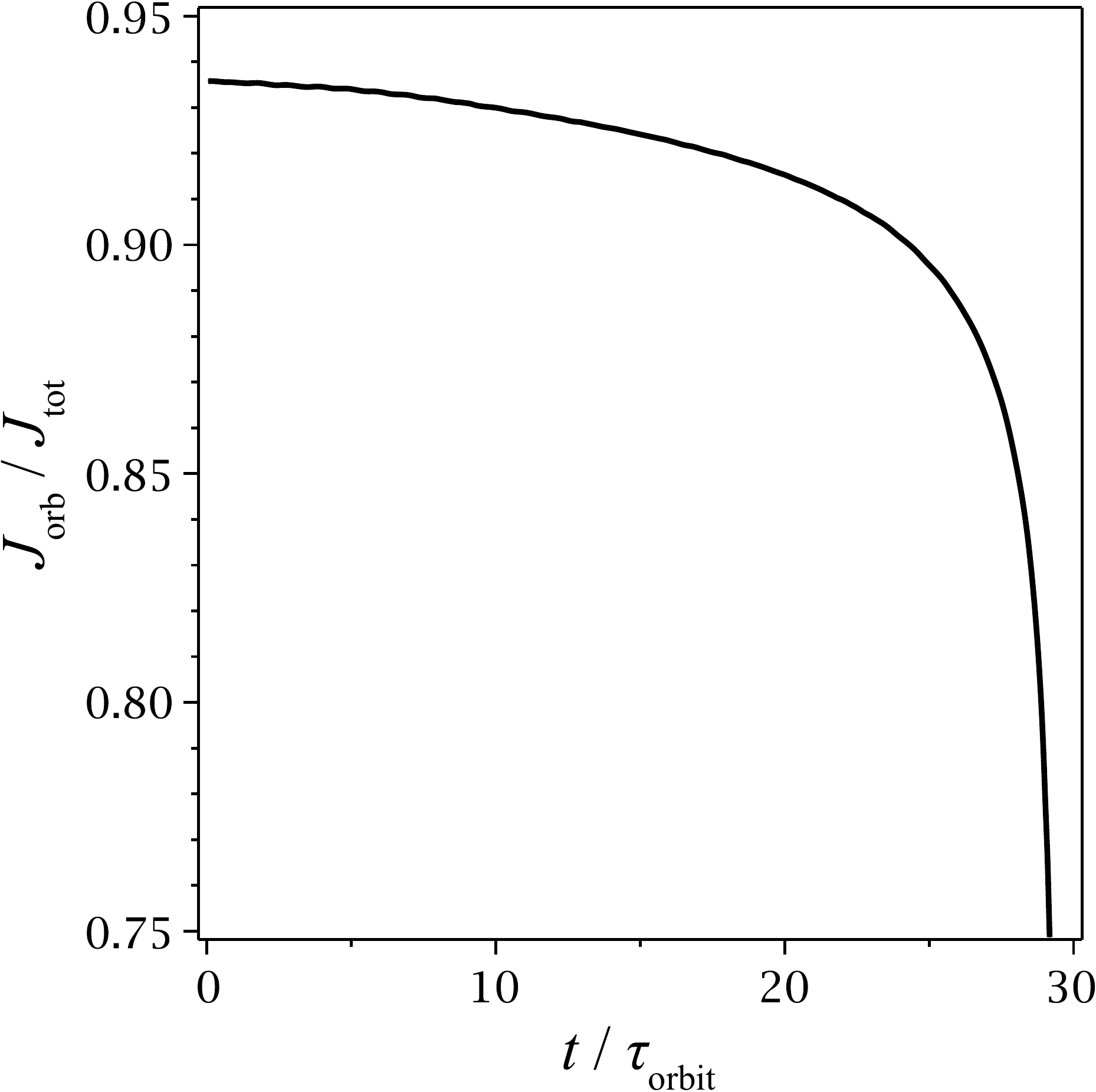}
}
\caption{Dynamical evolution of the $0.6 + 0.9\ M_\odot$ binary system. From
  left to right we have plotted the spin, disk and orbital angular 
momentum (normalized to the initial total angular momentum of the system:
$J_{\rm tot}$), respectively.}
\label{fig:WD06WD09_200K_ang_mom}
\end{figure}

The motion of the SPH particles within the Roche-type potential $\Psi$ (see 
Equation \ref{eq:psi_potential}) are shown in Figure \ref{fig:WD06WD09roche}. As expected from a direct impact 
configuration, the angular momentum carried across is efficiently converted 
into the spin of the accreting star (Figure \ref{fig:WD06WD09_200K_ang_mom}), with a small fraction being stored
in the accretion torus. Similar to the $0.2+0.8$ \Msun binary system, the
donor star is observed to remain close to synchronization during the entire
duration of the mass transfer phase.


\begin{figure}[!h]
\centerline{
 \includegraphics[height=2.5in]{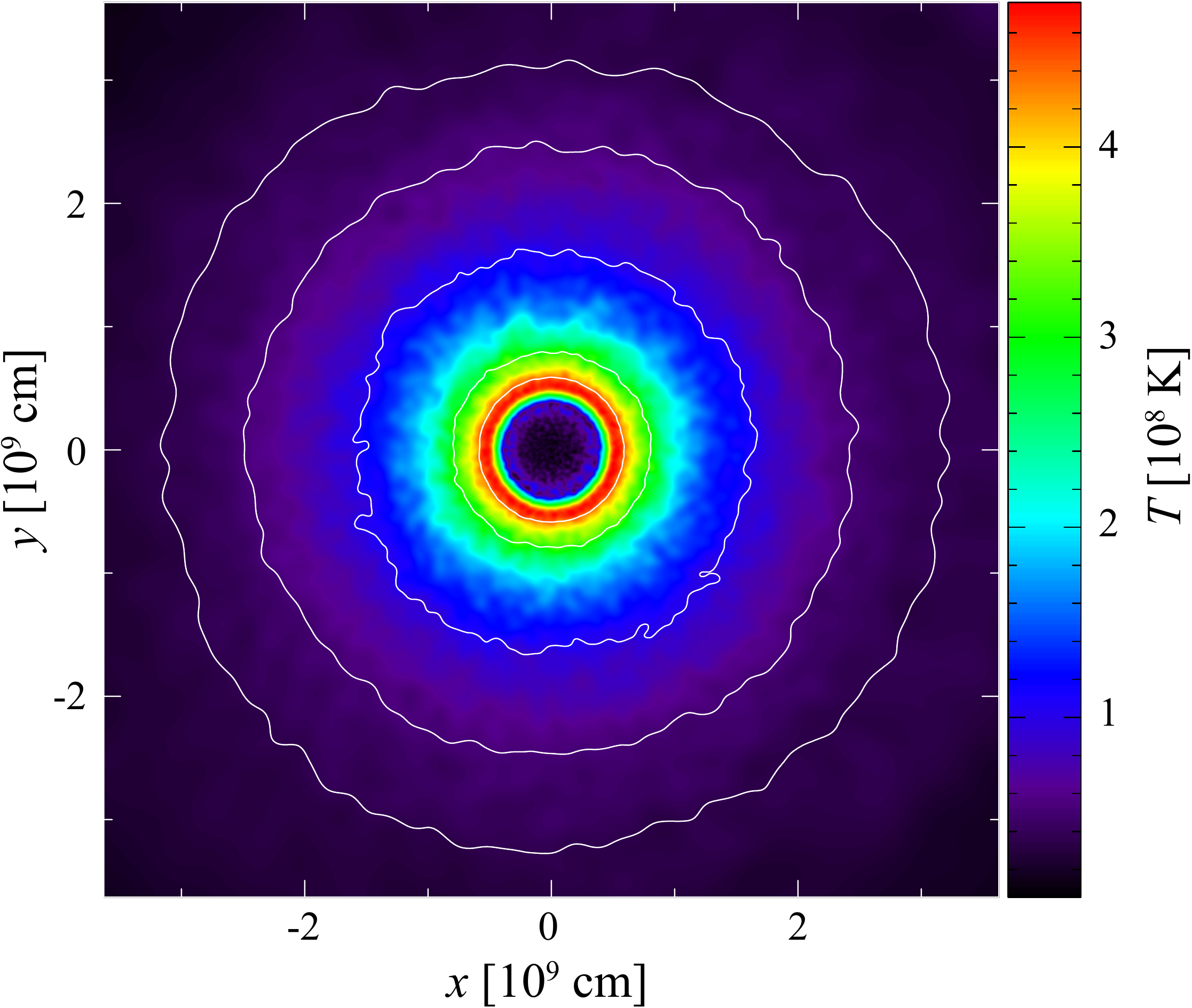}\hspace{0.5cm}
 \includegraphics[height=2.5in]{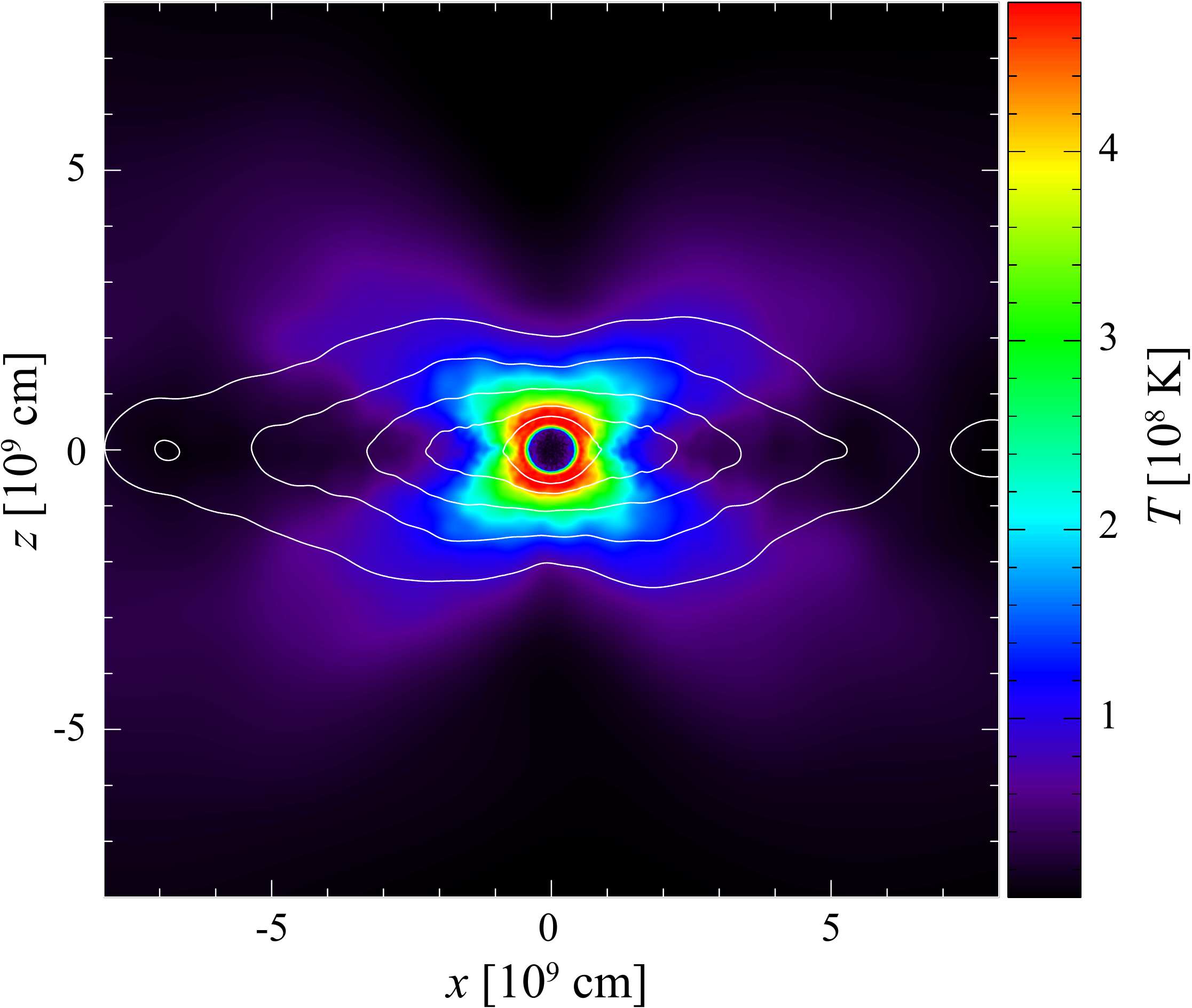}}
\caption{Densities and temperatures of the final remnants of the $0.6+0.9$ \Msun 
  case (P5): $xy$-plane (left) and $xz$-plane (right). 
  Color-coded is the temperature (in units of $10^8$ K), the 
overlaid white contours refer to $\log{\rho}$ ($\rho$ in \gcc). The contours
in the left panel show densities ranging from $\log_{10} \rho = 5$ (innermost
contour) to $\log_{10} \rho = 2$ (outermost contour) in steps of 0.75, while
the contours in the right panel range from $\log_{10} \rho = 5.6$ to $\log_{10} \rho = 3.6$ in steps of 0.5.} 
\label{fig:rho_T_remnant0609}
\end{figure}

The thermodynamical properties of the remnant core are displayed in Figure
\ref{fig:rho_T_remnant0609}. The peak temperatures now reach about $5 \times
10^8$ K at densities of about $10^6$ g cm$^{-3}$. Under these conditions,
carbon burning is observed to occur, but the rate of energy injection is not
large enough to be dynamically dominant. The structure and further evolution
of our double degenerate merger remnant products will be discussed in more
detail elsewhere.  

\subsection{Summary of all other simulations}
All the systems that we have simulated in this paper belong to the regime of direct impact, unstable
mass transfer. In all cases, we find the mass transfer to survive for many
orbits. For those close to the guaranteed stable disk regime  
(see Figure \ref{fig:marsh04}) the mass transfer phase extends for many tens
of orbits. Systems belonging to these categories are represented here by our
P1 ($0.2 + 0.8\ M_\odot$) and P2 ($0.3 + 1.1\ M_\odot$) simulations. The $0.3
+ 1.1\ M_\odot$ binary system shows a very similar evolution to the $0.2 +
0.8\ M_\odot$ system described above. The 
mass transfer phase is stabilized, i.e. the rate of mass transfer does not
increases exponentially but is observed to have a nearly constant mass
transfer rate for many orbital periods. 

The evolution of all other systems, which are far away from the limit of
stable disk regime, is similar evolution to that of the $0.6 + 0.9$ \Msun
binary system discussed above. In this case, a slow decrease in the separation
is accompanied by a severe mass transfer rate increase. During the evolution,
the donor star remains synchronized and the accretor is substantially 
spun up. 
All these systems lack stabilizing effects and are therefore prone to
Kelvin-Helmholtz-triggered He-detonations at the accretor surface that were
discussed in detail in 
\citep{guillochon10} (see Section \ref{sec:surfdet} for further details). Such systems
could be possible candidates for sub-Chandrasekhar double detonation scenarios
\citep{fink07, fink10}. 

For all systems, we observe pronounced spiral arms extending out of the disk which engulfs
the central remnant core. As we have discussed in Section \ref{sec:dependence_IC} and shown in
Figure \ref{fig:WD06WD09_remnant}, this feature may be much less pronounced in simulations that start 
from approximate ICs since they contain substantially less angular momentum. These cases
do not settle into a quasi-steady-state. Instead the 
torques from accretor and tidal tail continue shaping the dynamical evolution of the disk, see e.g. Figure \ref{fig:WD02WD08}.

\subsection{Comparison with other mass transfer calculations}
\label{sec:dsouza}
The simulations of \citeauthor{dsouza06} were the first to show that the
binary system does not merge within few orbital time scales, but instead can
survive for more than 30 orbital periods. These authors investigated the
evolution of a polytropic system with mass ratio $q=0.5$ starting from accurate initial
conditions constructed in the framework of the self-consistent field technique.
\citeauthor{dsouza06} found an increase in the orbital separation and a concordant decrease of the
mass transfer rate after about 20 orbital periods. After 32 orbits they had 
to stop their simulation due to the severe degradation of numerical conservation.
At this stage, the binary system showed no sign of a pending merger.
The authors experimented with a higher degree of contact of the same binary
and consequently a higher initial mass transfer rate by extracting angular
momentum at a rate of 1\% per orbit. The merger is avoided even if the system
is artificially driven together for three more orbits after it has reached contact. 
Only when driving is
applied throughout the simulation, the donor is observed to get disrupted. This occurs after about 8 orbits. 

We found that if we reduce the angular momentum at a 0.5\%/orbit rate,
the disruption process is 3 times shorter, compared when no driving is applied. For our $q=0.5$ 
case ($0.3 + 0.6$ \msun, P4), for as long as we have the possibility to compare, 
we see a similar behavior to that of \citeauthor{dsouza06} in both the orbital separation and the mass transfer
rate. At the end, the donor star is, however, disrupted after 45 orbits 
of mass transfer.

The same authors also studied the evolution of a $q=0.4$ system. They found 
a peak in the mass transfer rate after about 10 orbits followed by a rapid decrease. 
The authors stopped the simulation after 43 orbits, with no visible sign of a merger.
From our calculations, P3 is the closest to their mass ratio ($\approx 0.41$). Our P3 simulation, however, shows
a donor disruption after 31 orbits. This difference may be due to the different 
shock treatment and the different equation of state (Helmholtz vs. 5/3 polytrope).

\begin{figure}[!t]
\centerline{
\includegraphics[height=3.0in]{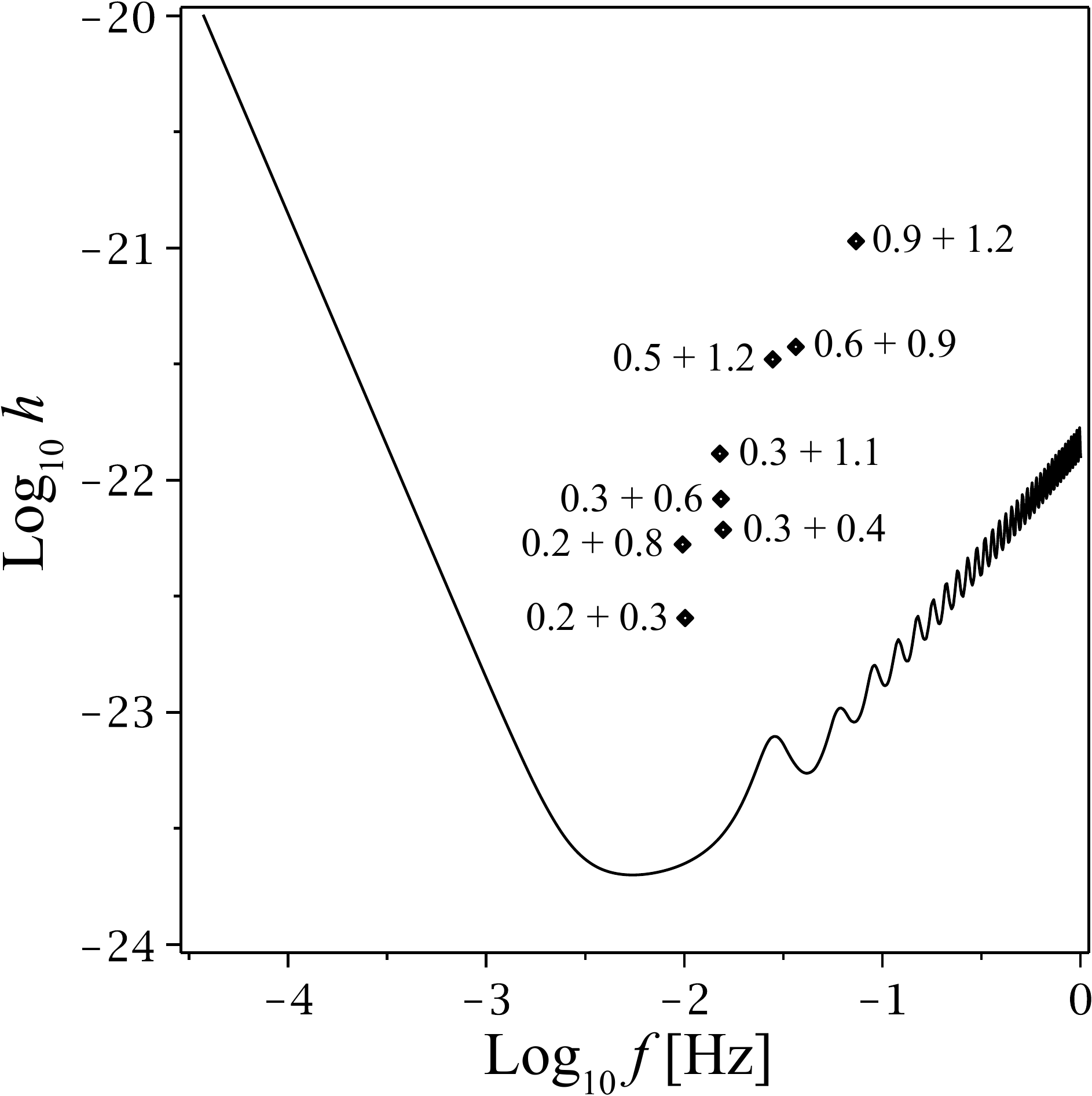}
}
\caption{Gravitational wave amplitudes of the studied double WD systems for an assumed
   distance of $10$ kpc. The overlaid curve shows the LISA sensitivity for a 
   one-year integration period that has been produced with the online sensitivity curve 
   generator from
   http://www.srl.caltech.edu/\~{}shane/sensitivity/}
\label{fig:lisa}
\end{figure}

\section{Gravitational Wave Signals}
\label{sec:gwave}
The Laser Interferometer Space Antenna (LISA) will provide the largest
observational sample of (interacting) double white dwarf binaries
\citep{farmer03,ruiter10}. Such systems enter the LISA observational window
(0.1 mHz - 100 mHz) when they reach a period $\approx$ 5 hr. They subsequently
secularly evolve through radiation reaction across the LISA band and are so
numerous to create a stochastic foreground. When a binary reaches a
gravitational wave frequency of a few mHz and is close enough to be resolved
individually, the recorded signal shows an intrinsic frequency evolution over
a typical one year observation time. Figure \ref{fig:lisa} shows the
gravitational wave amplitudes (for an assumed distance of 10 kpc) for  
all our production runs together we the LISA sensitivity curve. Since none of the orbital frequencies
change by more than 1 \% in the last year prior to merger, we chose this as a typical integration time. Provided
they are in the Milky Way, all of the investigated systems would be detectable by LISA.

We show that, depending primarily on the initial conditions, the frequency
evolution of the binary can be calculated incorrectly. Simulations that use
the approximate initial conditions, as discussed in Section
\ref{sec:dependence_IC}, severely underestimate the initial 
separation when mass transfer sets in. As a result, the binary only survives
for a few orbits resulting in a rapidly fading gravitational wave signal (see
e.g. Figure \ref{fig:GWs_06_09}). The accurate initial conditions, on the
other hand, produce binary systems in which the mass transfer phase is
dramatically extended by almost two orders of magnitude in time, resulting in
a gravitational wave signal with amplitude and frequency that remain
essentially constant up until merger. Population synthesis models have shown
that these binaries are more numerous than dynamically stable systems for a
given mass ratio \citep{nelemans01a,ruiter10}. As we show that these binaries
can survive at small separation for hundreds of orbital periods (keep in mind
our mass transfer duration estimates are --due to finite numerical
resolution-- strict lower limits on the durations realized in nature), their
associated gravitational wave signal should be included when calculating the
LISA gravitational wave foreground (although expected to below LISA's
sensitivity at these high frequencies).

\begin{figure}[!h]
\centerline{
 \includegraphics[height=2.5in]{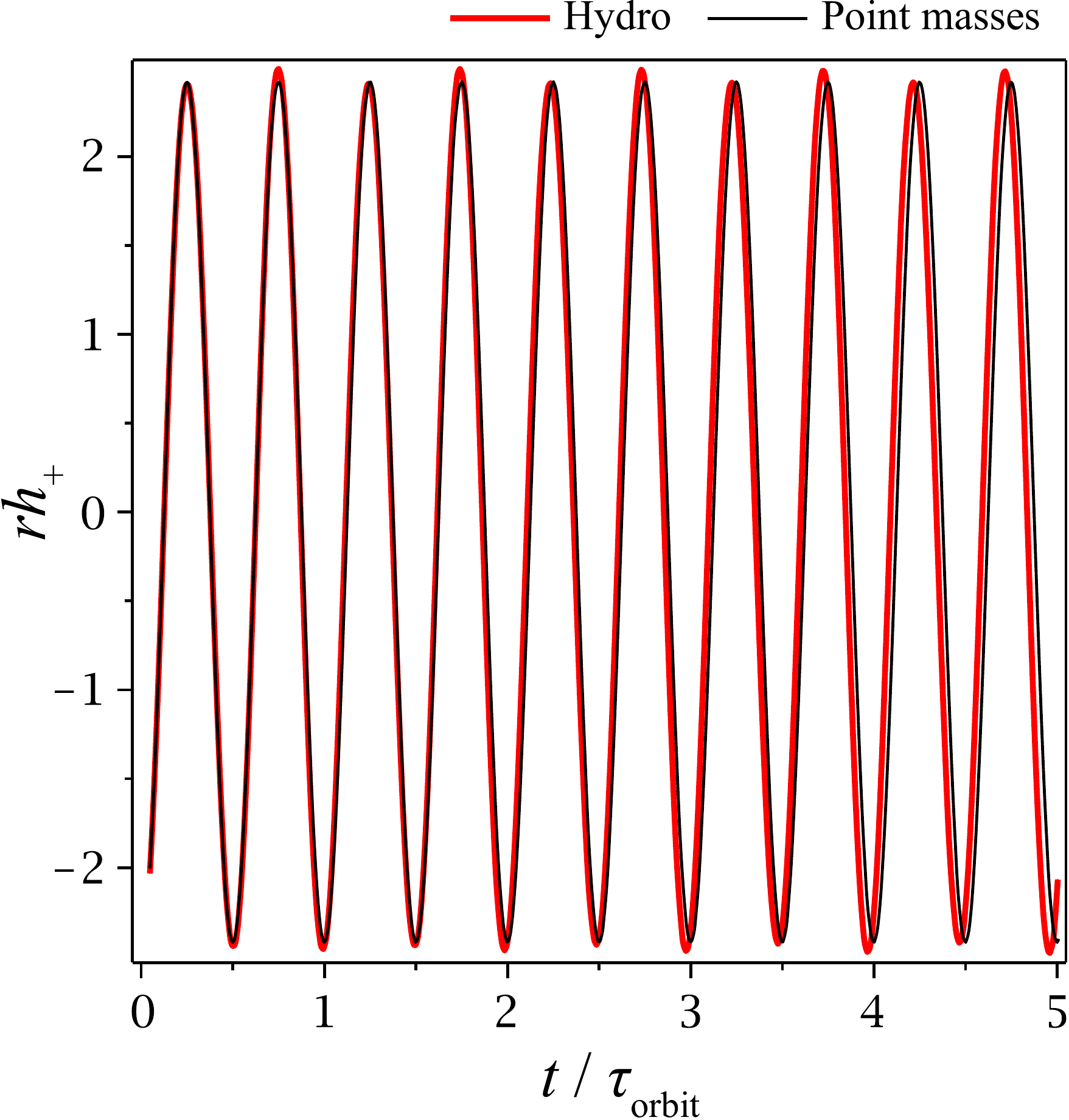}\hspace{0.5cm}
 \includegraphics[height=2.5in]{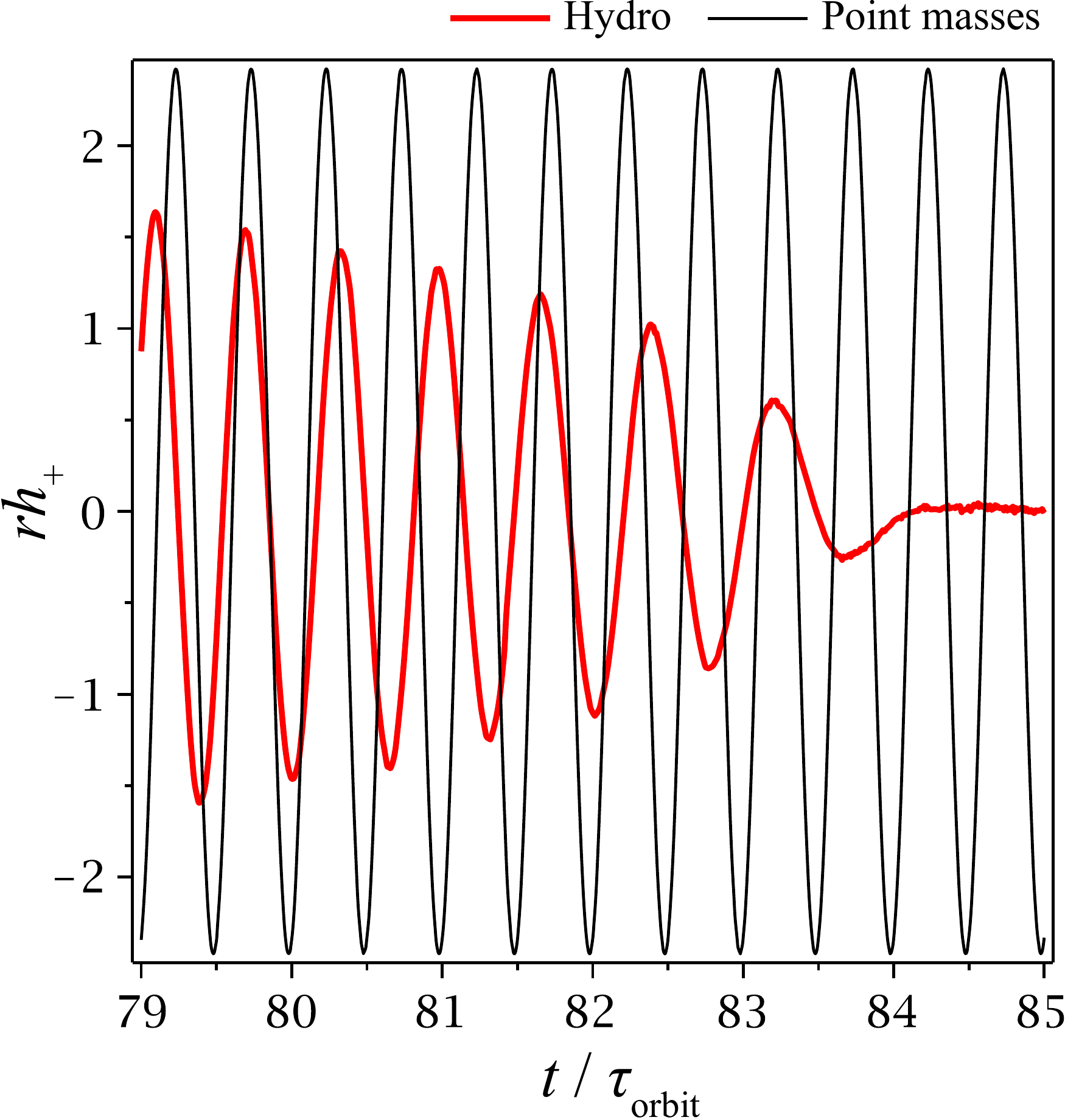}
}
\caption{0.2 He and 0.8 \Msun CO WDs: comparison of the gravitational wave amplitude 
   $h_+$ of the simulation with the point mass
 limit. $r$ is the distance to the observer.}
\label{fig:GW_pm_hydro_WD02WD08}
\end{figure}

\begin{figure}[!h]
\centerline{
 \includegraphics[height=2.5in]{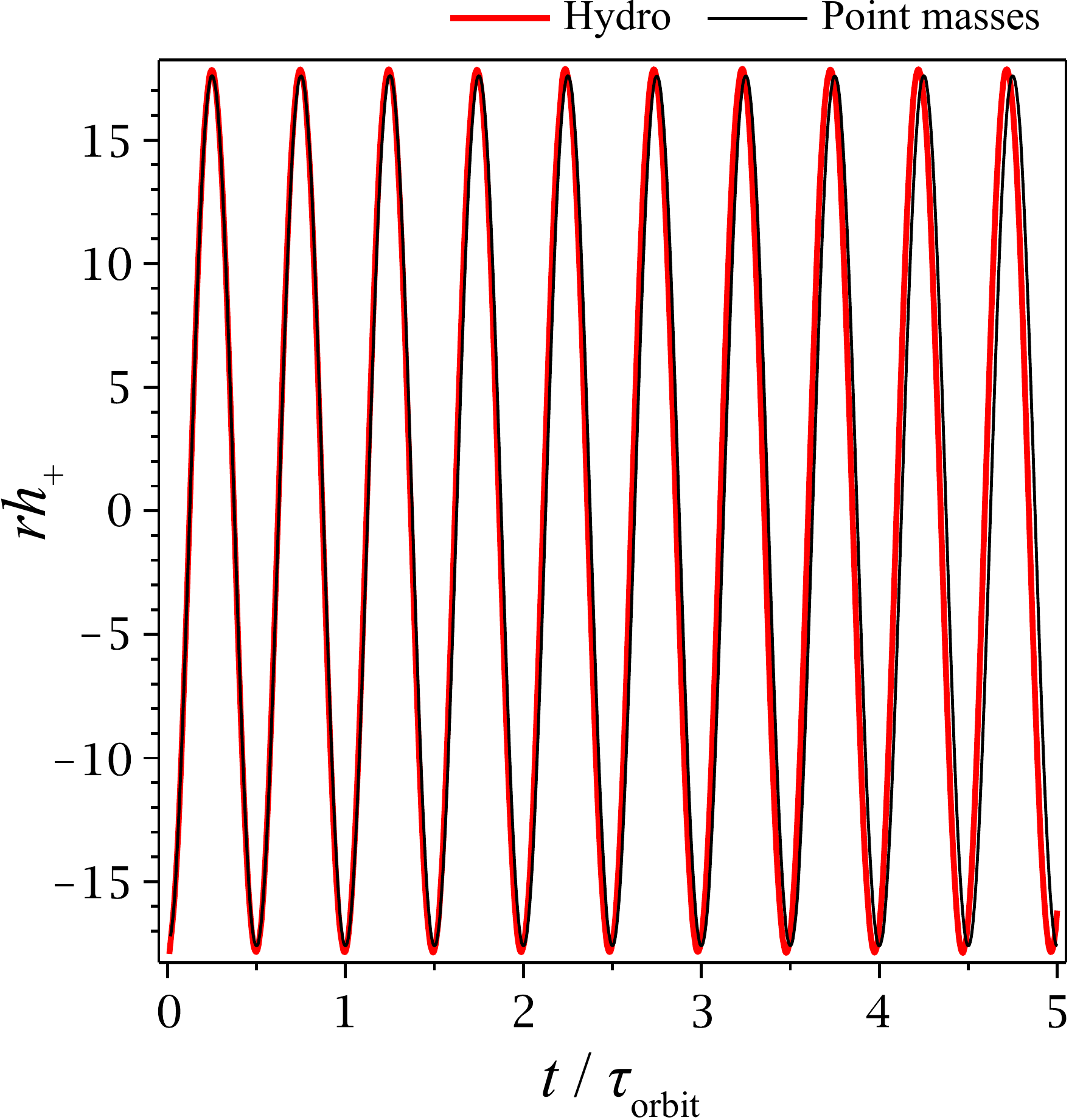}\hspace{0.5cm}
 \includegraphics[height=2.5in]{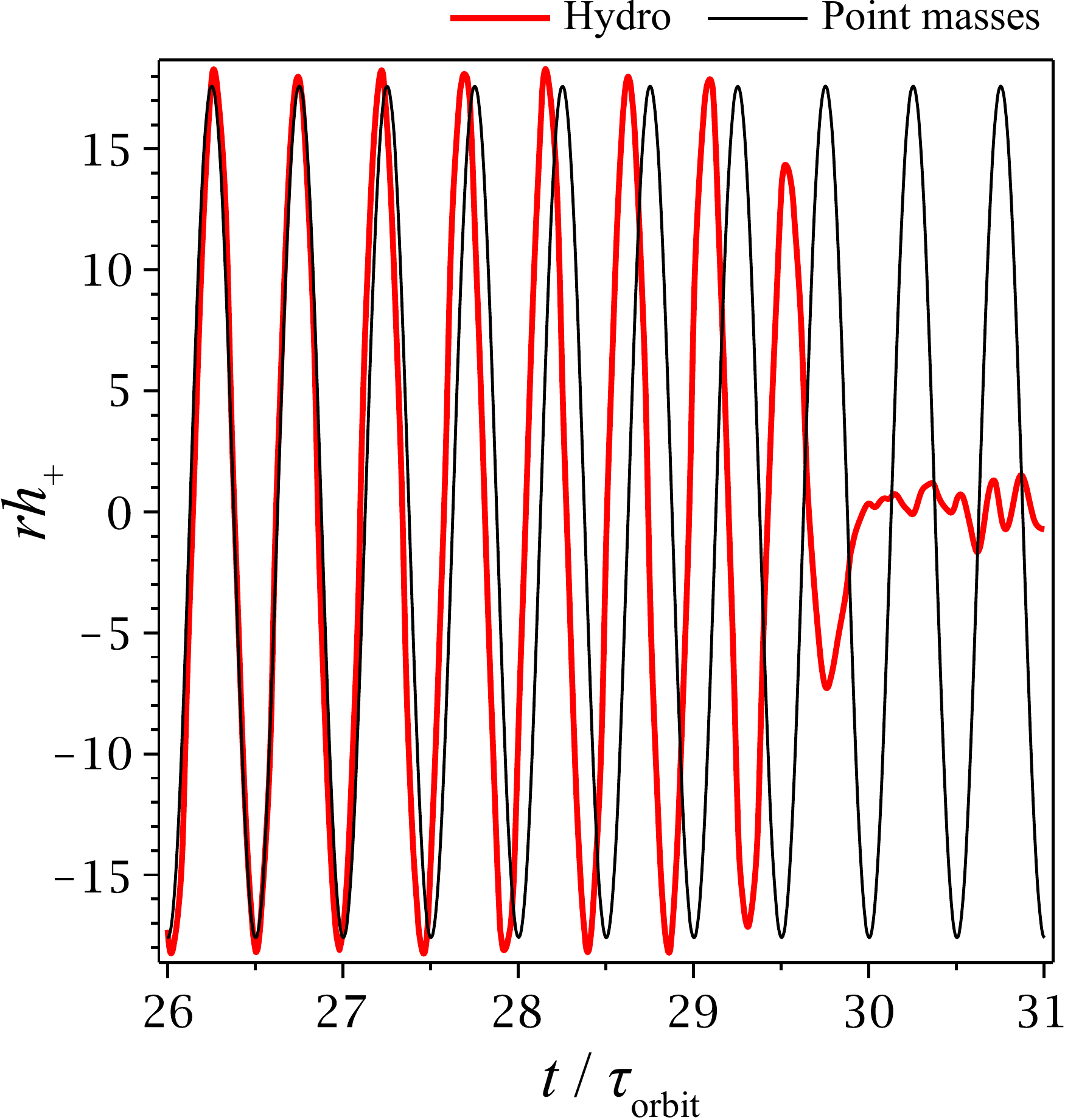}
}
\caption{CO WDs with $0.6 + 0.9$ \msun: comparison of the gravitational wave amplitude 
   $h_+$ of the full hydrodynamic simulation with the point mass
 limit. $r$ is the distance to the observer and $t$ is the time measured
 in units of the initial orbital period.}
\label{fig:GW_pm_hydro_WD06WD09}
\end{figure}

With the use of accurate initial conditions as well as realistic shock heating
treatment, the frequency evolution of the binary can thus be explored in
detail. Depending primarily on the mass ratio, as we have argued here, the
resulting gravitational wave signal  
can be significantly different from that predicted in the point mass limit,
particularly for marginally unstable systems. In such systems,  
the gravitational wave signal carries the imprint of the orbital oscillation,
as illustrated by the evolution of the marginally unstable 0.2 + 0.8 \Msun
system in Figure \ref{fig:GW_pm_hydro_WD02WD08}. 

For systems that are far from the stable disk limit, we find the point mass approximation to provide a better description of our simulated 
gravitational wave signals. This is mainly because the oscillations in the
orbital separation are much less pronounced for these systems. Contrary 
to the $0.2+0.8$ \msun\; case, the gravitational wave amplitude derived for
the $0.6 + 0.9$ \msun\; binary system remains close to the point mass
prediction up to almost complete disruption, as seen in Figure
\ref{fig:GW_pm_hydro_WD06WD09}.  

\section{Surface Detonation Criteria in White Dwarf Binaries} \label{sec:surfdet} 
During the final phase of mass transfer prior to merger, the rate of mass
exchanged approaches the mass of the donor divided by the orbital timescale,  
\be
\dot{M}_2 \approx \left({M_2 \over R_2}\right)^{3/2} \sqrt{G \over \mu_1} \approx 0.1 M_{\odot}\, {\rm s}^{-1},
\ee
where $\mu_1$ is the reduced mass of the accretor. As shown in
\cite{guillochon10}, Kelvin-Helmholtz instabilities develop within the
accretion stream once the accretor has built up a torus of material acquired
from the donor, and this leads to the formation of dense knots of material
that can periodically strike the accretor's surface. This striking action can
lead to the initiation of a detonation within the torus about the accretor,
assuming that matter acquired from the donor mostly consists of helium. 

The question of whether knots are capable of detonating the accretor's helium
layer depends on the ram pressure applied by individual knots, and by the
initial conditions of the helium torus at the time of compression. From the
SPH simulations presented in this paper we have an accurate record of the
time-evolution of the accretion stream's density at L1 and the conditions at
the base of the helium torus. As shown in our FLASH simulations (Figure
\ref{fig:burning_overdensity}), a standing shock develops between the stream
and the torus, equalizing the pressure at the stream-torus interface, and
reducing the velocity of the fluid flow in the torus relative to the accretion
stream to subsonic velocities. As the interaction is subsonic and no heat
transport processes can operate effectively on a dynamical timescale, the
accretion stream smoothly evolves relative to the background pressure, and
thus maintains constant entropy. The exterior pressure experienced by the
accretion stream is given by the pressure within the torus. 

\begin{figure}[!h]
\centerline{
\includegraphics[height=2.1in]{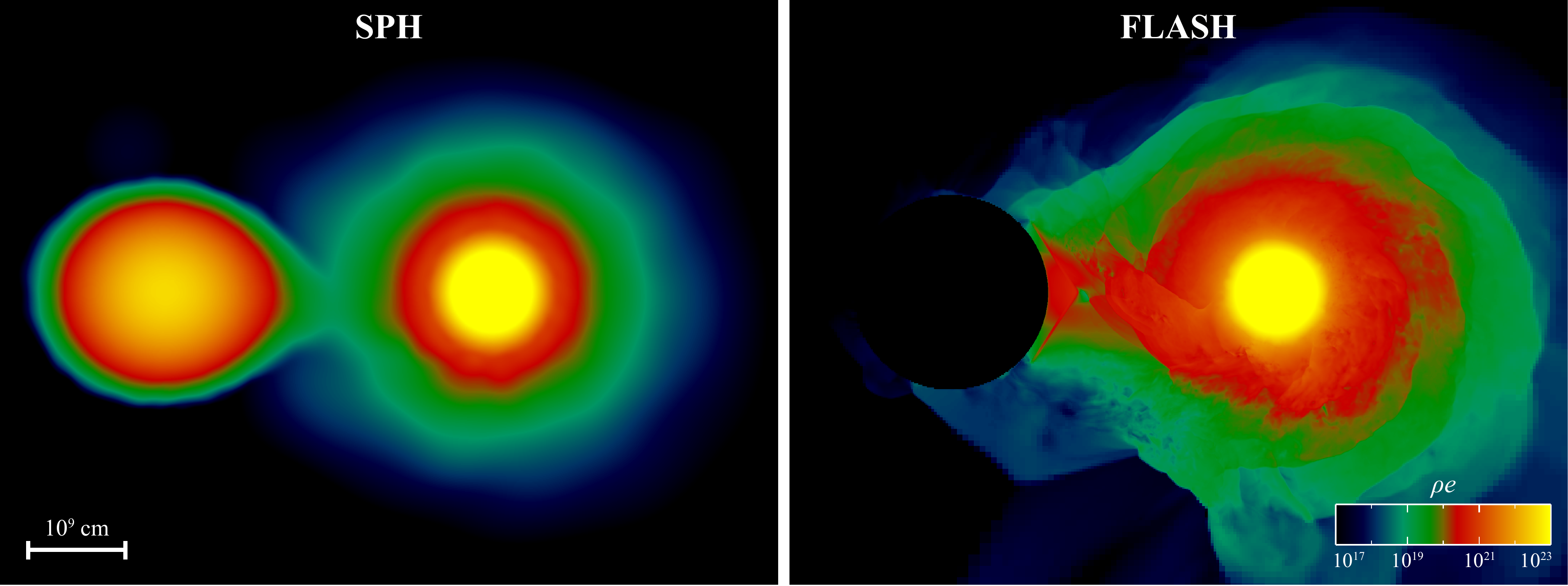}
}
\caption{Side-by-side comparison of the fluid internal energy $\rho e$ for one
  of the SPH calculations featured in this work (G1) and one of the FLASH
  calculations of \citep[run Ba]{guillochon10} for a 0.45 + 0.9 \Msun
  binary. The SPH calculation fully resolves both the donor and the accretor,
  which allows for the question of binary stability to be addressed
  self-consistently. However, the SPH calculation has limited resolution and
  is unable to sufficiently resolve the stream-torus interface, artificially
  suppressing the instabilities and the standing shocks that are observed to
  develop in the FLASH calculation. As a result of not resolving the standing
  shocks properly, the SPH calculation shows somewhat less heating in the torus than
  the FLASH calculation. On the other hand, the stream boundary condition used
  in the FLASH simulation is idealized to lie on the Roche surface of the
  donor, with a simplified initial velocity profile normalized to the sound
  speed of the donor, which does not perfectly represent the actual flow of
  material leaving L1.} 
\label{fig:sph_flash_compare}
\end{figure}

Given the initial density $\rho_{\rm L1}$ and temperature $T_{\rm L1}$ of the
stream material as it leaves L1, the pressure at the base of the helium torus
$P_{\rm tor}$, and an equation of state that parameterizes the entropy $s$ and
pressure $P$ as functions of density and temperature, we can determine the
density of the stream when it reaches the base of the helium torus by solving
the following system of equations for $\rho_{\rm knot}$ and $T_{\rm knot}$, 
\begin{align}
s(\rho_{\rm L1}, T_{\rm L1}) &= s(\rho_{\rm knot}, T_{\rm knot})\label{eq:ent1}\\
P(\rho_{\rm tor}, T_{\rm tor}) &= P(\rho_{\rm knot}, T_{\rm knot}\label{eq:prs1}),
\end{align}
where ``tor'' refers to the base of the helium torus and ``knot'' refers to
the stream post-compression. For all of the SPH simulations presented in this
paper, the maximum increase in density achieved within the stream as it falls
from L1 to the accretor's surface is $\gtrsim 10$. The small initial entropy
of the stream ($T_{\rm L1} \sim 10^5$ K) and minimal shock-heating means that
the final temperature reached within the stream is substantially smaller than
the temperature required for dynamical burning, and thus burning is not
expected to be important within the stream itself. 

\begin{figure}[!h]
\centerline{
\includegraphics[height=3.0in]{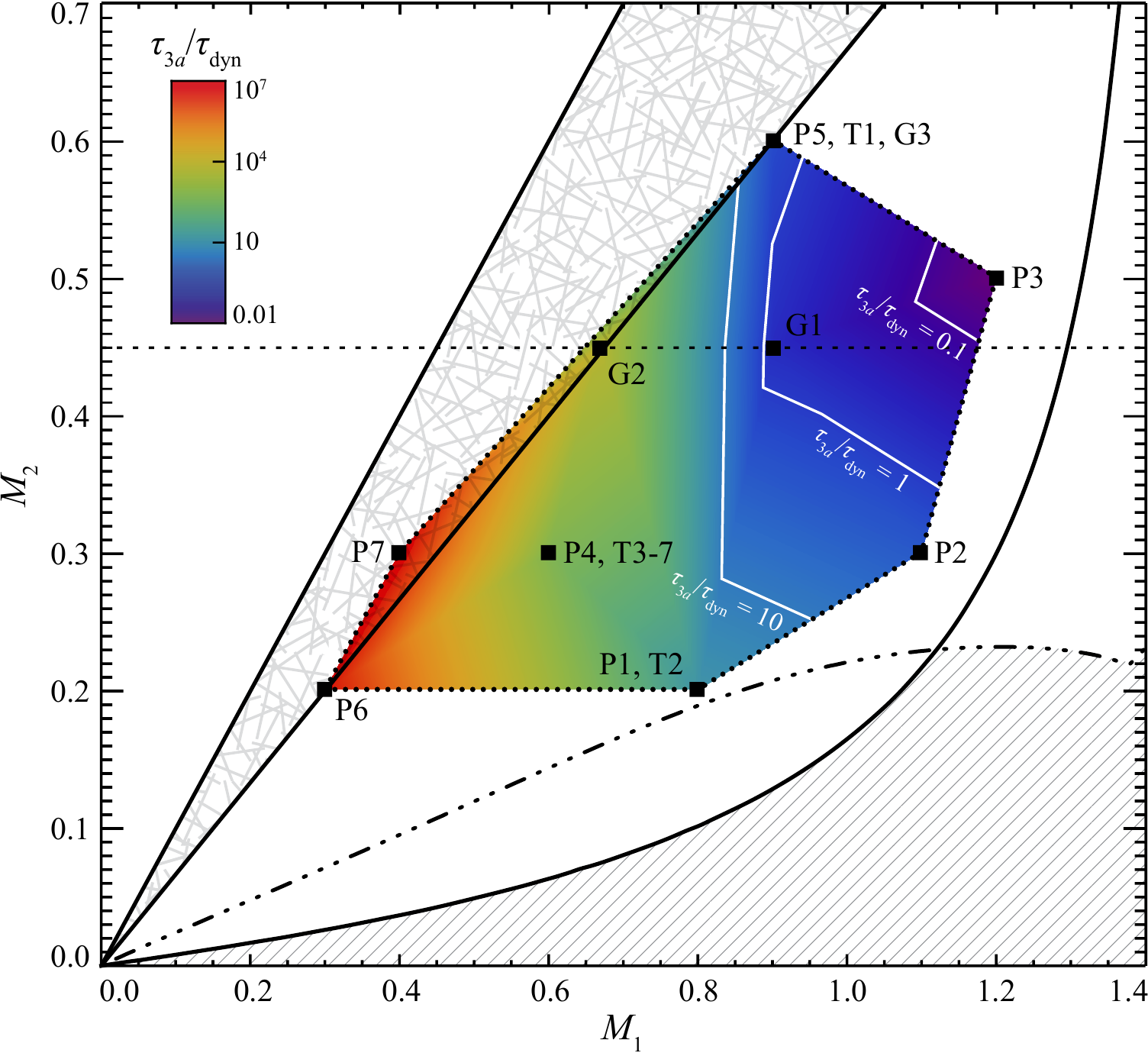}
}
\caption{Ratio of the triple-alpha burning timescale $\tau_{3\alpha}$ to the
  dynamical timescale $\tau_{\rm dyn}$ for different combinations of donor and
  accretor masses. The figure shows the same stability criteria as Figure
  \ref{fig:marsh04}, sans labels for clarity. The colored region shows the
  ratio of the two timescales, with the white contours showing where
  $\tau_{3\alpha} = 0.1 \tau_{\rm dyn}$, $\tau_{3\alpha} = \tau_{\rm dyn}$,
  and $\tau_{3\alpha} = 10\tau_{\rm dyn}$. The dotted line shows the convex
  hull of all simulations performed in which the donor can possess a
  significant amount of helium in its outer layers. For the two FLASH
  simulations performed in \cite{guillochon10} in which a surface detonation
  was observed (G1 and G3), the ratio of timescales is of order unity, whereas
  the simulation in which no detonation was observed (G2), $\tau_{3\alpha}$ is
  significantly larger than $\tau_{\rm dyn}$.} 
\label{fig:burning_overdensity}
\end{figure}

With our estimate for $\rho_{\rm knot}$, we can estimate the increase in
density and temperature of the torus after it is compressed by the stream. As
in \cite{guillochon10}, we solve for the trajectory of a test particle
released at L1 with initial velocity given by the sound speed of the donor to
estimate $v_\perp$, the component of the velocity perpendicular to the
accretor's surface. Given the density of the stream $\rho_{\rm knot}$, we can
determine the ram pressure of the stream 
\be
P_{\rm ram} = \rho_{\rm knot} v_\perp^2. 
\ee
We can then use $P_{\rm ram}$ and solve a set equations similar to Equations
(\ref{eq:ent1}) and (\ref{eq:prs1}) to determine the post-compression state of
the torus  

\begin{align}
s(\rho_{\rm tor}, T_{\rm tor}) &= s(\rho_{\rm comp}, T_{\rm comp})\label{eq:ent2}\\
P_{\rm ram} &= P(\rho_{\rm comp}, T_{\rm comp}\label{eq:prs2}),
\end{align}
where ``comp'' refers to the conditions in the helium torus after
compression. This enables us to calculate the triple-alpha burning timescale
$\tau_{3\alpha}$ and compare it to the dynamical timescale $\tau_{\rm dyn}$ at
the accretor's surface to determine if runaway burning can be achieved,
leading in turn to a detonation.  

In Figure \ref{fig:burning_overdensity} we show the curves for which
$\tau_{\rm dyn} = \tau_{3\alpha}$ and $10 \tau_{\rm dyn} = \tau_{3\alpha}$
within the convex hull of the SPH simulations presented in this paper. For the
two simulations which exhibited surface detonations in \cite{guillochon10} the
two timescales are comparable, whereas $t_{\rm dyn}$ is far shorter than
$\tau_{3\alpha}$ for the simulation that showed no surface detonation. The SPH
simulations show that the conditions for dynamical burning being triggered on
the surface of the accretor are achieved for accretor masses larger than $\sim
0.8 M_\odot$, with a cutoff at low donor masses as the circularization radius
of the accretion stream begins to exceed the radius of the accretor. Such
systems could be possible candidates for sub-Chandrasekhar double detonation
scenarios, as argued in \cite{guillochon10}. 

\section{Summary}
\label{sec:summary}
The onset of (numerically resolvable) mass transfer in double degenerate binary 
systems and its impact on the orbital evolution forms the main theme of our
paper. We find for all investigated cases that mass transfer ensues for at
least dozens of orbits,  
even for systems which according to the analysis of \cite{marsh04}
are guaranteed to be unstable. Due to the finite
numerical resolution, the duration of the mass transfer phase calculated here
should be considered as a strict lower limit. 

Part of our motivation has been to settle a previous debate concerning the
duration of mass transfer  
in such systems. Almost all of the earlier SPH simulations had found quick
disruption of the donor  
star within a few orbital periods after the onset of mass transfer, while
recent grid-based simulations  
\citep{dsouza06} found a longer lived mass transfer phase. Using the same
assumptions and  
initial conditions we have reproduced essentially all previous results. In
addition we have compared our  
results to those obtained by using the StarCrash code developed by Faber and 
Rasio and found excellent agreement when using the same initial conditions and   
equation of state. 

Concerning the mass transfer duration debate, we find that the reported
discrepancies can be accounted for by the following two effects. 
The first is the self-consistent inclusion of entropy production in shocks,
which we find to have a major impact on the orbital dynamics. 
In direct-impact systems, we find the realistic inclusion of shock heating to
produce an extended, high-entropy halo around the 
accretor. When ignored, a smooth incorporation of the transferred material
into the stellar  
surface is observed. Since substantial amount of angular momentum
can be stored in the shocked material (which is provided at the expense of the
orbital angular momentum), the realistic inclusion of shocks yields shorter
lived mass transfer phases. 
These results confirm the earlier concerns raised by \cite{fryer08}.

The second and even more important reason is attributed to the initialization
procedure of the binary in the simulations. The results of simulations
constructed using  
approximate corotating initial conditions have been compared with those using
carefully constructed binaries and are found to be markedly different.  
Simulations that use the approximate initial conditions underestimate the
initial separation at which mass transfer sets in by about 15\%, which yields
a binary that only survives for a few orbits and thus a rapidly fading
gravitational wave signal. This is because once mass transfer sets in, the
orbital evolution is no longer driven by angular momentum losses due to
gravitational wave emission but instead is dominated by the redistribution of
angular momentum in the system. The use of approximate initial conditions
systematically underestimates the number of orbits and overestimates the
gravitational wave peak frequencies and amplitudes by about 20 \%. In
contrast, the accurate initial conditions produce a binary system in which the
mass transfer phase is extended by almost two orders of magnitude in time,
resulting in a gravitational wave signal with amplitude and frequency that
remain essentially constant up until merger.  
The mass transfer also shows a unique oscillatory signal that is most
pronounced for binary systems that are near the disk forming limit. 

The use of accurate initial conditions and a correct treatment of shock
heating allows us to better characterize the time evolution of the
temperature, density, and angular momentum, which are important when
considering thermonuclear events that may occur during the mass transfer phase
and/or after merger. Approximate initial conditions, for example, tend to
significantly overestimate 
the densities and temperatures of the final merger remnant. This fact may have
adverse consequences for possible type Ia candidate systems. 
Our treatment allows us to also accurately identify when surface detonations
may occur in the lead-up to the merger.  
The results of this study have been used as boundary conditions for FLASH
simulations that focus on  
the fate of the low-density accreted material and that are described in a
companion paper \citep{guillochon10}. 
This combined study has shown that contrary to earlier beliefs, the mass
transfer can be very eventful.  
We have shown that for the case of unstable, direct impact He accretion onto the accretor
a hot ($>5 \times 10^8$ K) helium torus builds up in which the incoming accretion stream triggers 
Kelvin-Helmholtz instabilities. Such Kelvin-Helmholtz ``knots'' repeatedly impact the accretor's helium 
surface and, once the triple-alpha time scale is shorter than the local dynamical time scale, 
can trigger violent surface explosions. This explosion launches shocks into
the accretor's interior that can subsequently ignite the carbon oxygen
core. This could  
possibly be the long sought-after explosion mechanism how to trigger supernova
explosion in a double degenerate 
scenario which, together with the promising rates, could make double
degenerate mergers a viable (sub-luminous) type Ia 
progenitor system.

\acknowledgments We thank Lars Bildsten, Marcus Br\"uggen, Philipp
Podsiadlowski, John Fregeau, Dan Kasen, Gijs Nelemans and Ken 
Shen for very useful discussions. We acknowledge support from
DFG grant RO 3399 (M.D. and S.R.), the David and Lucille Packard Foundation
(JG and ER-R), NSF grant: PHY-0503584 (ER-R), and the NASA Earth and Space
Science Fellowship (JG). 

\bibliographystyle{apj}
\bibliography{apj-jour,Mass_transfer_DD}

\begin{thebibliography}{51}
\expandafter\ifx\csname natexlab\endcsname\relax\def\natexlab#1{#1}\fi

\bibitem[{{Balsara}(1995)}]{balsara95}
{Balsara}, D.~S. 1995, Journal of Computational Physics, 121, 357

\bibitem[{{Benz} {et~al.}(1990){Benz}, {Cameron}, {Press}, \&
  {Bowers}}]{benz90b}
{Benz}, W., {Cameron}, A.~G.~W., {Press}, W.~H., \& {Bowers}, R.~L. 1990, \apj,
  348, 647

\bibitem[{{Clayton} {et~al.}(2007){Clayton}, {Geballe}, {Herwig}, {Fryer}, \&
  {Asplund}}]{clayton07}
{Clayton}, G.~C., {Geballe}, T.~R., {Herwig}, F., {Fryer}, C., \& {Asplund}, M.
  2007, \apj, 662, 1220

\bibitem[{{D'Souza} {et~al.}(2006){D'Souza}, {Motl}, {Tohline}, \&
  {Frank}}]{dsouza06}
{D'Souza}, M.~C.~R., {Motl}, P.~M., {Tohline}, J.~E., \& {Frank}, J. 2006,
  \apj, 643, 381

\bibitem[{{Eggleton}(1983)}]{eggleton83}
{Eggleton}, P.~P. 1983, \apj, 268, 368

\bibitem[{{Farmer} \& {Phinney}(2003)}]{farmer03}
{Farmer}, A.~J. \& {Phinney}, E.~S. 2003, \mnras, 346, 1197

\bibitem[{{Fink} {et~al.}(2007){Fink}, {Hillebrandt}, \& {R{\"o}pke}}]{fink07}
{Fink}, M., {Hillebrandt}, W., \& {R{\"o}pke}, F.~K. 2007, \aap, 476, 1133

\bibitem[{{Fink} {et~al.}(2010){Fink}, {R{\"o}pke}, {Hillebrandt},
  {Seitenzahl}, {Sim}, \& {Kromer}}]{fink10}
{Fink}, M., {R{\"o}pke}, F.~K., {Hillebrandt}, W., {Seitenzahl}, I.~R., {Sim},
  S.~A., \& {Kromer}, M. 2010, \aap, 514, A53

\bibitem[{{Frank} {et~al.}(2002){Frank}, {King}, \& {Raine}}]{frank02}
{Frank}, J., {King}, A., \& {Raine}, D.~J. 2002, {Accretion Power in
  Astrophysics: Third Edition}, ed. {Frank, J., King, A., \& Raine, D.~J.}

\bibitem[{{Fryer} \& {Diehl}(2008)}]{fryer08}
{Fryer}, C.~L. \& {Diehl}, S. 2008, in Astronomical Society of the Pacific
  Conference Series, Vol. 391, Hydrogen-Deficient Stars, ed. {A.~Werner \&
  T.~Rauch}, 335

\bibitem[{{Fryxell} {et~al.}(2000){Fryxell}, {Olson}, {Ricker}, {Timmes},
  {Zingale}, {Lamb}, {MacNeice}, {Rosner}, {Truran}, \& {Tufo}}]{fryxell00}
{Fryxell}, B., {Olson}, K., {Ricker}, P., {Timmes}, F.~X., {Zingale}, M.,
  {Lamb}, D.~Q., {MacNeice}, P., {Rosner}, R., {Truran}, J.~W., \& {Tufo}, H.
  2000, \apjs, 131, 273

\bibitem[{{Guerrero} {et~al.}(2004){Guerrero}, {Garc{\'{\i}}a-Berro}, \&
  {Isern}}]{guerrero04}
{Guerrero}, J., {Garc{\'{\i}}a-Berro}, E., \& {Isern}, J. 2004, \aap, 413, 257

\bibitem[{{Guillochon} {et~al.}(2010){Guillochon}, {Dan}, {Ramirez-Ruiz}, \&
  {Rosswog}}]{guillochon10}
{Guillochon}, J., {Dan}, M., {Ramirez-Ruiz}, E., \& {Rosswog}, S. 2010, \apjl,
  709, L64

\bibitem[{{Han} {et~al.}(2003){Han}, {Podsiadlowski}, {Maxted}, \&
  {Marsh}}]{han03}
{Han}, Z., {Podsiadlowski}, P., {Maxted}, P.~F.~L., \& {Marsh}, T.~R. 2003,
  \mnras, 341, 669

\bibitem[{{Hicken} {et~al.}(2007){Hicken}, {Garnavich}, {Prieto}, {Blondin},
  {DePoy}, {Kirshner}, \& {Parrent}}]{hicken07}
{Hicken}, M., {Garnavich}, P.~M., {Prieto}, J.~L., {Blondin}, S., {DePoy},
  D.~L., {Kirshner}, R.~P., \& {Parrent}, J. 2007, \apjl, 669, L17

\bibitem[{{Hix} {et~al.}(1998){Hix}, {Khokhlov}, {Wheeler}, \&
  {Thielemann}}]{hix98}
{Hix}, W.~R., {Khokhlov}, A.~M., {Wheeler}, J.~C., \& {Thielemann}, F. 1998,
  \apj, 503, 332

\bibitem[{{Howell} {et~al.}(2006){Howell}, {Sullivan}, {Nugent}, {Ellis},
  {Conley}, {Le Borgne}, {Carlberg}, {Guy}, {Balam}, {Basa}, {Fouchez}, {Hook},
  {Hsiao}, {Neill}, {Pain}, {Perrett}, \& {Pritchet}}]{howell06}
{Howell}, D.~A., {Sullivan}, M., {Nugent}, P.~E., {Ellis}, R.~S., {Conley},
  A.~J., {Le Borgne}, D., {Carlberg}, R.~G., {Guy}, J., {Balam}, D., {Basa},
  S., {Fouchez}, D., {Hook}, I.~M., {Hsiao}, E.~Y., {Neill}, J.~D., {Pain}, R.,
  {Perrett}, K.~M., \& {Pritchet}, C.~J. 2006, \nat, 443, 308

\bibitem[{{Iben} \& {Tutukov}(1984)}]{iben84}
{Iben}, Jr., I. \& {Tutukov}, A.~V. 1984, \apjs, 54, 335

\bibitem[{{Iben} \& {Tutukov}(1986)}]{iben86}
---. 1986, \apj, 311, 753

\bibitem[{{Lee} \& {Ramirez-Ruiz}(2007)}]{lee07}
{Lee}, W.~H. \& {Ramirez-Ruiz}, E. 2007, New Journal of Physics, 9, 17

\bibitem[{{Lor{\'e}n-Aguilar} {et~al.}(2009){Lor{\'e}n-Aguilar}, {Isern}, \&
  {Garc{\'{\i}}a-Berro}}]{loren09}
{Lor{\'e}n-Aguilar}, P., {Isern}, J., \& {Garc{\'{\i}}a-Berro}, E. 2009, \aap,
  500, 1193

\bibitem[{{Marsh} {et~al.}(2004){Marsh}, {Nelemans}, \& {Steeghs}}]{marsh04}
{Marsh}, T.~R., {Nelemans}, G., \& {Steeghs}, D. 2004, \mnras, 350, 113

\bibitem[{{Monaghan}(2005)}]{monaghan05}
{Monaghan}, J.~J. 2005, Reports on Progress in Physics, 68, 1703

\bibitem[{{Monaghan} \& {Gingold}(1983)}]{monaghan83}
{Monaghan}, J.~J. \& {Gingold}, R.~A. 1983, Journal of Computational Physics,
  52, 374

\bibitem[{{Morris} \& {Monaghan}(1997)}]{morris97}
{Morris}, J.~P. \& {Monaghan}, J.~J. 1997, J. Comput. Phys., 136, 41

\bibitem[{{Motl} {et~al.}(2007){Motl}, {Frank}, {Tohline}, \&
  {D'Souza}}]{motl07}
{Motl}, P.~M., {Frank}, J., {Tohline}, J.~E., \& {D'Souza}, M.~C.~R. 2007,
  \apj, 670, 1314

\bibitem[{{Motl} {et~al.}(2002){Motl}, {Tohline}, \& {Frank}}]{motl02}
{Motl}, P.~M., {Tohline}, J.~E., \& {Frank}, J. 2002, \apjs, 138, 121

\bibitem[{{Napiwotzki}(2009)}]{napiwotzki09}
{Napiwotzki}, R. 2009, Journal of Physics Conference Series, 172, 012004

\bibitem[{{Nelemans} {et~al.}(2001{\natexlab{a}}){Nelemans}, {Portegies Zwart},
  {Verbunt}, \& {Yungelson}}]{nelemans01b}
{Nelemans}, G., {Portegies Zwart}, S.~F., {Verbunt}, F., \& {Yungelson}, L.~R.
  2001{\natexlab{a}}, \aap, 368, 939

\bibitem[{{Nelemans} {et~al.}(2001{\natexlab{b}}){Nelemans}, {Yungelson},
  {Portegies Zwart}, \& {Verbunt}}]{nelemans01a}
{Nelemans}, G., {Yungelson}, L.~R., {Portegies Zwart}, S.~F., \& {Verbunt}, F.
  2001{\natexlab{b}}, \aap, 365, 491

\bibitem[{{New} \& {Tohline}(1997)}]{new97}
{New}, K.~C.~B. \& {Tohline}, J.~E. 1997, \apj, 490, 311

\bibitem[{{Paczy{\'n}ski}(1967)}]{paczynski67}
{Paczy{\'n}ski}, B. 1967, Acta Astronomica, 17, 287

\bibitem[{{Paczy{\'n}ski}(1971)}]{paczynski71}
---. 1971, \araa, 9, 183

\bibitem[{{Pakmor} {et~al.}(2010){Pakmor}, {Kromer}, {R{\"o}pke}, {Sim},
  {Ruiter}, \& {Hillebrandt}}]{pakmor10}
{Pakmor}, R., {Kromer}, M., {R{\"o}pke}, F.~K., {Sim}, S.~A., {Ruiter}, A.~J.,
  \& {Hillebrandt}, W. 2010, \nat, 463, 61

\bibitem[{{Rasio} \& {Shapiro}(1992)}]{rasio92}
{Rasio}, F.~A. \& {Shapiro}, S.~L. 1992, \apj, 401, 226

\bibitem[{{Rasio} \& {Shapiro}(1994)}]{rasio94}
---. 1994, \apj, 432, 242

\bibitem[{{Rasio} \& {Shapiro}(1995)}]{rasio95}
---. 1995, \apj, 438, 887

\bibitem[{{Rosswog}(2005)}]{rosswog05a}
{Rosswog}, S. 2005, \apj, 634, 1202

\bibitem[{{Rosswog}(2007)}]{rosswog07a}
---. 2007, \mnras, 376, L48

\bibitem[{{Rosswog}(2009)}]{rosswog09b}
---. 2009, New Astronomy Reviews, 53, 78

\bibitem[{{Rosswog} {et~al.}(2008){Rosswog}, {Ramirez-Ruiz}, {Hix}, \&
  {Dan}}]{rosswog08b}
{Rosswog}, S., {Ramirez-Ruiz}, E., {Hix}, W.~R., \& {Dan}, M. 2008, Computer
  Physics Communications, 179, 184

\bibitem[{{Rosswog} {et~al.}(2004){Rosswog}, {Speith}, \& {Wynn}}]{rosswog04b}
{Rosswog}, S., {Speith}, R., \& {Wynn}, G.~A. 2004, \mnras, 351, 1121

\bibitem[{{Ruiter} {et~al.}(2010){Ruiter}, {Belczynski}, {Benacquista},
  {Larson}, \& {Williams}}]{ruiter10}
{Ruiter}, A.~J., {Belczynski}, K., {Benacquista}, M., {Larson}, S.~L., \&
  {Williams}, G. 2010, \apj, 717, 1006

\bibitem[{{Saio} \& {Jeffery}(2000)}]{saio00}
{Saio}, H. \& {Jeffery}, C.~S. 2000, \mnras, 313, 671

\bibitem[{{Segretain} {et~al.}(1997){Segretain}, {Chabrier}, \&
  {Mochkovitch}}]{segretain97}
{Segretain}, L., {Chabrier}, G., \& {Mochkovitch}, R. 1997, \apj, 481, 355

\bibitem[{{Silverman} {et~al.}(2011){Silverman}, {Ganeshalingam}, {Li},
  {Filippenko}, {Miller}, \& {Poznanski}}]{silverman11}
{Silverman}, J.~M., {Ganeshalingam}, M., {Li}, W., {Filippenko}, A.~V.,
  {Miller}, A.~A., \& {Poznanski}, D. 2011, \mnras, 410, 585

\bibitem[{{Solheim}(2010)}]{solheim10}
{Solheim}, J. 2010, \pasp, 122, 1133

\bibitem[{{Timmes} \& {Swesty}(2000)}]{timmes00a}
{Timmes}, F.~X. \& {Swesty}, F.~D. 2000, \apjs, 126, 501

\bibitem[{{Webbink}(1984)}]{webbink84}
{Webbink}, R.~F. 1984, \apj, 277, 355

\bibitem[{{Yoon} {et~al.}(2007){Yoon}, {Podsiadlowski}, \& {Rosswog}}]{yoon07a}
{Yoon}, S., {Podsiadlowski}, P., \& {Rosswog}, S. 2007, \mnras, 380, 933

\bibitem[{{Yuan} {et~al.}(2007){Yuan}, {Akerlof}, {Miller}, {Quimby}, {Peters},
  {Thorstensen}, {Baltay}, {Bauer}, {Rabinowitz}, {Scalzo}, {Rigaudier},
  {Pecontal}, {Buton}, {Copin}, {Gangler}, {Smadja}, {Tao}, {Antilogus},
  {Bailey}, {Pain}, {Pereira}, {Wu}, {Aldering}, {Aragon}, {Bongard},
  {Childress}, {Loken}, {Nugent}, {Perlmutter}, {Runge}, {Thomas}, {Weaver},
  {Birchall}, {Cough}, {Holtzman}, {Rau}, {Kasliwal}, \& {Gal-Yam}}]{yuan07}
{Yuan}, R.~F., {Akerlof}, C., {Miller}, J., {Quimby}, R., {Peters}, C.,
  {Thorstensen}, J., {Baltay}, C., {Bauer}, A., {Rabinowitz}, D., {Scalzo}, R.,
  {Rigaudier}, G., {Pecontal}, E., {Buton}, C., {Copin}, Y., {Gangler}, E.,
  {Smadja}, G., {Tao}, C., {Antilogus}, P., {Bailey}, S., {Pain}, R.,
  {Pereira}, R., {Wu}, C., {Aldering}, G., {Aragon}, C., {Bongard}, S.,
  {Childress}, M., {Loken}, S., {Nugent}, P., {Perlmutter}, S., {Runge}, K.,
  {Thomas}, R.~C., {Weaver}, B.~A., {Birchall}, D., {Cough}, J., {Holtzman},
  J., {Rau}, A., {Kasliwal}, M., \& {Gal-Yam}, A. 2007, The Astronomer's
  Telegram, 1212, 1

\end{thebibliography}

\end{document}